\documentclass[ reprint, aip, jcp, amsmath, amssymb ]{revtex4-2}

\usepackage{xcolor}
\usepackage{amsmath}
\usepackage{graphicx}
\usepackage{dsfont}
\usepackage{tikz}
\usepackage{titletoc}
\usepackage[ colorlinks=true, linkcolor=blue, citecolor=blue, urlcolor=blue, linktoc=all]{hyperref}

\usepackage[normalem]{ulem}
\definecolor{revgreen}{HTML}{008000}
\newif\ifshowrevisions
\showrevisionstrue
\ifshowrevisions
    
    \DeclareRobustCommand{\rdel}[1]{\ifmmode{\color{revgreen}\text{\sout{$#1$}}}\else{\color{revgreen}\sout{#1}}\fi}
\else
    
    \DeclareRobustCommand{\rdel}[1]{}
\fi

\definecolor{cavity}{HTML}{3cb371}
\definecolor{phonons}{HTML}{800080}

\allowdisplaybreaks[1]

\begin{document}

\title{Competing pathways in polaritonic chemistry and general harmonic environments}

\author{Jonas Vinther}
\email[]{jonas.vinther@nbi.ku.dk}
\affiliation{Department of Computer Science, University of Copenhagen, DK}
\affiliation{Niels Bohr Institute, University of Copenhagen, DK}

\author{Leonardo A. Cunha}
\affiliation{\mbox{Center for Computational Quantum Physics, Flatiron Institute, 162 5th Avenue, New York, New York 10010, USA}}
\affiliation{\mbox{Department of Chemistry and Biochemistry, Bates College, Lewiston, Maine 04240, USA}}

\author{Johannes Flick}
\email[]{jflick@ccny.cuny.edu}
\affiliation{\mbox{Center for Computational Quantum Physics, Flatiron Institute, 162 5th Avenue, New York, New York 10010, USA}}
\affiliation{\mbox{Department of Physics, City College of New York, New York, New York 10031, USA}}
\affiliation{\mbox{Department of Physics, The Graduate Center, City University of New York, New York, New York 10016, USA}}


\begin{abstract}
    Recent experiments suggest that strong light–matter coupling between molecules and cavity modes can significantly modify chemical reaction rates and yields.
    Despite multiple proposed mechanisms, a clear microscopic understanding of how vibrational strong coupling alters reaction kinetics remains incomplete. 
    Here, we employ a general quantum-dynamical framework for analyzing competing reaction pathways in polaritonic chemistry by considering reaction coordinates with multiple relevant transition frequencies and general harmonic environments.    
    We extend on a tensor-network formulation of the hierarchical equations of motion (HEOM) method to compute probability currents between vibrational states, leveraging a concise formulation for effective spectral densities and a controlled compression of Markovian bath modes into system-level jump operators.
    This allows us to decompose the resulting state-to-state probability currents, revealing how cavity-induced changes in individual transitions propagate through multistep reaction networks. We find that these local modifications can alter kinetic bottlenecks and selectively redistribute reaction flux between competing pathways, changing branching ratios under irreversible conditions without substantially affecting the equilibrium yield in the reversible regime. 
    Finally, for coupled reaction-coordinate, spectator-mode, and cavity environments, we find structured, asymmetric rate profiles arising from interference between hybridized decay channels. Together, these results provide a mechanistic framework for understanding 
    pathway-specific chemical reactivity under vibrational strong coupling.
\end{abstract}

\maketitle


\section{Introduction}
Hybrid light-matter states, known as polaritons, have received increased attention partly due to the novel phenomena that can emerge when molecules and materials are embedded in structured electromagnetic environments \cite{xu2026vacuumdressedsuperconductivitynbnobserved, Keren2026-jz}.
In polaritonic chemistry \cite{Ebbesen2023, doi:10.1021/acs.chemrev.2c00895, Xiang2024}, for instance, an ensemble of molecules is strongly coupled to the harmonic modes of an optical cavity. Even in the dark, these strong light-matter interactions have been observed to alter the reaction rate and chemical yield of the molecular processes involved. 
Polaritonic chemistry has evolved from modifying photochemical reaction landscapes under strong coupling~\cite{https://doi.org/10.1002/anie.201107033} to controlling ground-state chemical reactivity through vibrational strong coupling (VSC)~\cite{Shalabney2015-su, https://doi.org/10.1002/anie.201605504}. 
These early demonstrations paved the way to a flurry of subsequent experiments \cite{doi:10.1126/science.aau7742, https://doi.org/10.1002/anie.201905407, https://doi.org/10.1515/nanoph-2019-0340, doi:10.1126/science.ade7147, 10.1039/d1sc04707h, https://doi.org/10.1002/anie.202409528} and form part of a broader effort to exploit polaritonic states to control molecular processes, including intermolecular energy transfer \cite{Coles2014, https://doi.org/10.1002/anie.201703539}, long-range energy transport \cite{https://doi.org/10.1002/adma.202002127}, photochemical stability \cite{doi:10.1126/sciadv.aas9552}, and selective emission from molecular isomers \cite{doi:10.1126/sciadv.abj0997}. A recent review surveys these emerging functionalities \cite{10.1039/d5cc04712a}. 

For ground-state chemical reactivity, experiments have reported a diverse range of responses to VSC.
Reaction suppression has been reported by strongly coupling cavity modes to reactant vibrations\cite{https://doi.org/10.1002/anie.201605504, https://doi.org/10.1002/anie.201915632} or additionally to product or cooperative reactant-solvent modes\cite{doi:10.1126/science.ade7147}. Reaction acceleration has been observed under cooperative coupling of overlapping vibrations of the reacting molecules and the surrounding solvent, both in Fabry–Pérot cavities \cite{https://doi.org/10.1002/anie.201905407, 10.1039/d1sc04707h} and using resonances supported by non-local metasurfaces \cite{https://doi.org/10.1002/anie.202409528}.
Temperature-dependent kinetic studies of both suppressed and accelerated reactions have further revealed compensating changes in apparent activation parameters under cavity coupling\cite{https://doi.org/10.1515/nanoph-2019-0340, 10.1039/d1sc04707h}. 
It should be noted, as a testament to the difficulty of bridging the gap between theory and experiments, that the reproducibility of some reported rate enhancements remains contested \cite{10.1063/5.0077549}, with subsequent analyses identifying optical and kinetic-fitting artifacts that may contribute to discrepancies between studies \cite{https://doi.org/10.1002/nap2.70143}.

In addition to modifying overall reaction rates, VSC has been reported to alter product selectivity. Previous work observed a redistribution of products arising from competing pathways~\cite{doi:10.1126/science.aau7742}. Further studies showed that coupling different vibrations of the same molecule biased the product distribution in opposite directions, even though both coupling conditions slowed the overall reaction~\cite{https://doi.org/10.1002/anie.202013465}. Alongside these changes in kinetic selectivity, shifts in chemical equilibrium have also been reported, with the direction of the shift depending on the coupled vibrational mode~\cite{https://doi.org/10.1002/anie.202002527}.

Together, these findings motivate microscopic descriptions that resolve how coupling at distinct vibrational frequencies affects both molecular transitions and equilibrium populations, such that the conditions under which cavity coupling can control reaction rates and product distributions, can be identified.



In an attempt to interpret this range of observed cavity-modified reactivity, theoretical approaches have drawn on transition-state theory\cite{Truhlar1996-rm} and dynamical barrier-crossing theories\cite{Hanggi1990-yf}, such as Grote-Hynes\cite{Li2021-zg} and Pollak-Grabert-Hänggi theory\cite{Lindoy2022-sf, Mandal2023}. Despite the ability to account for environmental effects, these approaches are fundamentally concerned with the energy barrier between reactant and product populations and their classical implementations do not explicitly resolve transitions between quantized vibrational states. Consequently, some classical models predict rate-modification profiles that are substantially broader than reported experimentally, or whose extrema occur near the barrier frequency rather than the relevant molecular vibrational frequencies \cite{lindoy_quantum_2023}. Comparisons with quantum simulations have further demonstrated limitations of approximate treatments of molecular and cavity dynamics in reproducing resonant rate modifications \cite{10.1063/5.0220908}.

Quantum dynamical treatments have demonstrated that the sharp resonance in VSC-modified reaction rates can be recovered when the molecular, cavity, and environmental degrees of freedom are treated quantum mechanically \cite{lindoy_quantum_2023, ying_resonance_2023, YingTaylorHuo+2024+2601+2615, Ying2024, Vega2025, 10.1063/5.0248419, 10.1063/5.0324649}. Complementary theoretical analyses based on effective environmental descriptions and Fermi's Golden Rule have provided microscopic interpretations of this resonance and its dependence on cavity properties \cite{ying_resonance_2023, YingTaylorHuo+2024+2601+2615, Ying2024, Vega2025}. 
Tensor-network approaches have extended quantum simulations to models of collective effects \cite{LindoyMandalReichman+2024+2617+2633, 10.1063/5.0248419, 10.1063/5.0324649}, 
yet, establishing the connection between such microscopic models and collective experimental systems nevertheless remains an open challenge \cite{annurev:/content/journals/10.1146/annurev-physchem-082624-023219}.
Additional theoretical efforts seek to relate changes in the equilibrium chemical composition to the thermodynamics of the coupled light-matter system \cite{Sun2024-tp}.


An important next step is to determine how frequency-selective modifications of individual transitions propagate through reactions involving distinct reactant and product vibrational spectra or multiple accessible products. 
Furthermore, extending the mechanistic insights provided by Fermi’s golden rule through a nonperturbative analysis of state-to-state probability currents, resolved into system and environmental contributions, would further clarify how individual vibrational transitions translate into changes in overall reaction rates and product distributions.

Here, we expand upon a tensor-network formulation of the hierarchical equations of motion to analyze reaction coordinates with independently adjustable reactant and product well frequencies, including a model with two competing product wells. We combine quantum dynamical simulations with effective spectral densities for general harmonic environments to extract probability currents between vibrational states and decompose these into system and environmental contributions. This allows us to identify how cavity-induced modifications of individual transitions alter kinetic bottlenecks and redistribute reactant consumption between competing pathways. By examining both reaction dynamics and equilibrium populations of the cavity- and bath-coupled system, we distinguish changes in kinetic selectivity from changes in thermodynamic preference.

This paper is organized as follows. Section~\ref{sec:theory} formulates the model of the reaction coordinate and its harmonic environment. 
It also presents the reaction rate, defined in terms of system observables, and the probability currents, which serve as a diagnostic tool for the reaction mechanism.
Section~\ref{sec:methodology} describes our method for performing the numerically exact simulations of the dynamics. The results are presented and discussed in {Sec.}~\ref{sec:results}, where we address potential energy surfaces with multiple relevant transition frequencies and competing pathways and harmonic environments with interfering decay channels. 
Our conclusions are summarized in Sec.~\ref{sec:conclusion}.

\begin{figure}[h]
    \centering
    \includegraphics[angle=-90, trim=1cm 0cm 1cm 0cm, clip, width=0.8\linewidth]{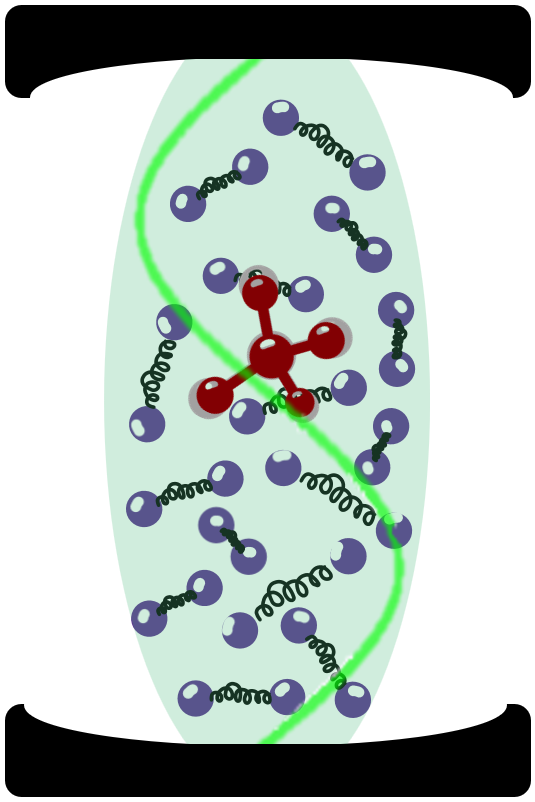}
    \caption{
    An example setup consists of a single molecule (red), comprising a reaction coordinate, and may also include spectator vibrational modes. The molecule is embedded in a solvent environment represented by a phonon bath (purple). The molecular and solvent degrees of freedom are confined within an optical cavity, represented by a single cavity mode (green).}
    \label{fig:SchematicSetup}
\end{figure}

\section{\label{sec:theory}Theory}
\subsection{Model and effective harmonic environment}
We employ a complete quantum description of a reaction coordinate (the system) on a one dimensional potential energy surface (PES) in a harmonic environment (the bath). A schematic for the physical setup is presented in Fig.~\ref{fig:SchematicSetup}.

The microscopic model that we consider, is given by the Pauli-Fierz Hamiltonian in the length gauge \cite{10.1021/acs.chemrev.2c00788, lindoy_quantum_2023, Vega2025}:
\begin{align}
\label{eq:hamiltonian}
    H &= H_R + H_Q + H_c + H^\text{CL}_R + H^\text{CL}_Q + H^\text{CL}_c\\
    H_R &= \frac{\hat{P}^2}{2M} + V(\hat{R})\\
    H_Q &= \frac{\hat{P}_Q^2}{2} + \frac{\omega_Q^2}{2}\left(\hat{Q} + \frac{c_Q}{\omega_Q^2}\hat{R}\right)^2\\
    H_c &= \frac{\hat{p}_c^2}{2} + \frac{\omega_c^2}{2}\left(\hat{q}_c +\sqrt{\frac{2}{\omega_c}}\eta_c\left(\eta_R\hat{R} + \eta_Q\hat{Q}\right)\right)^2
\end{align}
This Hamiltonian describes a reaction coordinate $\hat{R}$, coupled to the displacement $\hat{Q}$ of a spectator mode, both of which are coupled to a single cavity mode with displacement $\hat{q}_c$. Furthermore, each of these coordinates are assumed to experience their own harmonic environment, as described by the Caldeira-Leggett model of dissipation \cite{CALDEIRA1983374}, via the inclusion of $H^\text{CL}_{(R,Q,c)}$ in Eq.~\ref{eq:hamiltonian}. These are of the form:
\begin{equation}
    H^\text{CL}_R = \sum_{{j}}\frac{\hat{p}_{{j}}^2}{2}+\frac{\omega_{{j}}^2}{2}\left(\hat{x}_{{j}}-\frac{c_{{j}}}{\omega_{{j}}^2}\hat{R}\right)^2
\end{equation}

The reaction coordinate (RC) is the only anharmonic degree of freedom in the Hamiltonian, Eq.~\ref{eq:hamiltonian}, and all the other harmonic degrees of freedom may be referred to as the bath. Consequently, if we assume that the RC is initially in a product state with the bath and that the bath is in a thermal state, the influence of the bath is fully described by the bath correlation function (BCF) \cite{FEYNMAN1963118, 10.1093/acprof:oso/9780199213900.001.0001}:
\begin{align}
    C(t) &= \int_{-\infty}^\infty\frac{d\omega}{2\pi}S_\beta(\omega)e^{-i\omega t}
    =\int_{-\infty}^\infty\frac{d\omega}{2\pi}\frac{2J(\omega)e^{-i\omega t}}{1-e^{-\beta\omega}}
\end{align}
where $S_\beta(\omega)$ is the noise spectrum, $\beta$ is the inverse temperature and $J(\omega)$ is the spectral density of the full harmonic environment.

\begin{figure}[h]
    \centering
    \includegraphics[width=0.8\linewidth]{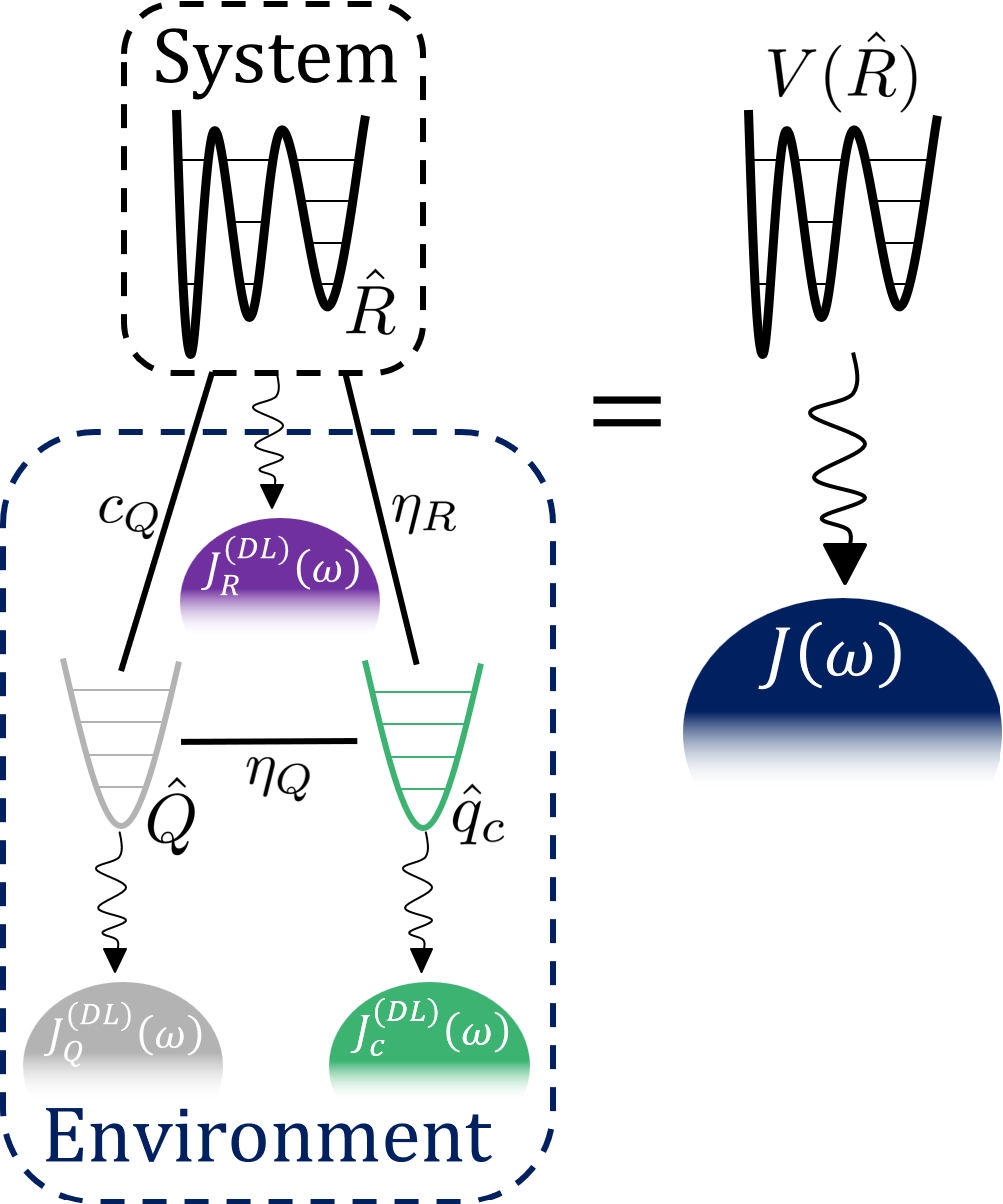}
    \caption{
    Schematic for the efficient modeling of the harmonic degrees of freedom in terms of the spectral density for the full harmonic environment. Supplementary Section~\ref{sec:GeneralSDDerivation} derives and presents the formula for arbitrary harmonic environments. The system refers to the reaction coordinate $\hat{R}$, here visualized in terms of the PES $V(\hat{R})$ for the triple well system treated in Sec.~\ref{subsec:competingpathways}.
    The spectator mode has coordinate $\hat{Q}$ and the cavity $\hat{q}_c$. The couplings between them are denoted by $c_Q$, $\eta_R$ and $\eta_Q$ and $J^{(DL)}_{(R,Q,c)}(\omega)$ refer to the individual baths which we choose to be of the Drude-Lorentz form.
    }
    \label{fig:ModelDichotomy}
\end{figure}

In Supplementary Section~\ref{sec:GeneralSDDerivation} we provide a succinct formula for $J(\omega)$ for arbitrary harmonic environments. It relies on the inversion of a matrix that can be formed directly from the parameters of the Hamiltonian, thus mitigating the algebraic labor that earlier methods required \cite{10.1063/1.449017, Vega2025}. 

Figure~\ref{fig:ModelDichotomy} schematically shows the abstraction from the microscopic harmonic environment in question to the full spectral density $J(\omega)$.
With our approach, Eq.~\ref{eq:hamiltonian} immediately leads to:
\begin{align}
\label{eq:spectraldensity}
J(\omega) &= J_R^{(DL)}(\omega) + \operatorname{Im}\left[\pmb{c}^TM^{-1}(\omega) \pmb{c}\right]\\
    \pmb{c} &=
    \begin{pmatrix}
        \sqrt{2\omega_c^3}\eta_R\eta_c\\
        c_Q + 2\omega_c\eta_c^2\eta_R\eta_Q
    \end{pmatrix}\\
    M(\omega) &=
    \begin{pmatrix}
        \Gamma_c(\omega) & -\sqrt{2\omega_c^3}\eta_Q\eta_c \\
        -\sqrt{2\omega_c^3}\eta_Q\eta_c & \Gamma_Q(\omega)- 2\omega_c\eta_c^2\eta_Q^2\\
    \end{pmatrix}\\
    \Gamma_{(c,Q)}(\omega)&= \omega^2 - \omega_{(c,Q)}^2-\left(i+\frac{\omega}{\gamma_{(c,Q)}}\right)J_{(c,Q)}^{(DL)}(\omega)
\end{align}
For simplicity, we have assumed that the spectral densities of the individual harmonic environments, described by $H^\text{CL}_{(R,Q,c)}$, are all of the Drude-Lorentz form:
\begin{equation}
    J^{(DL)}_n(\omega)=\frac{2\lambda_n\gamma_n\omega}{\omega^2+\gamma_n^2}
    \quad\text{with}\quad
    n\in\{R,Q,c\}
\end{equation}
Finally, we consider the cavity bath to be Markovian, and consider the limit $\gamma_c\rightarrow\infty$ with $\frac{2\lambda_c}{\gamma_c}\rightarrow\alpha(\omega_c)=(1-e^{-\beta\omega_c})/\tau_c$~\cite{ying_resonance_2023}.

\begin{figure}[h]
    \centering
    \includegraphics[width=1\linewidth,
    trim=0cm 0.2cm 0.2cm 0cm,  
    clip
    ]{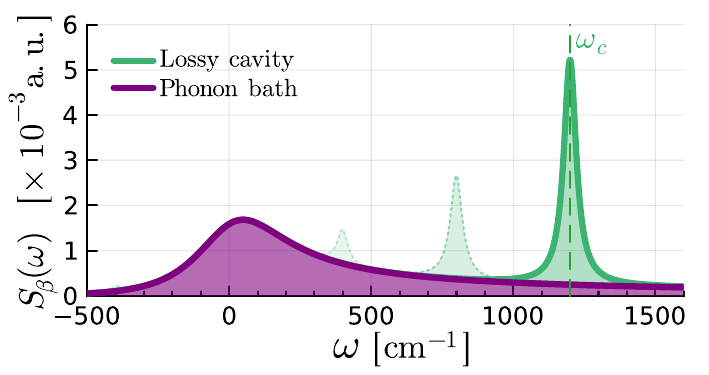}
    \caption{The noise spectrum when there is no spectator mode ($c_Q=\eta_Q=0$). The effect of the cavity is captured by the peak centered around $\omega_c$ (Eq.~\ref{eq:SimpleSD}). The height of this peak scales with $\sim\omega_c^2$, as dictated by the Pauli-Fierz Hamiltonian. For visualizations of noise spectra with the full spectral density of Eq.~\ref{eq:spectraldensity}, see Fig.~\ref{fig:NoiseSpectrums}{.}}
    \label{fig:SimpleNoiseSpectrum}
\end{figure}

In the case of no spectator mode ($c_Q=\eta_Q=0$), the harmonic environment reduces to that of the phonon bath, modeled by the spectral density $J_R^{(DL)}(\omega)$, plus the lossy cavity \cite{ying_resonance_2023}:
\begin{equation}
\label{eq:SimpleSD}
    J(\omega) = J_R^{(DL)}(\omega) +\frac{2\omega_c^3\eta_R^2\eta_c^2\alpha(\omega_c)\omega}{(\omega^2-\omega_c^2)^2+\alpha^2(\omega_c)\omega^2}.
\end{equation}
 The resulting noise spectrum is visualized for three different cavity frequencies in Fig.~\ref{fig:SimpleNoiseSpectrum}.

As alluded to in Fig.~\ref{fig:ModelDichotomy}, the cavity-modified kinetics are essentially governed by two scalar {functions}: the potential energy surface $V(\hat{R})$ and the effective spectral density for the full harmonic environment $J(\omega)$.
The former accounts for the specifics of the reaction while the latter contains all the harmonic degrees of freedom such as lossy cavity modes, spectator modes and the vibrations of the solvent.
Thus, the simple spectral density (Eq.~\ref{eq:SimpleSD}) will be used to isolate the effects of the cavity when studying PES driven effects (Sec.~\ref{subsec:PES-driven effects}). The full spectral density (Eq.~\ref{eq:spectraldensity}) will be applied to study how the alignment between the reaction coordinate and spectator mode polarities, encoded in $\eta_R$ and $\eta_Q$ respectively, affects the reaction rate profile (Sec.~\ref{subsec:Environment-driven effects}).

We parameterize the potential energy surface $V(\hat{R})$ in terms of piecewise parabolic functions, providing a convenient framework for studying independent well frequencies and competing pathways. Explicit expressions for $V(\hat{R})$ are found in Supplementary Section~\ref{sec:RxNPES}. While the simulations are carried out in the eigenbasis of the system Hamiltonian $H_R$, the analysis is performed in a basis where select eigenstates have been hybridized to construct localized states. Our approach for constructing the rotation between these two bases relies on diagonalizing $\hat{R}$ in the subspace of the selected states (see Supplementary Section~\ref{sec:StateLocalization}). 



\subsection{{Observables and mechanistic analysis}}
\label{subsec:Observables}

The reaction rate is defined in terms of expectation values of projection operators onto a reactant region and product regions \cite{10.1063/1.445581, 10.1063/1.2772265}. We define these explicitly, in terms of the eigenstates $\hat{R}|R\rangle = R|R\rangle$.
The double well PESs are centered around $R=0$ with a reactant region in the left well $\hat{P}_\text{left}=\int_{-\infty}^0dR|R\rangle\langle R|$ and a product region in the right well $\hat{P}_\text{right}=\mathds{1}-\hat{P}_\text{left}$.
The forward reaction rate $k_\rightarrow$ is obtained under the ansatz of the rate equation:
\begin{equation}
\label{eq:RateEquation}
        \dot{P}_\text{left}(t)=-P_\text{left}(t)k_\rightarrow + (1 - P_\text{left}(t))k_\leftarrow
\end{equation}
by obtaining the time-series $P_\text{left}(t) = \text{Tr}\left[\hat{P}_\text{left}\rho(t)\right]$ and solving for $k_\rightarrow$. The backwards reaction rate is denoted by $k_\leftarrow$ and its proportionality to $k_\rightarrow$ is linked to the equilibrium distribution of the reaction coordinate $\frac{k_\rightarrow}{k_\leftarrow}=\frac{P^{eq}_\text{right}}{P^{eq}_\text{left}}$, with e.g. $P^{eq}_\text{left}=\lim_{t\rightarrow\infty}P_\text{left}(t)=\frac{\text{Tr}\left[\hat{P}_\text{left}e^{-\beta H}\right]}{\text{Tr}\left[e^{-\beta H}\right]}$ \cite{10.1116/5.0073853}.
In the case of PESs with more than two wells, we define a projection operator for each well and solve for the reaction rate between neighboring wells. Further details are found in Supplementary Section~\ref{sec:ReactionRateAnalysis}. 

In order to study the dynamics in more detail, we decompose the rate of the population of the vibrational states in terms of probability currents associated with the transitions from state $n$ to state $m$ \cite{PhysRevE.93.012128, schumacher2016probabilitycurrentthermodynamicsopen}:
\begin{equation}
\label{eq:probcurrents}
    \dot{\rho}_{nn}(t) = -\sum_m j_{n\rightarrow m}(t)
\end{equation}
We further decompose each current $j_{n\rightarrow m}$ in terms of the contributions from the system Hamiltonian and contributions from the harmonic environment. We will mainly use this diagnostic tool in the case of the simple spectral density Eq.~\ref{eq:SimpleSD}, such that we make the distinction between the following three contributions:
\begin{equation}
\label{eq:probabilitycurrents}
    j_{n\rightarrow m}(t) = j_{n\rightarrow m}^\text{Tunneling}(t) + j_{n\rightarrow m}^\text{Solvent}(t) + j_{n\rightarrow m}^\text{Cavity}(t)
\end{equation}
This allows us to directly interpret how the cavity affects the quantum dynamics and furthermore the role of the tunneling dynamics and the phonon bath. 

If a basis of localized well states is used, we can relate the rate equation to the total probability current that goes from the reactant region to the product region (Supplementary Section~\ref{sec:ProbCurrentsAsReactionRate}):
\begin{equation}
\label{eq:ReactionRateFromCurrents}
    \sum_n ^\text{reactant} \sum_m^\text{product} j_{n\rightarrow m}(t)\approx k_\rightarrow P_\text{left}(t) - k_\leftarrow P_\text{right}(t)
\end{equation}
where the sum $n$ runs over states localized in the left well and $m$ indexes states localized in the right well. At early times where $P_\text{left}(t)\approx1$ and $P_\text{right}(t)\approx0$, the forward reaction rate is directly related to the total probability current going from reactant to product. Eq.~\ref{eq:ReactionRateFromCurrents} also extends naturally to PESs with more than two wells.

In the perturbative limit, the bath contributions to Eq.~\ref{eq:probcurrents} can be reduced to the Pauli-equation{:}
\begin{equation}
\label{eq:PerturbativeCurrents}
    j_{n\rightarrow m}(t) \sim \Gamma_{n\rightarrow m} \rho_{nn}(t) - \Gamma_{m\rightarrow n} \rho_{mm}(t)
\end{equation}
with the transition rates $\Gamma_{n\leftrightharpoons m}$ given by Fermi's Golden Rule (Supplementary Section~\ref{sec:FGR}).
These two reductions (Eq.~\ref{eq:ReactionRateFromCurrents} \&~\ref{eq:PerturbativeCurrents}) provide a simple perspective on how to relate the cavity modified noise spectrum to the cavity modified reaction rates, in a perturbative setting. Such a perspective has already been successfully applied in Refs.~\citenum{10.1063/1.5116800, ying_resonance_2023, YingTaylorHuo+2024+2601+2615, 10.1063/5.0231396, Ying2024, Vega2025}.
For the definition of the currents (Eq.~\ref{eq:probabilitycurrents}) in terms of the master equation presented in the next section (Eq.~\ref{eq:FPHEOM}), see Supplementary Section~\ref{sec:RateBetweenStates}.

\section{\label{sec:methodology}Methodology}
The time-series for Eq.~\ref{eq:RateEquation} is obtained from a numerically exact simulation of the open quantum system dynamics using the methodology known under the names of HEOM \cite{10.1063/5.0011599}, pseudomode \cite{PhysRevResearch.2.043058, PhysRevB.110.195148} and QD-MESS \cite{PhysRevLett.129.230601, xu2023universalframeworkquantumdissipationminimally, w3nw-hbjc}, among others \cite{10.1063/1.3159671, 10.1063/1.3490188, 10.1063/1.4866769, Nazir2018}.
The main principle behind these methods is to use Markovian embeddings to emulate the effect of a macroscopic bath through a reduced, few-body harmonic environment.

For the simulations in this study, the equations of {motion} for the few-body system $\varrho(t)$ {are}:
\begin{align}
    \dot{\varrho}(t) =&  
    -i\left[\hat{H}_R+\lambda \hat{R}^2,\varrho(t)\right] + \hat{L}\varrho(t)\hat{L}^\dagger - \frac{1}{2}\left\{\hat{L}^\dagger \hat{L},\varrho(t)\right\} \nonumber\\
    + &\sum_{k\in {K}}\Bigg\{
    -z_k\hat{a}_k^\dagger\hat{a}_k\varrho(t) - i\sqrt{d_k^*}\hat{R}\varrho(t) \hat{a}^\dagger_k \nonumber\\&\quad\quad\quad\quad-i\sqrt{d_k}\left(\hat{a}_k + \hat{a}_k^\dagger\right)\hat{R}\varrho(t)+ H.C.\Bigg\}
    \label{eq:FPHEOM}
\end{align}
where $H.C.$ refers to the hermitian conjugate of everything inside the brackets.
$\hat{a}_k$ is the lowering operator of the $k$'th auxiliary mode and $z_k$ and $d_k$ are complex parameters that {characterize} the effective few-body harmonic environment.
We have introduced the system jump operator $\hat{L}$, in order to more efficiently emulate the environment, as will be elaborated upon in the following. The so-called reorganization energy $\lambda$ is given by:
\begin{equation}
\label{eq:HEOMReorganizationEnergy}
    \lambda = -\operatorname{Im}\sum_{k\in {K}}\frac{d_k}{z_k}
\end{equation}
and the initial state and system density matrix is given by:
\begin{align}
    \varrho(0) =& \rho_S(0)\otimes \overbrace{|0\rangle\langle 0|\otimes\hdots \otimes |0\rangle\langle 0|}^{|{K}|\text{ times}} =: \rho_S(0)\otimes^{K} |0\rangle\langle 0|\\
    \rho_S(t) =&\text{Tr}_{K}\left[\mathds{1}\otimes^{K}|0\rangle\langle 0|\varrho(t)\right]
\end{align}
That is, the auxiliary modes are initialized in the vacuum and to obtain the reduced density matrix of the system $\rho_S(t)$, the auxiliary modes should be projected onto their vacuum.
While ${K}$ is the set that labels the effective bath modes of Eq.~\ref{eq:FPHEOM}, consider a similar effective harmonic environment whose modes are labeled by the larger set $K_\text{fit}$. 
These give rise to a bath correlation function (BCF) that is a sum of complex exponentials \cite{PhysRevLett.129.230601}. If this effective BCF resembles the original BCF:
\begin{equation}
\label{eq:BCFSumOfExponentials}
    C(t) \approx \sum_{k \in K_\text{fit}} d_k e^{-z_k t}
    \quad \quad
    d_k, z_k\in \mathds{C}
\end{equation}
the few-body system successfully emulates the macroscopic environment with deviations in observables being controlled by the decomposition in Eq.~\ref{eq:BCFSumOfExponentials} \cite{PhysRevLett.118.100401, huang2025unifiederrorboundsperturbations, Liu2025errorboundsopen}, barring errors from numerically solving the master equation. 
Thus, the starting point of our method is to decompose the original BCF to numerical precision, as a sum of complex exponentials. This is achieved using the AAA algorithm \cite{doi:10.1137/23M1570508, Driscoll2025-sm}. Details and general considerations regarding the method or fitting procedure can be found in Supplementary Section~\ref{sec:HEOMDetails}.
Once the $K_\text{fit}$ pairs of $z_k$ and $d_k$ have been obtained, we reduce $K_\text{fit}$ to the smaller set ${K}$ in the following manner.

\subsubsection{Controlled compression of Markovian bath modes}
Modern introductions to HEOM-type master equations, such as Eq.~\ref{eq:FPHEOM}, generally do not include system jump operators $\hat{L}$~\cite{10.1063/5.0011599, PhysRevResearch.2.043058, PhysRevB.110.195148, PhysRevLett.129.230601, xu2023universalframeworkquantumdissipationminimally, w3nw-hbjc}. 
The inclusion of $\hat{L}$ in Eq.~\ref{eq:FPHEOM} {represents} our novel approach to compress the number of effective bath modes needed to accurately simulate the open system dynamics, in a controllable fashion. 
Essentially, some of the $K_\text{fit}$ harmonic oscillators resemble a Markovian bath and simulating them as an explicit degree of freedom is not necessary. 
We distinguish our approach from earlier proposals that utilize adiabatic elimination~\cite{Ishizaki2005-kr, Fay2022-bc, PhysRevA.86.012126, w3nw-hbjc} to address the short-comings of HEOM in computationally challenging regimes.
Here, we utilize the rigorous analysis underlying the Universal Lindblad Equation \cite{PhysRevB.102.115109}, to accurately and controllably account for the Markovian bath modes via the system jump operator:
\begin{align}
    \hat{L} &= \sum_{n,m}\sqrt{S_M(E_m - E_n)}R_{nm}|n\rangle\langle m|\\
    S_M(\omega) &= \sum_{k\in K_\text{M}}2\operatorname{Re}\left(\frac{d_k}{z_k-i\omega}\right)
\end{align}
In Supplementary Section~\ref{sec:EfficientMarkovCorrection}, we show how to control the error introduced by this modeling technique, including additional assumptions that simplify the implementation, such as using the eigendecomposition of $\hat{H}_R+\lambda \hat{R}^2=\sum_n E_n|n\rangle\langle n|$ for the calculation of $\hat{L}$. 
$K_\text{M}$ {denotes} the set of auxiliary bath modes that are deemed to be Markovian \textit{a priori}. For brevity, we quote the necessary but insufficient condition, that serves as a guiding principle for assigning bath modes to $K_\text{M}$:
\begin{equation}
\label{eq:MarkovianCondition}
    \sum_{k\in K_\text{M}}\frac{|d_k|}{\operatorname{Re}(z_k)^2} \ll 1
\end{equation}
This defines the set ${K}=K_\text{fit}\setminus K_\text{M}$ used in Eq.~\ref{eq:FPHEOM}. In terms of the BCF:
\begin{equation}
C(t)\approx\sum_{k\in K_\text{fit}} d_ke^{-z_kt}=\overbrace{\sum_{k\in K_\text{M}} d_ke^{-z_kt}}^\text{Markovian bath}+\overbrace{\sum_{k\in K}d_ke^{-z_kt}}^\text{Residual bath}
\end{equation}
Thus, we may fit the BCF to numerical precision, while compressing the Liouville space in Eq.~\ref{eq:FPHEOM} for the ensuing numerical simulations. 
We illustrate the accuracy of our implementation by comparing the reaction rates obtained from the compressed phonon bath to the uncompressed simulation. ${|}K_\text{fit}{|}=19$ is reduced to ${|}{K}{|}=2$ or $1$ with a relative error in the extracted reaction rate of $0.1\%$ and $1.2\%$, in line with the small value of the numerically calculated error bound summarized in  Supplementary Table~\ref{tab:MarkovianCompressionSummary}.
The dynamical simulations are carried out using the tensor network algorithm TDVP \cite{PhysRevLett.107.070601} implemented in ITensor \cite{ITensor, ITensor-r0.3}.

\section{\label{sec:results}Results and Discussions}

The model for the physical system is primarily characterized by the two scalar functions $V(\hat{R})$, for the PES, and $J(\omega)$, for the effective spectral density. 
Therefore, the results section is divided into two parts, which aim to elucidate the role of each in the resulting reaction kinetics.

\subsection{PES-driven effects}
\label{subsec:PES-driven effects}
In the following, the environment will be assumed to consist of a phonon bath and a lossy cavity mode (Eq.~\ref{eq:SimpleSD}). We utilize our parameterization of the PES to control and add additional transition frequencies to the reaction pathway. The probability currents between vibrational states, decomposed and assigned to their different contributors (Eq.~\ref{eq:probabilitycurrents}), will then be used to analyze the dynamics.

\subsubsection{Symmetric double well potential}
\label{subsec:Benchmark}
\begin{center}
\begin{figure}[b]
    \centering
    \includegraphics[width=1.0\linewidth,
    trim=0.0cm 1.2cm 1.1cm 1cm,  
    clip
    ]{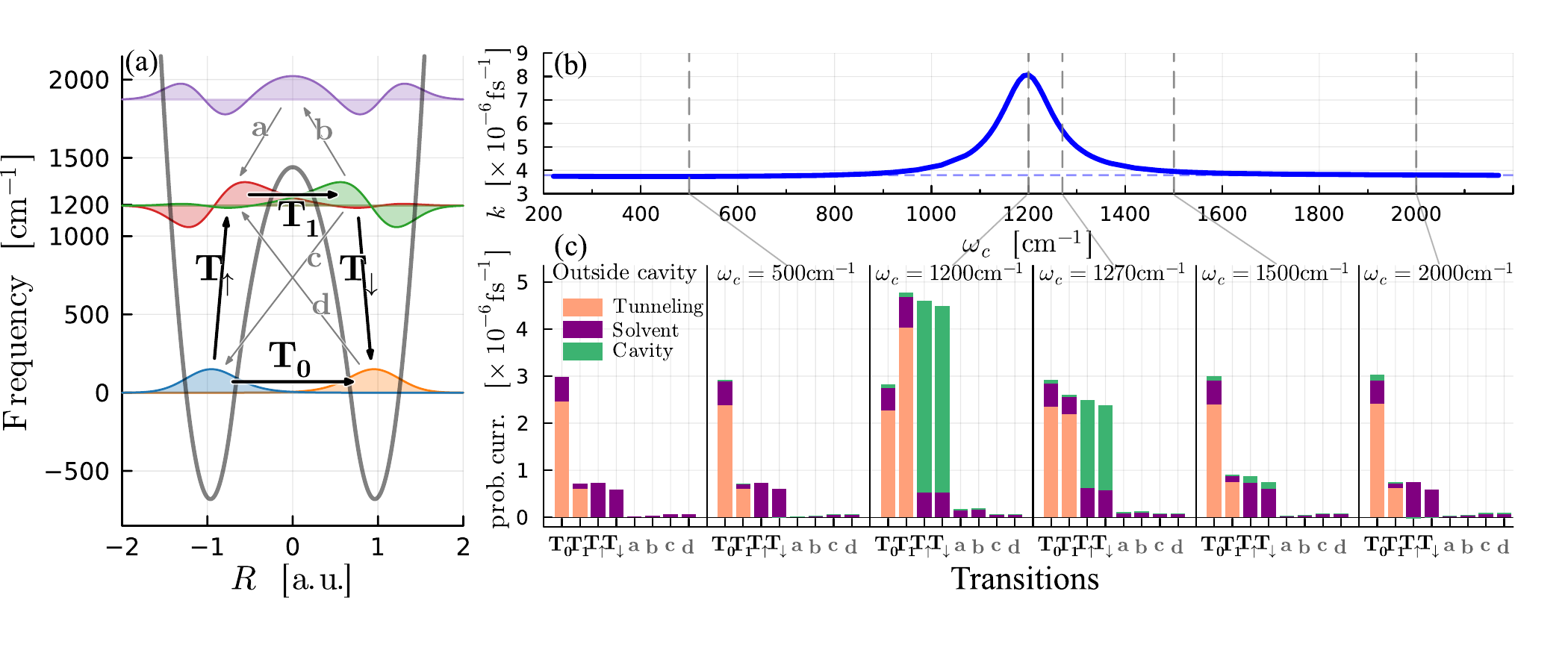}
    \caption{\textbf{(a)} The wavefunction of the localized states of the reaction coordinate visualized alongside the potential with the y-axis providing the energy difference to the ground state. The transitions between these states are labeled and visualized by arrows.
    \textbf{(b)} Reaction rate profile when varying the cavity frequency compared to the out-of-cavity reaction rate (horizontal dashed line). Due to the symmetry of the PES, the forward and backward reaction rates are equal $k:=k_\rightarrow=k_\leftarrow${.}
    \textbf{(c)} The probability current at $t=6\,\text{ps}$ for the labeled transitions as a function of the cavity frequency. The probability currents are decomposed in terms of contributions from the system Hamiltonian (tunneling) and from coupling to bath degrees of freedom (solvent and cavity).
    }
    \label{fig:EstablishMethod}
\end{figure}
\end{center}

We will begin by considering the simple symmetric double well, whose PES and vibrational states are shown in Fig.~\hyperref[fig:EstablishMethod]{\ref*{fig:EstablishMethod}a}.
Outside the cavity, the reaction coordinate only interacts with the phonon bath. The coupling strength between the reaction coordinate and the solvent places the reaction near the crossover between the tunneling-suppressed and energy-diffusion-limited regimes.  (Supplementary Section~\ref{sec:KramersTurnover}). Tunneling between the wells has almost been minimized by the solvent and the thermal fluctuations it provides to the reaction is minimal.

Still, as seen in Fig.~\hyperref[fig:EstablishMethod]{\ref*{fig:EstablishMethod}c}, the reaction outside the cavity is mainly driven by tunneling. Notice that the reaction rate, denoted by the horizontal line in Fig.~\hyperref[fig:EstablishMethod]{\ref*{fig:EstablishMethod}b}, is approximately given by the net current going across the barrier (Eq.~\ref{eq:ReactionRateFromCurrents}), which in this case is the sum of the probability currents for the $\text{T}_0$ and $\text{T}_1$ transitions\footnote{The same holds for the cavity-modified reaction rates, although, for the symmetric double well $\sum_n ^\text{reactant} \sum_m^\text{product} j_{n\rightarrow m}(t)\approx ke^{-2kt}$ which for the fasted dynamics ($k\approx8\times10^{-6}\text{fs}^{-1}$) evaluates to $0.9085k$ at $t=6\text{ps}$}. 
The transitions labeled $\text{a}$, $\text{b}$, $\text{c}$ and $\text{d}$ are included to provide the complete picture, despite their small values. 
Aside from an incoherent tunneling contribution to $\text{T}_0$, the thermal fluctuations of the phonon bath contribute to excitation-assisted tunneling of the reaction coordinate through the potential energy barrier, which is captured by the transitions denoted by $\text{T}_{\uparrow}$ and $\text{T}_{\downarrow}$.

The cavity modified reaction rate is shown in Fig.~\hyperref[fig:EstablishMethod]{\ref*{fig:EstablishMethod}b} and the related probability currents are shown in Fig.~\hyperref[fig:EstablishMethod]{\ref*{fig:EstablishMethod}c} at different values of the cavity frequency, spanning resonant and off-resonant tunings. When resonant with the transition frequency of the wells $\omega_c= 1200\,\text{cm}^{-1}$, the cavity is seen to greatly enhance both the excitation $\text{T}_\uparrow$ and relaxation $\text{T}_\downarrow$ processes. Furthermore, the tunneling through the top of the potential barrier $\text{T}_1$ is accordingly increased while the tunneling among the ground states $\text{T}_0$ is largely unaffected. Finally, when the cavity is tuned off resonance, its effect on the dynamics, as diagnosed by the probability currents, vanishes. 

\onecolumngrid
\newpage
\twocolumngrid

To elucidate these observations, we consider the time-scales of the system. Such an approach will also be helpful when we interpret the results of the coming sections.
First, consider the timescale of the transition labeled $\text{T}_1$. It must be much faster than $\text{T}_0$ since it constitutes tunneling the barrier at a point where it is narrower. Considering the transition $\text{T}_0$ itself provides a sizable portion of the reaction dynamics, despite its longer timescale, the transition $\text{T}_1$ must be far from being a rate-limiting step.
Conversely, the rate at which spectral weight reaches this excited state manifold is much slower. To leading order, we can estimate the rate by Fermi's Golden Rule (Eq.~\ref{eq:PerturbativeCurrents}) such that intra-well decay, due to the phonon bath, occurs at a rate $\sim\frac{2\lambda_R \gamma_R}{M\omega^2}$, while the excitation rate is related by detailed balance $\frac{2\lambda_R \gamma_R}{M\omega^2} e^{-\beta\omega}$ where $\omega$ is the frequency of the well\footnote{At the relevant transition frequencies $\omega >\gamma_R=200\,\text{cm}^{-1}$ the phonon bath spectral density becomes $\frac{2\lambda_R\gamma_R\omega}{\omega^2+\gamma_R^2}\approx\frac{2\lambda_R\gamma_R}{\omega}$ and the transition matrix element within a well is approximated as $|\langle 1|\hat{R}|0\rangle|^2\approx\frac{1}{2M\omega}$ with $|n\rangle$ the harmonic oscillator states. The noise spectrum evaluated at the transition frequency times the matrix element, yields the FGR rate which are the quoted expressions. Alternatively, see Supplementary Section~\ref{sec:FGR}}.
Thus, excitation is the rate-limiting step, whose rate for the given values is of the order\footnote{$\frac{\lambda \gamma}{\omega^2}e^{-\omega/T}\frac{2}{M}\approx\frac{83.6\,\text{cm}^{-1}\times200\,\text{cm}^{-1}}{(1200\,\text{cm}^{-1})^2}e^{-1200\,\text{cm}^{-1}/208.5\,\text{cm}^{-1}}\times 0.045\,\text{fs}^{-1}\approx1.66\times10^{-6}\,\text{fs}^{-1}$} $10^{-6}\,\text{fs}^{-1}$.
As a result, once the cavity is resonant with the molecular transition, the rate is strongly enhanced due to the increased value of the noise spectrum at the transition frequency \cite{ying_resonance_2023}. Effectively, the lossy cavity provides an additional thermalization channel, in close analogy with the Purcell effect \cite{PhysRevLett.123.153001}.




Having established that the probability-current analysis is well aligned with our understanding of this simple scenario, we will now focus on understanding how the complexity and asymmetries in the PES can affect the overall rate.

\subsubsection{Asymmetry in reactant and product well frequencies}
\label{subsec:asymmetric_wells}
\begin{figure}[h]
    \centering
    \includegraphics[width=1.0\linewidth,
    trim=0.1cm 0.33cm 0.38cm 0.02cm,  
    clip
    ]{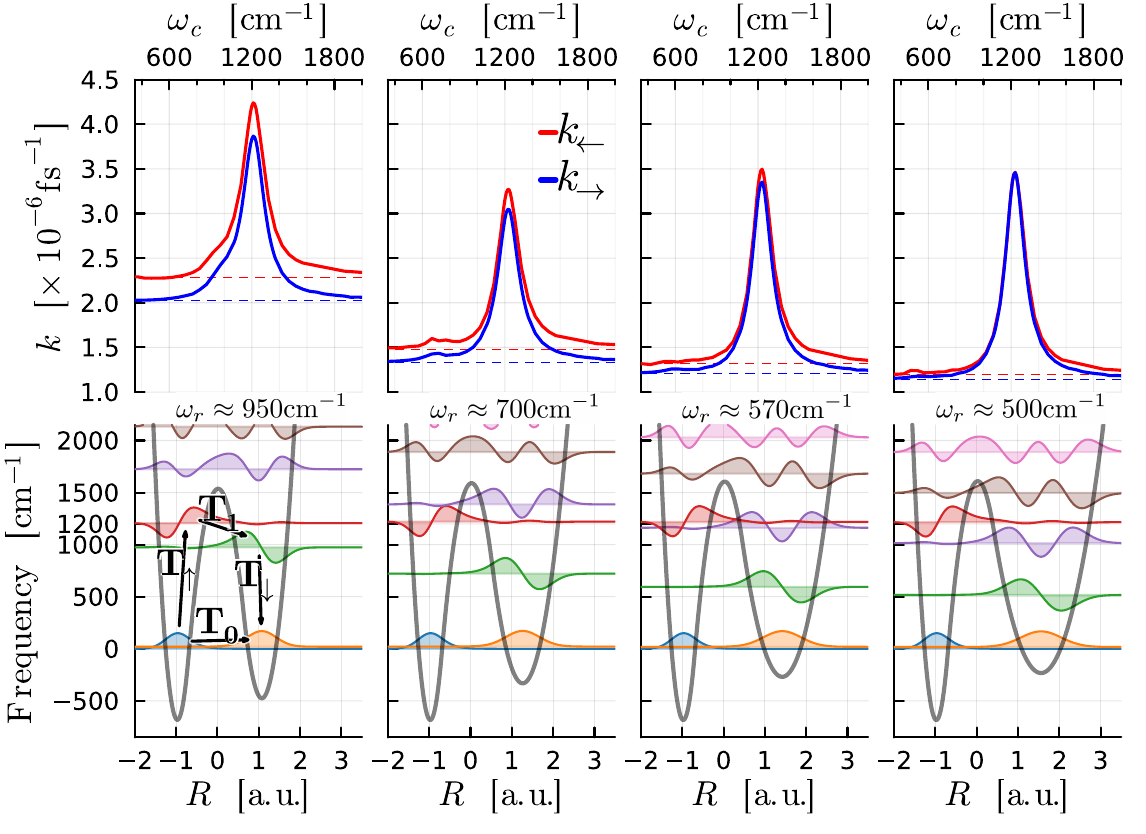}
    \caption{
    The bottom row shows four different potential energy surfaces, with transition frequencies in the right-well of around $\omega_r\in\left\{950,700,570,500\right\}\times\text{cm}^{-1}$.
    The top row shows the reaction rate from the left well to the right well in blue and the converse in red, for varying $\omega_c$, when coupled to the spectral density in Eq.~\ref{eq:SimpleSD}. The horizontal dashed lines correspond to the reaction rates outside the cavity.
    The probability currents of the system with $\omega_r\approx 950\,\text{cm}^{-1}$ are shown in Fig.~\ref{fig:asymmetric_currents}.
    }
    \label{fig:asymmetric_rates}
\end{figure}

Figure~\ref{fig:asymmetric_rates} shows both the forward and backward reaction rate profiles for four double well PESs, where the transition frequency of the right well has been lowered when compared to the symmetric PES.
The reaction rate outside the cavity is decreasing with the decreasing well frequency because the overlap between the ground states of each well becomes smaller.
This decreases the tunneling rate, which was the main reaction pathway for the symmetric PES (Fig.~\hyperref[fig:EstablishMethod]{\ref*{fig:EstablishMethod}c}). In the most asymmetric PES we consider, with a right well frequency of $\omega_r\approx 500\,\text{cm}^{-1}$, tunneling along $\text{T}_0$ is practically zero and the only reaction mechanism is the phonon bath inducing the transition over the barrier (see Supplementary Fig.~\ref{fig:all_asymmetric_currents}).

For all of the PESs, we see an asymmetry in the backward ($k_\leftarrow$ in red) vs. forward ($k_\rightarrow$ in blue) reaction rates. Their ratio is linked to the equilibrium populations in each well $\frac{k_\rightarrow}{k_\leftarrow}=\frac{P_\text{right}^{eq}}{P_\text{left}^{eq}}$ (Sec.~\ref{subsec:Observables}). The ratio persists when changing $\omega_c$, indicating that the cavity does not modify the equilibrium distribution.
We can compare $\frac{k_\rightarrow}{k_\leftarrow}$ to the theoretical prediction for the equilibrium state given by the \textit{mean force Gibbs state} \cite{10.1116/5.0073853}. To leading order in the bath coupling, the reduced density matrix for the system at equilibrium is the Gibbs state $\rho^{eq}_S\propto e^{-\beta H_R}$.
The ground states in both wells are only approximately degenerate, with a slight energetic preference for the left well ($\Delta E \approx20\,\text{cm}^{-1}$) and at $T=300\text{K}$ (corresponding to $\approx208.5\,\text{cm}^{-1}$) the population of the left well will be larger in the steady-state distribution, which puts $k_\leftarrow$ above $k_\rightarrow$ in magnitude.
When more low-energy states are introduced in the right well, by lowering its frequency, the asymmetry decreases as seen and eventually $P_\text{right}^{eq}$ will dominate, for which we also expect $k_\rightarrow \gg k_\leftarrow$. 
Going beyond leading order in the calculation of $\rho^{eq}_S$ \cite{PhysRevLett.127.250601} (Supplementary Section~\ref{sec:MeanForceGibbsState}), we see a small correction to the Gibbs state, which biases the left well. This can be understood in terms of the bath observing the reaction coordinate, which localizes the wavefunction and favors dispensing more spectral weight in the deeper (left) well. 
This effect is mainly due to the phonon bath, in the parameter regime considered. 
The presence of the cavity has a negligible effect on the equilibrium distribution and the weak dependence on $\omega_c$ does not exhibit resonant behavior.


The frequency of the right well $\omega_r$ {represents} a secondary relevant transition frequency, different from the left well's frequency at $\omega_l\approx1200\,\text{cm}^{-1}$. This shows up as a secondary peak in the reaction rate profiles located at $\omega_r$, whose magnitude, however, quickly vanishes when decreasing $\omega_r$. 
To explain this, we consider the reaction mechanism which goes through the top of the barrier. Once the excited states are reached, the population inside this manifold will oscillate back and {forth} between the two wells, before decaying into either.
Whether the excitation decays to the opposite well or returns to its original well determines whether a reaction takes place.
Thus, tuning the cavity to match the frequency of a well, which provides an additional thermalization channel, will only have an effect on the reaction rate if it changes the rate-limiting step of the overall process.
Excitations to the top of the potential energy barrier are still the rate-limiting step, so once the cavity is in resonance with one well, it means that the phonon-induced excitations of the other well becomes the rate-limiting step. Resultingly, the ratio between the magnitude of the primary peak and the secondary peak, relative to the baseline $k_0$ in the reaction rate profiles, are related by the timescales provided in Sec.~\ref{subsec:Benchmark}, such that for both the forward and backward reaction rates: 
$\frac{k(\omega_c=\omega_r)-k_0}{k(\omega_c=\omega_l)-k_0}\approx\frac{k_\leftarrow}{k_\rightarrow}\left(\frac{\omega_r}{\omega_l}\right)^2e^{-\beta(\omega_l-\omega_r)}$%
\footnote{Using that $k_\rightarrow=\frac{P^{eq}_\text{right}}{P^{eq}_\text{left}}k_\leftarrow$, the relative increase in the rate $k_\rightarrow(\omega_c=\omega_l)$ can be related to the phonon-induced excitations in the right well $k_\rightarrow(\omega_c=\omega_l)-k_0\sim\frac{P^{eq}_\text{right}}{P^{eq}_\text{left}}\frac{e^{-\beta\omega_r}}{\omega_r^2}$. Conversely, $k_\rightarrow(\omega_c=\omega_r)-k_0$ is related to the phonon-induced excitations in the left well, such that the ratio leads to the quoted result.}.
This explains the observed vanishing magnitude of the peak in $k$ at $\omega_r$ when decreasing $\omega_r$.

We also observe that the maximum relative enhancement increases as the right well is widened. 
This is attributable to the increased exposure to the phonon bath at lower frequencies. Once the cavity {helps} facilitate the large frequency transition, the phonon bath can more easily assist the transitions of the lower frequency well, leading to a larger relative enhancement of the reaction rate for decreasing $\omega_r$.

\begin{figure}[h]
    \centering
    \includegraphics[width=1.0\linewidth,
    trim=0.725cm 0.85cm 0.6cm 0.41cm,  
    clip
    ]{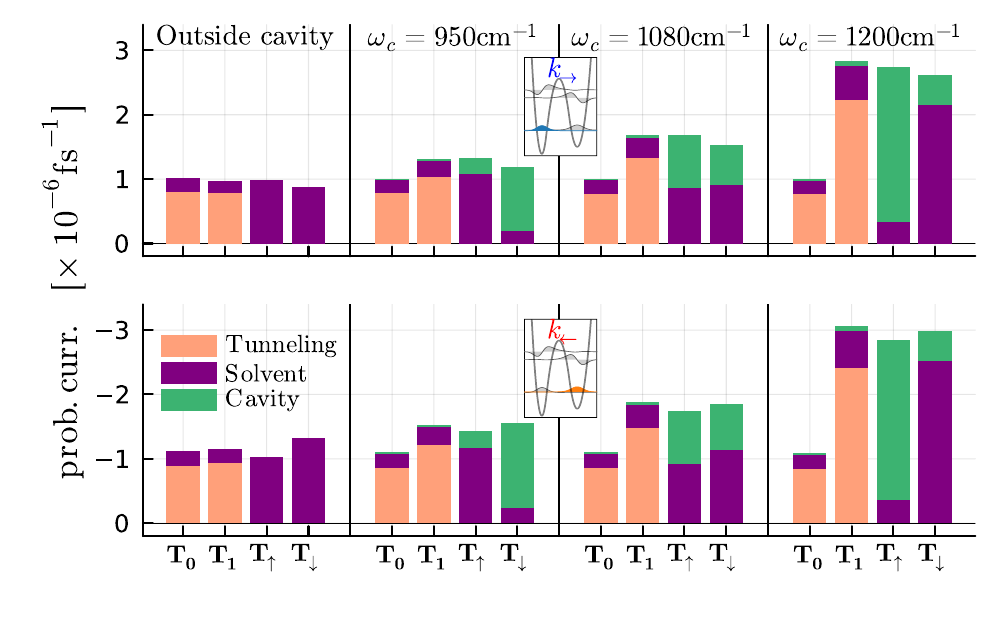}
    \caption{The associated probability currents for both the $k_\rightarrow$ (top) and $k_\leftarrow$ (bottom) reaction dynamics of the least asymmetric system in {Fig.}~\ref{fig:asymmetric_rates}.
    Probability currents for the other systems can be found in Supplementary Fig.~\ref{fig:all_asymmetric_currents}.
    }
    \label{fig:asymmetric_currents}
\end{figure}

Figure~\ref{fig:asymmetric_currents} shows the probability currents of the transitions labeled in Fig.~\ref{fig:asymmetric_rates}. Note that the probability currents are, by definition, anti-symmetric. Thus, if for a single dynamical simulation, we compared $\text{T}_\downarrow$ to its reversed partner, we would find that they only differ by a sign. However, here we compare two distinct dynamical simulations, that differ by which well the reaction coordinate wavefunction was initialized in. The currents for all systems shown in Fig.~\ref{fig:asymmetric_rates} are provided in Supplementary Fig.~\ref{fig:all_asymmetric_currents}.

Compared to the currents for the symmetric double well (Fig.~\hyperref[fig:EstablishMethod]{\ref*{fig:EstablishMethod}c}), the ground state tunneling rate $\text{T}_0$ is suppressed, as already noted. From the currents, it is also apparent that the transition at $1200\,\text{cm}^{-1}$, denoted $\text{T}_\uparrow$, is the rate-limiting step. Again, the tunneling along $\text{T}_1$ occurs at a faster time scale such that the net contribution simply follows what the thermodynamics of the wells dictates. At $\omega_c=\omega_r$ ($\approx950\,\text{cm}^{-1}$ for this PES) the cavity simply assumes the role of the phonon bath in enabling both excitation and relaxation transitions in the right well, as captured by $\text{T}_\downarrow$. How large a contribution either environmental {degree of freedom provides} to these transitions, is determined by the relative magnitude of their noise spectra (Fig.~\ref{fig:SimpleNoiseSpectrum}), in this perturbative regime. The more the cavity noise spectrum overlaps with the $\omega_l=1200\,\text{cm}^{-1}$ transition, the more the reaction dynamics are limited by the transitions in the right well, which occur on a faster time scale due to the $\sim\frac{1}{\omega^2}$ scaling of the approximate phonon-induced downward transition rate. This is apparent from the $\text{T}_\downarrow$ transitions in Fig.~\ref{fig:asymmetric_currents}, whose solvent-induced component grows significantly when $\omega_c=1200\,\text{cm}^{-1}$.



\subsubsection{{Competing pathways}}
\label{subsec:competingpathways}
\begin{figure*}[ht]
    \centering
    \includegraphics[width=1.0\linewidth,
    trim=0.2cm 1.7cm 2.6cm 2.3cm,  
    clip
    ]{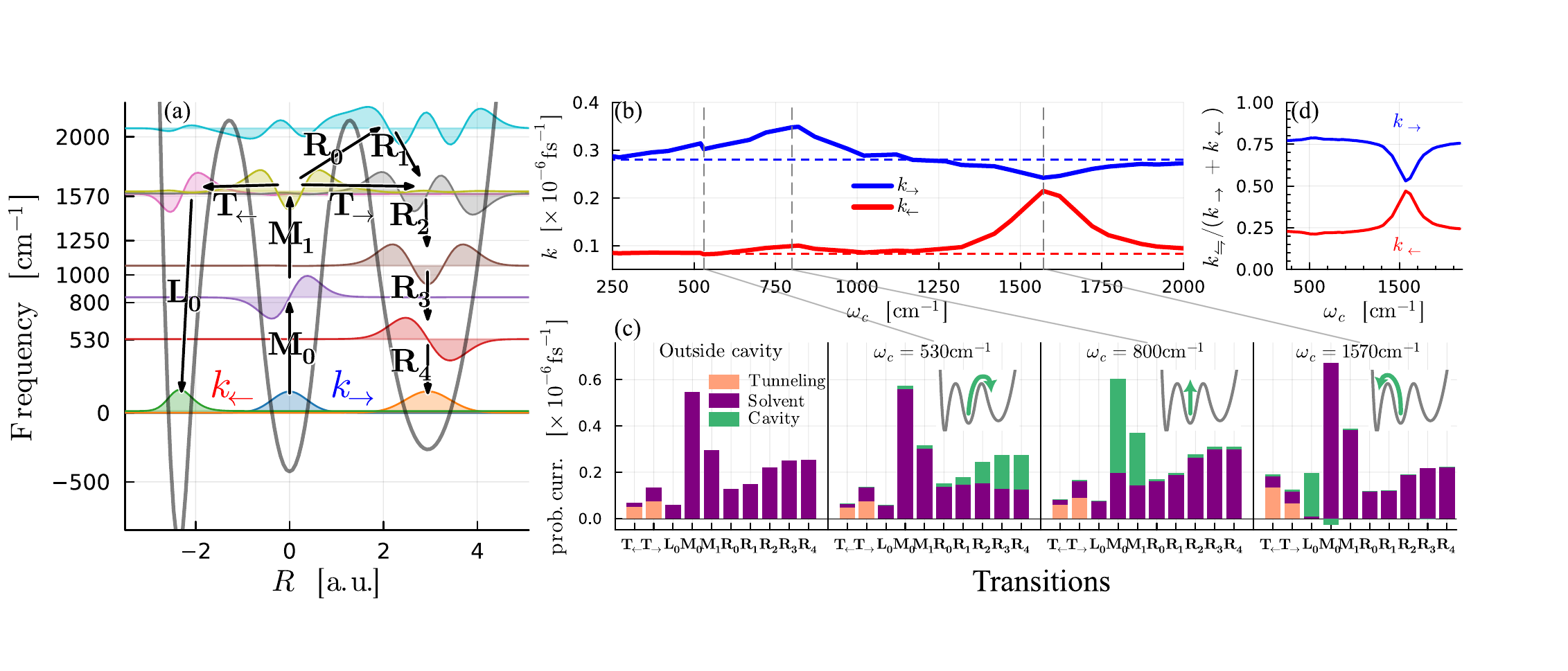}
    \caption{
    \textbf{(a)} Visualization of the triple well PES and the vibrational states. The middle well is the reactant region and $k_\leftarrow$ and $k_\rightarrow$ denote the reaction rate to the left and right product wells, respectively. The black arrows label the state-to-state transitions.
    \textbf{(b)} Shows the reaction rates at different cavity frequencies. The horizontal dashed lines denote the rates outside the cavity. 
    \textbf{(c)} Probability currents for the associated transitions, both outside and inside the cavity.
    \textbf{(d)} The inferred kinetic yield of the left product in red and the right product in blue for different cavity frequencies. This does not constitute the yield at equilibrium, which remains largely unaffected by the presence of the cavity.
    }
    \label{fig:competing_pathways}
\end{figure*}
Finally, we add a third relevant transition frequency to the dynamics. This will serve as our simplified model of a reaction coordinate containing two different product regions and thus two competing pathways. 
Figure~\hyperref[fig:competing_pathways]{\ref*{fig:competing_pathways}a} shows the PES and the associated vibrational states. We define the middle well to be the reactant region and the left and right well to be two distinct product regions. The three near-degenerate states at frequency $1570\,\text{cm}^{-1}$ are not eigenstates of $H_R$, but instead have been mixed to construct the localized states seen (Supplementary Section~\ref{sec:StateLocalization}).

Outside the cavity, the reaction rate from the reactant to the right product region $k_{\rightarrow}$ is about $3.36\times$ larger than the reaction rate to the left product well $k_\leftarrow$, as seen in Fig.~\hyperref[fig:competing_pathways]{\ref*{fig:competing_pathways}b}.
Informed by the probability currents in Fig.~\hyperref[fig:competing_pathways]{\ref*{fig:competing_pathways}c}, this difference in time scales can mainly be attributed to two factors. The right well ($\omega_r\approx530\,\text{cm}^{-1}$) samples the phonon bath at a favorable frequency, leading to the increase in $\text{T}_\rightarrow$ over $\text{T}_\leftarrow$.
Additionally, the right product well has an excited state, located in frequency just above $2000\,\text{cm}^{-1}$, that shares support with the reactant excited state leading to the phonon-assisted transition labeled $\text{R}_0$, adding to the $k_\rightarrow$ reaction rate.
Notice that $k_\leftarrow$ is approximately given by the probability current of the $\text{T}_\leftarrow$ transition and $k_\rightarrow$ by the $\text{T}_\rightarrow$ and $\text{R}_0$ transitions, in accordance with Eq.~\ref{eq:ReactionRateFromCurrents}.

Inside the cavity, there {exist} three distinct scenarios depending on which of the three wells the cavity is resonant with. Similar to the examples with double-well PESs, we generally observe that the cavity accelerates the dynamics in the resonant well
and that, due to the shape of the phonon bath, the higher frequency well generally hosts the rate-limiting steps.

When the cavity is resonant with the right product well $\omega_c=\omega_{r}$ the reaction rate $k_\rightarrow$ is slightly increased ($\sim1.12\times$) whereas $k_\leftarrow$ is unaffected. The probability currents reflect this observation, where the cavity is seen to help thermalize the $\text{R}_{0-4}$ transitions. 
Instead, when the cavity is resonant with the frequency of the reactant region $\omega_c=\omega_0\approx800\,\text{cm}^{-1}$ both reaction rates naturally increase, leading to an enhancement in $k_\leftarrow$ by $\sim1.2\times$ and in $k_\rightarrow$ by $\sim 1.25\times$ in line with the cavity enhanced dynamics of the $\text{M}_{0,1}$ transitions.


More interestingly, when the cavity is resonant with the left product well $\omega_c=\omega_l$, the rate $k_\leftarrow$ is increased by $\sim 2.6\times$ whereas $k_\rightarrow$ is decreased by $\sim 0.86\times$, making the two reaction rates almost equal. The increased rate of equilibration between the left well and the reactant well is expected, since the left-well frequency is resonant with the lossy cavity. The resulting cavity-induced bias toward the left well, reflected in the $\text{T}_\leftarrow$ and $\text{L}_0$ transitions, is accompanied by an increased excitation rate within the reactant well, as seen in the $\text{M}_{0,1}$ transitions. However, these effects do not fully compensate, so the excited reactant state feeding both product pathways is more depleted than without the cavity.
This leads to the decrease in the $\text{R}_{0-4}$ transitions and ultimately the $k_\rightarrow$ reaction rate. 
%
%
%

The fact that the cavity can modify the two reaction rates differently can lead to cavity modified yield of the reaction. However, whether this occurs depends on the reversibility of the reaction. The system we have considered is best described as a reversible parallel reaction. The population in the middle well can flow into both the left and right well and the population in these outer wells can also flow back into the middle well. This reversibility allows the system to equilibrate and as such, the yield of the reaction is determined by the equilibrium distribution. Using the mean force Gibbs state \cite{10.1116/5.0073853, PhysRevLett.127.250601} already employed for the double well PESs (Sec.~\ref{subsec:asymmetric_wells}) we can study how the cavity affects the equilibrium distribution.
Similar to the double well examples, the solvent itself induces a small bias towards the more narrow and deeper left well but the cavity provides no noticeable effect (Supplementary Section~\ref{sec:MFGTripleWell}).
Thus, for the setup in question, the presence of the cavity does not change the yield of left and right product but it affects the speeds at which the reactant equilibrates with the two products.

A change in reaction rates can still induce a change in the yield, if the reaction is more accurately described as an irreversible parallel reaction. This is the case, for example, if the left and right product wells are much lower in energy than the reactant well, when compared to the temperature. Then the backwards flow from the outer wells to the middle well would be negligible and the {kinetic }yield would be given by the branching ratios of the two forward reaction rates $P^{eq}_\text{left}=\frac{k_\leftarrow}{k_\rightarrow + k_\leftarrow}$ and $P^{eq}_\text{right}=\frac{k_\rightarrow}{k_\rightarrow + k_\leftarrow}$ (see Supplementary Section~\ref{sec:KineticAndThermodynamicControl} for a mathematical formulation of this discussion).
Thomas et al.~\cite{doi:10.1126/science.aau7742} provided the first experimental observation of such cavity-modified yield under kinetic control. Our model captures this effect, as seen in Fig.~\hyperref[fig:competing_pathways]{\ref*{fig:competing_pathways}d}, where the kinetic branching ratio is changed from approximately $1:3$ to $1:1$ when the cavity is resonant with the frequency of the left well\footnote{
The potential energy of the product regions could be lowered to incorporate the irreversibility requirement for the reaction to be kinetically controlled, placing our model more in line with the experimental study \cite{doi:10.1126/science.aau7742}.
Additionally, whether the cavity-modification is enhancive or suppressive in nature, is largely dependent on the harmonic environment, within these models \cite{lindoy_quantum_2023}. Thus, if we instead considered the coupling between the reaction coordinate and the cavity to be mediated by a spectator mode \cite{Vega2025}, we would expect the presence of the cavity to have a suppressive effect, similar to their observations.
}.
Thus, the cavity can change product yields under kinetic control, by modifying the relative rates of competing pathways, but it does not significantly alter the equilibrium product distribution for the thermodynamically controlled reversible reaction considered here.

\subsection{Environment-driven effects}
\label{subsec:Environment-driven effects}
All calculations in the preceding section employed the simple spectral density given in Eq. 
\ref{eq:SimpleSD}. We now turn our attention to the general spectral density, Eq.~\ref{eq:spectraldensity}, which describes a reaction coordinate interacting with a spectator mode, both of which are coupled to the cavity mode.
Simultaneously, we simplify the system by considering the symmetric double well seen in Fig.~\hyperref[fig:EstablishMethod]{\ref*{fig:EstablishMethod}a}.

How strongly the cavity couples to either the reaction coordinate $\eta_R$ or the spectator mode $\eta_Q$ 
depends on the alignment between the respective dipole moments and the cavity polarization.
We parameterize these couplings in terms of two angles: 
\begin{equation}
    \eta_Q {\propto}\cos(\phi)
    \quad\text{and}\quad
    \eta_R{\propto}\cos(\theta-\phi)
\end{equation}
where $\phi$ is the angle between the cavity polarization and the spectator mode's dipole moment and $\theta$ is the angle between the reaction-coordinate and spectator-mode dipoles.
This parameterization constitutes a reduced 2-dimensional orientational model in which the two molecular dipoles rotate rigidly, at a fixed relative angle $\theta$, within a plane containing the cavity polarization. 
To approximate the response of an ensemble of molecular orientations, we first calculate the orientationally averaged spectral density $\langle J(\omega)\rangle_\phi:=\frac{1}{2\pi}\int_0^{2\pi}J(\omega)d\phi$ and then use this averaged spectral density in the dynamical simulation. 



\begin{figure}[h]
    \centering
    \includegraphics[width=1.0\linewidth]{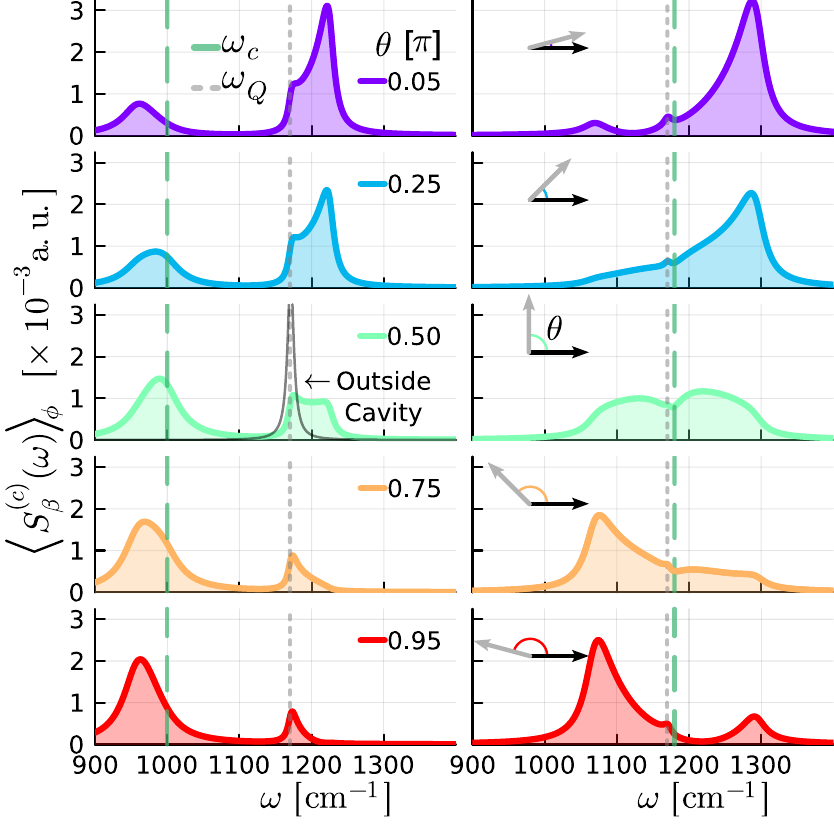}
    \caption{
    Visualization of the noise spectra with spectral density given by Eq.~\ref{eq:spectraldensity}, without the phonon bath, which has been averaged over the orientation of the molecule. The spectator mode frequency is $\omega_Q=1170\,\text{cm}^{-1}$ with a lifetime of about $500\,\text{fs}$. $\theta$ is the angle between the reaction coordinate and spectator mode dipole moments. Five different values of $\theta$ {are} shown vertically, while two different values of $\omega_c$ are shown horizontally. 
    }
    \label{fig:NoiseSpectrums} 
\end{figure}

These two simplifications allow us to isolate how the relative geometry of the two vibrational dipoles controls the effect of the cavity on the reaction rate.
Their validity in the parameter regime considered here is assessed in Supplementary Section~\ref{sec:DistributionOfRates}, by comparing the reduced orientational model with a 3-dimensional isotropic treatment and with an explicit account of orientation-resolved reaction rates.
These comparisons show that the reduced model retains the qualitative features of the cavity-induced rate changes.
In particular, the rate extracted from dynamics using $\langle J(\omega)\rangle_\phi$ closely approximates the effective rate obtained by fitting the ensemble reactant decay to a single exponential.
Likewise, the 2- and 3-dimensional orientational models yield similar results in this regime, supporting the use of the reduced model to examine the dependence on relative dipole geometry. 
The resulting noise spectra are visualized in Fig.~\ref{fig:NoiseSpectrums}, with explicit expressions found in Supplementary Section~\ref{sec:AngleAveragedSpectralDensity}.

We consider a spectator mode with frequency $\omega_Q=1170\,\text{cm}^{-1}$, slightly below the relevant system transition at around $1200\,\text{cm}^{-1}$. Outside the cavity, the damped spectator mode produces a Lorentzian feature in the noise spectrum. 
Introducing the cavity hybridizes the cavity and spectator mode, giving rise to features associated with two Rabi-split damped resonances distanced $\Omega_R\sim200\,\text{cm}^{-1}$ apart. Since the reaction coordinate also couples directly to the cavity, the resulting spectrum develops asymmetric and non-Lorentzian features that may be understood as reflecting interference between the two hybridized decay channels.
This structure depends on $\theta$, which controls the relative signs and magnitudes of the corresponding dipole couplings.

\begin{figure}[h]
    \centering
    \includegraphics[width=1.0\linewidth]{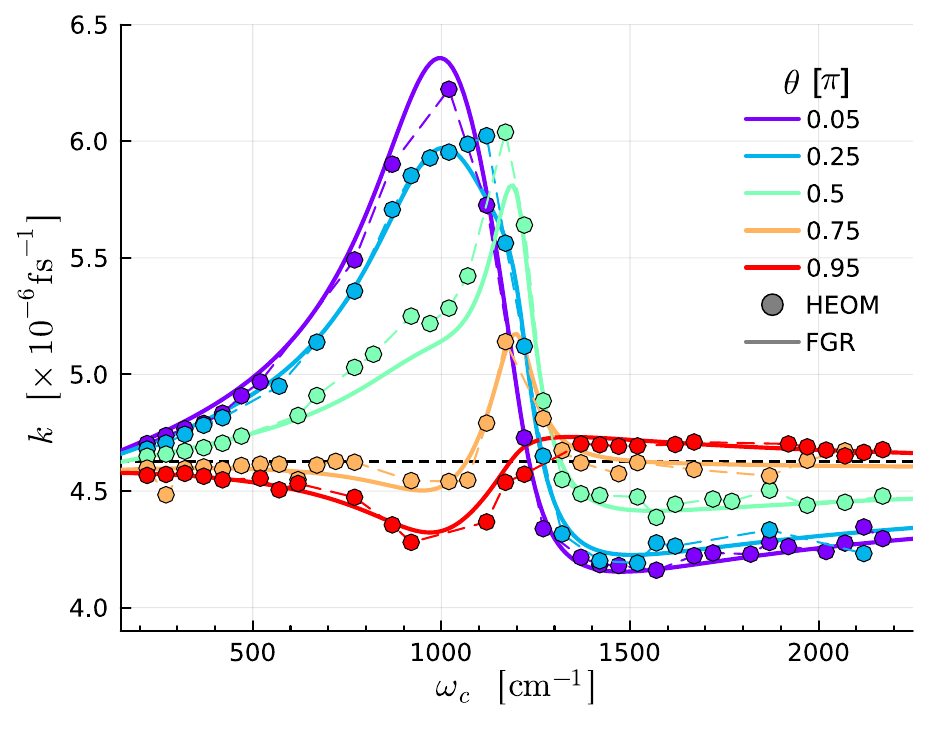}
    \caption{
    Numerically exact reaction rates compared to predictions calculated via Eq.~\ref{eq:FGR}. The Lorentzian has parameters ${\omega_0}=1205\,\text{cm}^{-1}$ and ${\Gamma}=30\,\text{cm}^{-1}$ and the approximate scaling factor was rescaled $\frac{1}{2}\rightarrow0.38$ to best match the results from HEOM.
    The horizontal dashed line {corresponds} to the reaction rate outside the cavity.
    }
    \label{fig:FanoRates}
\end{figure}

Using our method described in Sec.~\ref{sec:methodology}, we decompose, to great accuracy, the noise spectra in Fig.~\ref{fig:NoiseSpectrums}, for different cavity frequencies, and perform simulations of the dynamics in order to extract the reaction rate $k:=k_\rightarrow=k_\leftarrow$. Figure~\ref{fig:FanoRates} shows $k$ as a function of $\omega_c$ for different angles $\theta$\footnote{The HEOM results for some $\omega_c$ are omitted. This is because the automated AAA decomposition (Eq.~\ref{eq:BCFSumOfExponentials}) did not reach the specified accuracy of $10^{-10}$. HEOM results for any $\omega_c$ and $\theta$ can still be obtained with our method, in some cases one would just have to tune the settings for the AAA-fit by hand or choose a different method to achieve the decomposition.}. 
The black horizontal line denotes the reaction rate outside the cavity, while the circles show the cavity-modified rates. The solid lines represent the perturbative predictions obtained from the probability currents. We now derive the corresponding closed expression for $k$, drawing on the analysis of Fig.~\ref{fig:EstablishMethod}.

According to Eq.~\ref{eq:ReactionRateFromCurrents}, the reaction rate can be approximated by the total probability current across the potential barrier, $k\approx j_{\text{T}_0} + j_{\text{T}_1}$, for early times where $P_\text{left}(t)\approx1$.
The currents shown in Fig.~\hyperref[fig:EstablishMethod]{\ref*{fig:EstablishMethod}c} suggest that $j_{\text{T}_0}$ is largely unaffected by the cavity, whereas the cavity-induced change in $j_{\text{T}_1}$ originates from the cavity contribution to the $\text{T}_{\uparrow,\downarrow}$ transitions. We may therefore approximate the cavity-modified reaction rate as the rate independent of the cavity, $k_0$, plus the cavity dependent contribution to these transitions, e.g. $k\approx k_0+j_{\text{T}_\uparrow}^{(c)}$. 
In the perturbative limit, Eq.~\ref{eq:PerturbativeCurrents} allows the cavity contribution to be evaluated as $j_{\text{T}_\uparrow}^{(c)}\approx\Gamma_\uparrow^{(c)}\rho_{gg}-\Gamma_\downarrow^{(c)}\rho_{ee}$ where the rates $\Gamma_{\uparrow,\downarrow}^{(c)}$ are given by Fermi's Golden Rule and the subscripts in $\rho_{gg}$ and $\rho_{ee}$ refer to the reactant ground and excited state, respectively.
Despite the small value of $\rho_{ee}$ the product $\Gamma_\downarrow^{(c)}\rho_{ee}$ can be sizeable. Thus, we account for $\rho_{ee}$ through adiabatic elimination. As alluded to in the earlier discussions, the dynamics among the excited vibrational states is fast, so an excitation coming from the reactant ground state can decay into either the product or reactant well. Schematically, $\dot{\rho}_{ee}\approx \Gamma_\uparrow^{(c)} \rho_{gg}-2\Gamma_\downarrow^{(c)}\rho_{ee}$ and the intermediate steady state $\rho_{ee}\approx\frac{\Gamma_\uparrow^{(c)}}{2\Gamma_\downarrow^{(c)}}\rho_{gg}$ is quickly reached before back-excitations from the product ground state needs to be accounted for. Consequently, the apparent rate given by FGR is rescaled by a factor-half $j_{\text{T}_\uparrow}^{(c)}\approx\frac{\Gamma_\uparrow^{(c)}}{2}\rho_{gg}$. 
Together with $\rho_{gg}\approx 1$, we arrive at the following expression for the cavity modified reaction rate:
\begin{equation}
\label{eq:FGR}
k \approx k_0 + \frac{1}{2}|\langle e|\hat{R}|g\rangle|^2\int_{-\infty}^\infty e^{-\beta\omega}S^{(c)}_\beta(\omega)\mathcal{A}(\omega)d\omega 
\end{equation}
$S_\beta^{(c)}(\omega)$ refers to the cavity-modified component of the noise spectrum 
and $\mathcal{A}(\omega)=\frac{1}{\pi}\frac{\Gamma}{(\omega-\omega_0)^2+{\Gamma^2}}${, with center frequency $\omega_{0}$ and half-width $\Gamma$,} encodes the broadening of the transition frequency due to the cavity-independent component, in this case the phonon bath~\cite{ying_resonance_2023} (see also Supplementary Section~\ref{subsec:Broadening}). Similarly, $k_0$ refers to the phonon-only rate, not the outside-cavity rate, since the spectator mode is coupled to the cavity and thus contained in $S_\beta^{(c)}(\omega)$. $\langle e|\hat{R}|g\rangle$ is the matrix element that couples the ground and excited state of the reactant well and the factor $\frac{1}{2}$ is the approximate rescaling due to the decay of the excited states, as described above.
Such expressions for the reaction rate were successfully employed in {Refs.}~\citenum{ying_resonance_2023, Ying2024}, for the case of no spectator mode (${c_Q}=\eta_Q=0$) and in Ref.~\citenum{Vega2025}, for no direct coupling between the reaction coordinate and the cavity ($\eta_R=0$), with empirical rescaling factors introduced to obtain quantitative agreement with numerically exact results. Here, we have formulated a rationale in terms of probability currents, including a possible explanation for the required rescaling coefficient, and applied it to the case where all couplings are non-zero.

As seen in Fig.~\ref{fig:FanoRates}, the FGR predictions agree well with the HEOM simulations. Such agreement is not expected in general, since the assumptions underlying FGR require sufficiently weak bath coupling and a sufficiently smooth noise spectrum for the Born–Markov approximation to be applicable. 
Furthermore, the quantitative agreement seen in Fig~\ref{fig:FanoRates} is dependent on obtaining the appropriate $\mathcal{A}(\omega)$ and an accurately determined rescaling for the cavity-induced FGR contribution. 
Finally, for Eq.~\ref{eq:FGR} to be applicable in the first place, the system must also contain a well-defined rate-limiting step that can be identified.
Nonetheless, the qualitative form of the rate profiles in Fig.~\ref{fig:FanoRates} is governed by which parts of the structured noise spectrum overlap with the broadened transition function \(\mathcal{A}(\omega)\). If the spectator mode is far detuned from the relevant transition, \(|\omega_Q-{\omega_{0}}|\gg{\Gamma}\), the two hybridized channels do not contribute comparably within the transition window, and the asymmetric interference feature becomes less pronounced. The rate profile then approaches a simpler resonant cavity response, although its amplitude and specific shape can still depend on \(\theta\) 
when the cavity is tuned sufficiently close to the spectator mode to hybridize appreciably, \(|\omega_c-\omega_Q|\sim\Omega_R\sim200\,\text{cm}^{-1}\). 
This residual angular dependence reflects the way the dipole geometry redistributes spectral weight between the hybridized resonances. 

When the spectator mode lies within the broadened transition window, \(|\omega_Q-{\omega_{0}}|\lesssim\Gamma\), both cavity-spectator hybrid modes can contribute to the same rate-limiting transition, 
making the overlap with \(\mathcal{A}(\omega)\) sensitive to interference between the two channels. 

This interpretation is made more concrete by the fact that the rate curves in Fig.~\ref{fig:FanoRates} permit a unified phenomenological description in terms of a Fano-plus-Lorentzian profile\footnote{
Specifically, with $\tilde{\omega}_c$ a dimensionless cavity frequency that has been shifted and scaled appropriately and $a(\theta)$ the angle-dependent Fano-parameter, the ansatz:
\begin{equation*}
    \frac{k(\tilde{\omega}_c)}{k_0}\propto \frac{(a(\theta)+\tilde{\omega}_c)^2}{1+\tilde{\omega}_c^2}+b(\theta)\frac{\tilde{\Gamma}}{(\tilde{\omega}_c+\tilde{\omega}_0)^2+\tilde{\Gamma}^2}
\end{equation*}
can faithfully capture the rate profiles in Fig.~\ref{fig:FanoRates}, although it was not used for quantitative predictions. Obtaining the parameterization of $a(\theta)$ and $b(\theta)$ by reducing Eq.~\ref{eq:FGR} to this form was not immediately successful because naive pole-expansion of the terms in Eq.~\ref{eq:AngleAveraging} missed some of the observed features.}.
The Fano component captures the interference between competing hybridized channels within the broadened transition window \cite{10.1063/5.0324649}, while the Lorentzian component represents the remaining resonant contribution from the spectator mode's residual spectral weight.
In this form, the angle $\theta$ tunes the interference term through the Fano-parameter in addition to the magnitude of the residual Lorentzian contribution.



The rate profiles in Fig.~\ref{fig:FanoRates} are therefore organized by an effective Fano-plus-Lorentzian structure, reflecting the interplay between cavity-spectator hybridization, spectral overlap with the broadened transition, and geometry-dependent interference.
\section{Conclusion}
\label{sec:conclusion}
We have developed a general quantum-dynamical framework for understanding cavity-modified reaction kinetics in terms of the individual transitions that comprise a reaction pathway. The framework combines a concise expression for the effective spectral density of an arbitrary harmonic environment with a controlled compression of Markovian auxiliary modes into system-level jump operators, enabling efficient tensor-network simulations of HEOM dynamics while retaining a physically transparent description of the environment.

The probability-current decomposition provides the central diagnostic for connecting these dynamics to observable reaction rates. It shows that cavity resonance selectively modifies particular transitions and that the impact on the overall reaction depends on their role in the reaction network. In systems with multiple transition frequencies, this distinction explains why resonances can differ greatly in their effect on the total rate and, in competing pathways, can redirect reaction flux between products.

For the competing pathways, the cavity modifies the rates of population transfer and hence the timescale of equilibration between the wells, but does not modify the equilibrium distribution itself in the parameter regime considered here. The latter is governed by the mean force Gibbs state, whereas the cavity-induced changes in the transition rates redistribute flux along the way to equilibrium. Under irreversible conditions, where the final product distribution is instead determined kinetically by the relative pathway rates, this redistribution can translate directly into a cavity-modified product yield.

Finally, when spectator modes are included, hybridization and interference between environmental decay channels produce asymmetric, structured rate profiles that can be described in terms of Fano-plus-Lorentzian features. Taken together, these results show that cavity-modified reactivity is governed by how the cavity reshapes transition-resolved flux within a reaction network, rather than by a uniform modification of the reaction as a whole. 


An important direction for future work is to extend this picture beyond the perturbative regime explored here. In the present calculations, the probability currents are consistent with the transition-rate intuition provided by Fermi’s Golden Rule, while being obtained directly from the full HEOM dynamics rather than relying on a perturbative rate description. This makes the current-based analysis particularly promising for regimes where such rate theories become inadequate, as it provides a direct way to resolve how environmental effects propagate through the underlying reaction pathways. Extending the approach to more complex potential-energy landscapes, stronger system–environment and light–matter couplings, and more realistic models of the electromagnetic environment will help establish which of the mechanisms identified here persist beyond the present model systems. 
Importantly, the combination of general harmonic-environment representations, Markovian-mode compression, and tensor-network HEOM dynamics developed here provides a practical foundation for carrying out this analysis in increasingly complex settings.

\begin{acknowledgments}
JV gratefully acknowledges the support of the Flatiron Institute during the Pre-Doctoral Researcher Program.
The Flatiron Institute is a division of the Simons Foundation.
JV also {acknowledges} support from the Carlsberg Foundation, and the Novo Nordisk Foundation, Grant No. NNF22SA0081175, Quantum Computing Programme.

{\footnotesize \flushleft The data and the code for reproducing the figures can be accessed on GitHub \cite{githubrepo}
}
\end{acknowledgments}

\bibliography{ref}

@article{ying_resonance_2023,
    author = {Ying, Wenxiang and Huo, Pengfei},
    title = {Resonance theory and quantum dynamics simulations of vibrational polariton chemistry},
    journal = {The Journal of Chemical Physics},
    volume = {159},
    number = {8},
    pages = {084104},
    year = {2023},
    month = {08},
    issn = {0021-9606},
    doi = {10.1063/5.0159791},
    url = {https://doi.org/10.1063/5.0159791},
}

@Article{lindoy_quantum_2023,
author={Lindoy, Lachlan P.
and Mandal, Arkajit
and Reichman, David R.},
title={Quantum dynamical effects of vibrational strong coupling in chemical reactivity},
journal={Nature Communications},
year={2023},
month={May},
day={12},
volume={14},
number={1},
pages={2733},
issn={2041-1723},
doi={10.1038/s41467-023-38368-x},
url={https://doi.org/10.1038/s41467-023-38368-x}
}

@Article{Vega2025,
author={Vega, Sebastian Montillo
and Ying, Wenxiang
and Huo, Pengfei},
title={Theoretical Insights into the Resonant Suppression Effect in Vibrational Polariton Chemistry},
journal={Journal of the American Chemical Society},
year={2025},
month={Jun},
day={11},
publisher={American Chemical Society},
volume={147},
number={23},
pages={19727-19737},
issn={0002-7863},
doi={10.1021/jacs.5c03182},
url={https://doi.org/10.1021/jacs.5c03182}
}

@article{PhysRevE.93.012128,
  title = {Probability-current analysis of energy transport in open quantum systems},
  author = {Roden, Jan J. J. and Whaley, K. Birgitta},
  journal = {Phys. Rev. E},
  volume = {93},
  issue = {1},
  pages = {012128},
  numpages = {10},
  year = {2016},
  month = {Jan},
  publisher = {American Physical Society},
  doi = {10.1103/PhysRevE.93.012128},
  url = {https://link.aps.org/doi/10.1103/PhysRevE.93.012128}
}

@misc{schumacher2016probabilitycurrentthermodynamicsopen,
      title={Probability current and thermodynamics of open quantum systems}, 
      author={Benjamin Schumacher and Michael D. Westmoreland and Alexander New and Haifeng Qiao},
      year={2016},
      archivePrefix={arXiv},
      primaryClass={quant-ph},
      url={https://arxiv.org/abs/1607.01331}, 
      note={arXiv:1607.01331}
}

@article{PhysRevB.102.115109,
  title = {Universal {Lindblad} equation for open quantum systems},
  author = {Nathan, Frederik and Rudner, Mark S.},
  journal = {Phys. Rev. B},
  volume = {102},
  issue = {11},
  pages = {115109},
  numpages = {24},
  year = {2020},
  month = {Sep},
  publisher = {American Physical Society},
  doi = {10.1103/PhysRevB.102.115109},
  url = {https://link.aps.org/doi/10.1103/PhysRevB.102.115109}
}

@misc{agerskov2026markovianquantummasterequations,
      title={Markovian quantum master equations are exponentially accurate in the weak coupling regime}, 
      author={Johannes Agerskov and Frederik Nathan},
      year={2026},
      archivePrefix={arXiv},
      primaryClass={quant-ph},
      url={https://arxiv.org/abs/2603.04504}, 
      note={arXiv:2603.04504}
}

@article{https://doi.org/10.1002/anie.201107033,
author = {Hutchison, James A. and Schwartz, Tal and Genet, Cyriaque and Devaux, Eloïse and Ebbesen, Thomas W.},
title = {Modifying Chemical Landscapes by Coupling to Vacuum Fields},
journal = {Angewandte Chemie International Edition},
volume = {51},
number = {7},
pages = {1592-1596},
doi = {https://doi.org/10.1002/anie.201107033},
url = {https://onlinelibrary.wiley.com/doi/abs/10.1002/anie.201107033},
year = {2012}
}

@article{https://doi.org/10.1002/anie.201605504,
author = {Thomas, Anoop and George, Jino and Shalabney, Atef and Dryzhakov, Marian and Varma, Sreejith J. and Moran, Joseph and Chervy, Thibault and Zhong, Xiaolan and Devaux, Eloïse and Genet, Cyriaque and Hutchison, James A. and Ebbesen, Thomas W.},
title = {Ground-State Chemical Reactivity under Vibrational Coupling to the Vacuum Electromagnetic Field},
journal = {Angewandte Chemie International Edition},
volume = {55},
number = {38},
pages = {11462-11466},
doi = {https://doi.org/10.1002/anie.201605504},
url = {https://onlinelibrary.wiley.com/doi/abs/10.1002/anie.201605504},
year = {2016}
}

@Article{Coles2014,
author={Coles, David M.
and Somaschi, Niccolo
and Michetti, Paolo
and Clark, Caspar
and Lagoudakis, Pavlos G.
and Savvidis, Pavlos G.
and Lidzey, David G.},
title={Polariton-mediated energy transfer between organic dyes in a strongly coupled optical microcavity},
journal={Nature Materials},
year={2014},
month={Jul},
day={01},
volume={13},
number={7},
pages={712-719},
issn={1476-4660},
doi={10.1038/nmat3950},
url={https://doi.org/10.1038/nmat3950}
}

@article{https://doi.org/10.1002/anie.201703539,
author = {Zhong, Xiaolan and Chervy, Thibault and Zhang, Lei and Thomas, Anoop and George, Jino and Genet, Cyriaque and Hutchison, James A. and Ebbesen, Thomas W.},
title = {Energy Transfer between Spatially Separated Entangled Molecules},
journal = {Angewandte Chemie International Edition},
volume = {56},
number = {31},
pages = {9034-9038},
doi = {https://doi.org/10.1002/anie.201703539},
url = {https://onlinelibrary.wiley.com/doi/abs/10.1002/anie.201703539},
year = {2017}
}

@article{
doi:10.1126/sciadv.aas9552,
author = {Battulga Munkhbat  and Martin Wersäll  and Denis G. Baranov  and Tomasz J. Antosiewicz  and Timur Shegai },
title = {Suppression of photo-oxidation of organic chromophores by strong coupling to plasmonic nanoantennas},
journal = {Science Advances},
volume = {4},
number = {7},
pages = {eaas9552},
year = {2018},
doi = {10.1126/sciadv.aas9552},
URL = {https://www.science.org/doi/abs/10.1126/sciadv.aas9552}}

@article{https://doi.org/10.1002/anie.201915632,
author = {Hirai, Kenji and Takeda, Rie and Hutchison, James A. and Uji-i, Hiroshi},
title = {Modulation of {Prins} Cyclization by Vibrational Strong Coupling},
journal = {Angewandte Chemie International Edition},
volume = {59},
number = {13},
pages = {5332-5335},
doi = {https://doi.org/10.1002/anie.201915632},
url = {https://onlinelibrary.wiley.com/doi/abs/10.1002/anie.201915632},
year = {2020}
}

@article{https://doi.org/10.1002/adma.202002127,
author = {Hou, Shaocong and Khatoniar, Mandeep and Ding, Kan and Qu, Yue and Napolov, Alexander and Menon, Vinod M. and Forrest, Stephen R.},
title = {Ultralong-Range Energy Transport in a Disordered Organic Semiconductor at Room Temperature Via Coherent Exciton-Polariton Propagation},
journal = {Advanced Materials},
volume = {32},
number = {28},
pages = {2002127},
doi = {https://doi.org/10.1002/adma.202002127},
url = {https://advanced.onlinelibrary.wiley.com/doi/abs/10.1002/adma.202002127},
year = {2020}
}

@article{https://doi.org/10.1002/anie.202013465,
author = {Sau, Abhijit and Nagarajan, Kalaivanan and Patrahau, Bianca and Lethuillier-Karl, Lucas and Vergauwe, Robrecht M. A. and Thomas, Anoop and Moran, Joseph and Genet, Cyriaque and Ebbesen, Thomas W.},
title = {Modifying {Woodward–Hoffmann} Stereoselectivity Under Vibrational Strong Coupling},
journal = {Angewandte Chemie International Edition},
volume = {60},
number = {11},
pages = {5712-5717},
doi = {https://doi.org/10.1002/anie.202013465},
url = {https://onlinelibrary.wiley.com/doi/abs/10.1002/anie.202013465},
year = {2021}
}

@article{
doi:10.1126/sciadv.abj0997,
author = {Sitakanta Satapathy  and Mandeep Khatoniar  and Divya K. Parappuram  and Bin Liu  and George John  and Johannes Feist  and Francisco J. Garcia-Vidal  and Vinod M. Menon },
title = {Selective isomer emission via funneling of exciton polaritons},
journal = {Science Advances},
volume = {7},
number = {44},
pages = {eabj0997},
year = {2021},
doi = {10.1126/sciadv.abj0997},
URL = {https://www.science.org/doi/abs/10.1126/sciadv.abj0997}}

@article{
doi:10.1126/science.aau7742,
author = {A. Thomas  and L. Lethuillier-Karl  and K. Nagarajan  and R. M. A. Vergauwe  and J. George  and T. Chervy  and A. Shalabney  and E. Devaux  and C. Genet  and J. Moran  and T. W. Ebbesen },
title = {Tilting a ground-state reactivity landscape by vibrational strong coupling},
journal = {Science},
volume = {363},
number = {6427},
pages = {615-619},
year = {2019},
doi = {10.1126/science.aau7742},
URL = {https://www.science.org/doi/abs/10.1126/science.aau7742}}

@article{https://doi.org/10.1002/anie.201905407,
author = {Lather, Jyoti and Bhatt, Pooja and Thomas, Anoop and Ebbesen, Thomas W. and George, Jino},
title = {Cavity Catalysis by Cooperative Vibrational Strong Coupling of Reactant and Solvent Molecules},
journal = {Angewandte Chemie International Edition},
volume = {58},
number = {31},
pages = {10635-10638},
doi = {https://doi.org/10.1002/anie.201905407},
url = {https://onlinelibrary.wiley.com/doi/abs/10.1002/anie.201905407},
year = {2019}
}

@article{https://doi.org/10.1515/nanoph-2019-0340,
author = {Thomas, Anoop and Jayachandran, Anjali and Lethuillier-Karl, Lucas and Vergauwe, Robrecht M.A. and Nagarajan, Kalaivanan and Devaux, Eloise and Genet, Cyriaque and Moran, Joseph and Ebbesen, Thomas W.},
title = {Ground state chemistry under vibrational strong coupling: dependence of thermodynamic parameters on the Rabi splitting energy},
journal = {Nanophotonics},
volume = {9},
number = {2},
pages = {249-255},
doi = {https://doi.org/10.1515/nanoph-2019-0340},
url = {https://onlinelibrary.wiley.com/doi/abs/10.1515/nanoph-2019-0340},
year = {2020}
}

@article{
doi:10.1126/science.ade7147,
author = {Wonmi Ahn  and Johan F. Triana  and Felipe Recabal  and Felipe Herrera  and Blake S. Simpkins },
title = {Modification of ground-state chemical reactivity via light–matter coherence in infrared cavities},
journal = {Science},
volume = {380},
number = {6650},
pages = {1165-1168},
year = {2023},
doi = {10.1126/science.ade7147},
URL = {https://www.science.org/doi/abs/10.1126/science.ade7147}}

@article{10.1039/d1sc04707h,
    author = {Lather, Jyoti and Thabassum, Ahammad N. K. and Singh, Jaibir and George, Jino},
    title = {Cavity catalysis: modifying linear free-energy relationship under cooperative vibrational strong coupling},
    journal = {Chemical Science},
    volume = {13},
    number = {1},
    pages = {195-202},
    year = {2022},
    month = {01},
    issn = {2041-6520},
    doi = {10.1039/d1sc04707h},
    url = {https://doi.org/10.1039/d1sc04707h},
}

@article{https://doi.org/10.1002/anie.202409528,
author = {Verdelli, Francesco and Wei, Yu-Chen and Joseph, Kripa and Abdelkhalik, Mohamed S. and Goudarzi, Masoumeh and Askes, Sven H. C. and Baldi, Andrea and Meijer, E. W. and Gomez Rivas, Jaime},
title = {Polaritonic Chemistry Enabled by Non-Local Metasurfaces},
journal = {Angewandte Chemie International Edition},
volume = {63},
number = {48},
pages = {e202409528},
doi = {https://doi.org/10.1002/anie.202409528},
url = {https://onlinelibrary.wiley.com/doi/abs/10.1002/anie.202409528},
year = {2024}
}

@Article{Ebbesen2023,
author={Ebbesen, Thomas W.
and Rubio, Angel
and Scholes, Gregory D.},
title={Introduction: Polaritonic Chemistry},
journal={Chemical Reviews},
year={2023},
month={Nov},
day={08},
publisher={American Chemical Society},
volume={123},
number={21},
pages={12037-12038},
issn={0009-2665},
doi={10.1021/acs.chemrev.3c00637},
url={https://doi.org/10.1021/acs.chemrev.3c00637}
}

@article{doi:10.1021/acs.chemrev.2c00895,
author = {Bhuyan, Rahul and Mony, J{\"u}rgen and Kotov, Oleg and Castellanos, Gabriel W. and Gómez Rivas, Jaime and Shegai, Timur O. and B{\"o}rjesson, Karl},
title = {The Rise and Current Status of Polaritonic Photochemistry and Photophysics},
journal = {Chemical Reviews},
volume = {123},
number = {18},
pages = {10877-10919},
year = {2023},
doi = {10.1021/acs.chemrev.2c00895},
note = {PMID: 37683254},
URL = {https://doi.org/10.1021/acs.chemrev.2c00895},
}

@article{annurev:/content/journals/10.1146/annurev-physchem-082624-023219,
   author = "Ying, Wenxiang and Mondal, M. Elious and Koessler, Eric R. and Vega, Sebastian Montillo and Huo, Pengfei",
   title = "Collective Effects in Polariton Chemistry and Photophysics", 
   journal= "Annual Review of Physical Chemistry",
   year = "2026",
   volume = "77",
   number = "Volume 77, 2026",
   pages = "201-224",
   doi = "https://doi.org/10.1146/annurev-physchem-082624-023219",
   url = "https://www.annualreviews.org/content/journals/10.1146/annurev-physchem-082624-023219",
   publisher = "Annual Reviews",
   issn = "1545-1593",
   type = "Journal Article",
  }

@article{10.1039/d5cc04712a,
    author = {Singh, Jaibir and Garg, Pallavi and Lather, Jyoti and George, Jino},
    title = {Stirring chemical reaction landscapes through strong light–matter interactions},
    journal = {Chemical Communications},
    volume = {61},
    number = {93},
    pages = {18289-18301},
    year = {2025},
    month = {12},
    issn = {1359-7345},
    doi = {10.1039/d5cc04712a},
    url = {https://doi.org/10.1039/d5cc04712a},
}

@Article{Mandal2023,
author={Mandal, Arkajit
and Taylor, Michael A.D.
and Weight, Braden M.
and Koessler, Eric R.
and Li, Xinyang
and Huo, Pengfei},
title={Theoretical Advances in Polariton Chemistry and Molecular Cavity Quantum Electrodynamics},
journal={Chemical Reviews},
year={2023},
month={Aug},
day={23},
publisher={American Chemical Society},
volume={123},
number={16},
pages={9786-9879},
issn={0009-2665},
doi={10.1021/acs.chemrev.2c00855},
url={https://doi.org/10.1021/acs.chemrev.2c00855}
}

@Article{Xiang2024,
author={Xiang, Bo
and Xiong, Wei},
title={Molecular Polaritons for Chemistry, Photonics and Quantum Technologies},
journal={Chemical Reviews},
year={2024},
month={Mar},
day={13},
publisher={American Chemical Society},
volume={124},
number={5},
pages={2512-2552},
issn={0009-2665},
doi={10.1021/acs.chemrev.3c00662},
url={https://doi.org/10.1021/acs.chemrev.3c00662}
}

@article{LindoyMandalReichman+2024+2617+2633,
url = {https://doi.org/10.1515/nanoph-2024-0026},
title = {Investigating the collective nature of cavity-modified chemical kinetics under vibrational strong coupling},
author = {Lachlan P. Lindoy and Arkajit Mandal and David R. Reichman},
pages = {2617--2633},
volume = {13},
number = {14},
journal = {Nanophotonics},
doi = {10.1515/nanoph-2024-0026},
year = {2024},
lastchecked = {2026-07-20}
}

@article{10.1063/5.0220908,
    author = {Ke, Yaling and Richardson, Jeremy O.},
    title = {Quantum nature of reactivity modification in vibrational polariton chemistry},
    journal = {The Journal of Chemical Physics},
    volume = {161},
    number = {5},
    pages = {054104},
    year = {2024},
    month = {08},
    issn = {0021-9606},
    doi = {10.1063/5.0220908},
    url = {https://doi.org/10.1063/5.0220908},
}

@article{YingTaylorHuo+2024+2601+2615,
url = {https://doi.org/10.1515/nanoph-2023-0685},
title = {Resonance theory of vibrational polariton chemistry at the normal incidence},
author = {Wenxiang Ying and Michael A. D. Taylor and Pengfei Huo},
pages = {2601--2615},
volume = {13},
number = {14},
journal = {Nanophotonics},
doi = {10.1515/nanoph-2023-0685},
year = {2024},
lastchecked = {2026-07-20}
}

@article{10.1063/5.0231396,
    author = {Lai, Yifan and Ying, Wenxiang and Huo, Pengfei},
    title = {Non-equilibrium rate theory for polariton relaxation dynamics},
    journal = {The Journal of Chemical Physics},
    volume = {161},
    number = {10},
    pages = {104109},
    year = {2024},
    month = {09},
    issn = {0021-9606},
    doi = {10.1063/5.0231396},
    url = {https://doi.org/10.1063/5.0231396},
}

@Article{Ying2024,
author={Ying, Wenxiang
and Huo, Pengfei},
title={Resonance theory of vibrational strong coupling enhanced polariton chemistry and the role of photonic mode lifetime},
journal={Communications Materials},
year={2024},
month={Jun},
day={28},
volume={5},
number={1},
pages={110},
issn={2662-4443},
doi={10.1038/s43246-024-00551-y},
url={https://doi.org/10.1038/s43246-024-00551-y}
}

@article{10.1063/5.0248419,
    author = {Ke, Yaling},
    title = {Stochastic resonance in vibrational polariton chemistry},
    journal = {The Journal of Chemical Physics},
    volume = {162},
    number = {6},
    pages = {064702},
    year = {2025},
    month = {02},
    issn = {0021-9606},
    doi = {10.1063/5.0248419},
    url = {https://doi.org/10.1063/5.0248419},
}

@article{10.1063/5.0324649,
    author = {Ke, Yaling},
    title = {Cooperative chemical reactions in optical cavities: A complex interplay of mode hybridization, timescale balance, and pathway interference},
    journal = {The Journal of Chemical Physics},
    volume = {164},
    number = {17},
    pages = {174103},
    year = {2026},
    month = {05},
    issn = {0021-9606},
    doi = {10.1063/5.0324649},
    url = {https://doi.org/10.1063/5.0324649},
}

@article{10.1063/1.5116800,
    author = {Lawrence, Joseph E. and Fletcher, Theo and Lindoy, Lachlan P. and Manolopoulos, David E.},
    title = {On the calculation of quantum mechanical electron transfer rates},
    journal = {The Journal of Chemical Physics},
    volume = {151},
    number = {11},
    pages = {114119},
    year = {2019},
    month = {09},
    issn = {0021-9606},
    doi = {10.1063/1.5116800},
    url = {https://doi.org/10.1063/1.5116800},
}

@article{10.1063/1.445581,
    author = {Miller, William H. and Schwartz, Steven D. and Tromp, John W.},
    title = {Quantum mechanical rate constants for bimolecular reactions},
    journal = {The Journal of Chemical Physics},
    volume = {79},
    number = {10},
    pages = {4889-4898},
    year = {1983},
    month = {11},
    issn = {0021-9606},
    doi = {10.1063/1.445581},
    url = {https://doi.org/10.1063/1.445581},
}

@article{10.1063/1.2772265,
    author = {Craig, Ian R. and Thoss, Michael and Wang, Haobin},
    title = {Proton transfer reactions in model condensed-phase environments: Accurate quantum dynamics using the multilayer multiconfiguration time-dependent Hartree approach},
    journal = {The Journal of Chemical Physics},
    volume = {127},
    number = {14},
    pages = {144503},
    year = {2007},
    month = {10},
    issn = {0021-9606},
    doi = {10.1063/1.2772265},
    url = {https://doi.org/10.1063/1.2772265},
}

@article{10.1021/acs.chemrev.2c00788,
    author = {Ruggenthaler, Michael and Sidler, Dominik and Rubio, Angel},
    title = {Understanding
Polaritonic Chemistry from Ab Initio
Quantum Electrodynamics},
    journal = {Chemical Reviews},
    volume = {123},
    number = {19},
    pages = {11191-11229},
    year = {2023},
    month = {09},
    issn = {0009-2665},
    doi = {10.1021/acs.chemrev.2c00788},
    url = {https://doi.org/10.1021/acs.chemrev.2c00788},
}

@article{CALDEIRA1983374,
title = {Quantum tunnelling in a dissipative system},
journal = {Annals of Physics},
volume = {149},
number = {2},
pages = {374-456},
year = {1983},
issn = {0003-4916},
doi = {https://doi.org/10.1016/0003-4916(83)90202-6},
url = {https://www.sciencedirect.com/science/article/pii/0003491683902026},
author = {A.O Caldeira and A.J Leggett}
}

@article{FEYNMAN1963118,
title = {The theory of a general quantum system interacting with a linear dissipative system},
journal = {Annals of Physics},
volume = {24},
pages = {118-173},
year = {1963},
issn = {0003-4916},
doi = {https://doi.org/10.1016/0003-4916(63)90068-X},
url = {https://www.sciencedirect.com/science/article/pii/000349166390068X},
author = {R.P Feynman and F.L Vernon}
}

@article{PhysRevB.110.195148,
  title = {Quasi-{Lindblad} pseudomode theory for open quantum systems},
  author = {Park, Gunhee and Huang, Zhen and Zhu, Yuanran and Yang, Chao and Chan, Garnet Kin-Lic and Lin, Lin},
  journal = {Phys. Rev. B},
  volume = {110},
  issue = {19},
  pages = {195148},
  numpages = {12},
  year = {2024},
  month = {Nov},
  publisher = {American Physical Society},
  doi = {10.1103/PhysRevB.110.195148},
  url = {https://link.aps.org/doi/10.1103/PhysRevB.110.195148}
}

@book{10.1093/acprof:oso/9780199213900.001.0001,
    author = {Breuer, Heinz-Peter and Petruccione, Francesco},
    title = {The Theory of Open Quantum Systems},
    publisher = {Oxford University Press},
    year = {2007},
    month = {01},
    isbn = {9780199213900},
    doi = {10.1093/acprof:oso/9780199213900.001.0001},
    url = {https://doi.org/10.1093/acprof:oso/9780199213900.001.0001},
}

@article{10.1063/1.449017,
    author = {Garg, Anupam and Onuchic, José Nelson and Ambegaokar, Vinay},
    title = {Effect of friction on electron transfer in biomolecules},
    journal = {The Journal of Chemical Physics},
    volume = {83},
    number = {9},
    pages = {4491-4503},
    year = {1985},
    month = {11},
    issn = {0021-9606},
    doi = {10.1063/1.449017},
    url = {https://doi.org/10.1063/1.449017},
}

@article{10.1116/5.0073853,
    author = {Trushechkin, A. S. and Merkli, M. and Cresser, J. D. and Anders, J.},
    title = {Open quantum system dynamics and the mean force Gibbs state},
    journal = {AVS Quantum Science},
    volume = {4},
    number = {1},
    pages = {012301},
    year = {2022},
    month = {03},
    issn = {2639-0213},
    doi = {10.1116/5.0073853},
    url = {https://doi.org/10.1116/5.0073853},
}

@article{10.1063/5.0011599,
    author = {Tanimura, Yoshitaka},
    title = {Numerically “exact” approach to open quantum dynamics: The hierarchical equations of motion ({HEOM})},
    journal = {The Journal of Chemical Physics},
    volume = {153},
    number = {2},
    pages = {020901},
    year = {2020},
    month = {07},
    issn = {0021-9606},
    doi = {10.1063/5.0011599},
    url = {https://doi.org/10.1063/5.0011599},
}

@article{PhysRevResearch.2.043058,
  title = {Generalized theory of pseudomodes for exact descriptions of non-Markovian quantum processes},
  author = {Pleasance, Graeme and Garraway, Barry M. and Petruccione, Francesco},
  journal = {Phys. Rev. Res.},
  volume = {2},
  issue = {4},
  pages = {043058},
  numpages = {12},
  year = {2020},
  month = {Oct},
  publisher = {American Physical Society},
  doi = {10.1103/PhysRevResearch.2.043058},
  url = {https://link.aps.org/doi/10.1103/PhysRevResearch.2.043058}
}

@article{w3nw-hbjc,
  title = {Colloquium: Simulating non-Markovian dynamics in open quantum systems},
  author = {Xu, Meng and Vadimov, Vasilii and Stockburger, J. T. and Ankerhold, J.},
  journal = {Rev. Mod. Phys.},
  volume = {98},
  issue = {2},
  pages = {021002},
  numpages = {26},
  year = {2026},
  month = {May},
  publisher = {American Physical Society},
  doi = {10.1103/w3nw-hbjc},
  url = {https://link.aps.org/doi/10.1103/w3nw-hbjc}
}

@article{PhysRevLett.129.230601,
  title = {Taming Quantum Noise for Efficient Low Temperature Simulations of Open Quantum Systems},
  author = {Xu, Meng and Yan, Yaming and Shi, Qiang and Ankerhold, J. and Stockburger, J. T.},
  journal = {Phys. Rev. Lett.},
  volume = {129},
  issue = {23},
  pages = {230601},
  numpages = {7},
  year = {2022},
  month = {Nov},
  publisher = {American Physical Society},
  doi = {10.1103/PhysRevLett.129.230601},
  url = {https://link.aps.org/doi/10.1103/PhysRevLett.129.230601}
}

@misc{xu2023universalframeworkquantumdissipationminimally,
      title={A Universal Framework for Quantum Dissipation:Minimally Extended State Space and Exact Time-Local Dynamics}, 
      author={Meng Xu and Vasilii Vadimov and Malte Krug and J. T. Stockburger and J. Ankerhold},
      year={2023},
      archivePrefix={arXiv},
      primaryClass={quant-ph},
      url={https://arxiv.org/abs/2307.16790}, 
      note={arXiv:2307.16790}
}

@article{10.1063/1.3490188,
    author = {Chin, Alex W. and Rivas, {\'A}ngel and Huelga, Susana F. and Plenio, Martin B.},
    title = {Exact mapping between system-reservoir quantum models and semi-infinite discrete chains using orthogonal polynomials},
    journal = {Journal of Mathematical Physics},
    volume = {51},
    number = {9},
    pages = {092109},
    year = {2010},
    month = {09},
    issn = {0022-2488},
    doi = {10.1063/1.3490188},
    url = {https://doi.org/10.1063/1.3490188},
}

@Inbook{Nazir2018,
author="Nazir, Ahsan
and Schaller, Gernot",
editor="Binder, Felix
and Correa, Luis A.
and Gogolin, Christian
and Anders, Janet
and Adesso, Gerardo",
title="The Reaction Coordinate Mapping in Quantum Thermodynamics",
bookTitle="Thermodynamics in the Quantum Regime: Fundamental Aspects and New Directions",
year="2018",
publisher="Springer International Publishing",
address="Cham",
pages="551--577",
isbn="978-3-319-99046-0",
doi="10.1007/978-3-319-99046-0_23",
url="https://doi.org/10.1007/978-3-319-99046-0_23"
}

@article{10.1063/1.4866769,
    author = {Woods, M. P. and Groux, R. and Chin, A. W. and Huelga, S. F. and Plenio, M. B.},
    title = {Mappings of open quantum systems onto chain representations and Markovian embeddings},
    journal = {Journal of Mathematical Physics},
    volume = {55},
    number = {3},
    pages = {032101},
    year = {2014},
    month = {03},
    issn = {0022-2488},
    doi = {10.1063/1.4866769},
    url = {https://doi.org/10.1063/1.4866769},
}

@article{10.1063/1.3159671,
    author = {Hughes, Keith H. and Christ, Clara D. and Burghardt, Irene},
    title = {Effective-mode representation of non-Markovian dynamics: A hierarchical approximation of the spectral density. I. Application to single surface dynamics},
    journal = {The Journal of Chemical Physics},
    volume = {131},
    number = {2},
    pages = {024109},
    year = {2009},
    month = {07},
    issn = {0021-9606},
    doi = {10.1063/1.3159671},
    url = {https://doi.org/10.1063/1.3159671},
}

@article{PhysRevLett.118.100401,
  title = {Open Systems with Error Bounds: Spin-Boson Model with Spectral Density Variations},
  author = {Mascherpa, F. and Smirne, A. and Huelga, S. F. and Plenio, M. B.},
  journal = {Phys. Rev. Lett.},
  volume = {118},
  issue = {10},
  pages = {100401},
  numpages = {5},
  year = {2017},
  month = {Mar},
  publisher = {American Physical Society},
  doi = {10.1103/PhysRevLett.118.100401},
  url = {https://link.aps.org/doi/10.1103/PhysRevLett.118.100401}
}

@misc{huang2025unifiederrorboundsperturbations,
      title={Unified error bounds for perturbations of non-Markovian open quantum systems in Gaussian environments}, 
      author={Zhen Huang and Yuanran Zhu and Gunhee Park and Lin Lin},
      year={2025},
      archivePrefix={arXiv},
      primaryClass={quant-ph},
      url={https://arxiv.org/abs/2411.08741}, 
      note={arXiv:2411.08741v2}
}

@article{Liu2025errorboundsopen,
  doi = {10.22331/q-2025-10-28-1896},
  url = {https://doi.org/10.22331/q-2025-10-28-1896},
  title = {Error {B}ounds for {O}pen {Q}uantum {S}ystems with {H}armonic {B}osonic {B}ath},
  author = {Liu, Kaizhao and Lu, Jianfeng},
  journal = {{Quantum}},
  issn = {2521-327X},
  publisher = {{Verein zur F{\"{o}}rderung des Open Access Publizierens in den Quantenwissenschaften}},
  volume = {9},
  pages = {1896},
  month = oct,
  year = {2025}
}

@article{doi:10.1137/23M1570508,
author = {Driscoll, Tobin A. and Nakatsukasa, Yuji and Trefethen, Lloyd N.},
title = {{AAA} Rational Approximation on a Continuum},
journal = {SIAM Journal on Scientific Computing},
volume = {46},
number = {2},
pages = {A929-A952},
year = {2024},
doi = {10.1137/23M1570508},
URL = {https://doi.org/10.1137/23M1570508}
}

@misc{Driscoll2025-sm,
  title     = "Rational function approximation in Julia",
  author    = "Driscoll, Tobin A",
  publisher = "Zenodo",
  year      =  2023,
  url       =  {https://doi.org/10.5281/zenodo.8355791},
  note      =  {doi: 10.5281/zenodo.8355791, version used: v0.1.4}
}

@article{ITensor,
	title={{The ITensor Software Library for Tensor Network Calculations}},
	author={Matthew Fishman and Steven R. White and E. Miles Stoudenmire},
	journal={SciPost Phys. Codebases},
	pages={4},
	year={2022},
	publisher={SciPost},
	doi={10.21468/SciPostPhysCodeb.4},
	url={https://scipost.org/10.21468/SciPostPhysCodeb.4},
}

@article{ITensor-r0.3,
	title={{Codebase release 0.3 for ITensor}},
	author={Matthew Fishman and Steven R. White and E. Miles Stoudenmire},
	journal={SciPost Phys. Codebases},
	pages={4-r0.3},
	year={2022},
	publisher={SciPost},
	doi={10.21468/SciPostPhysCodeb.4-r0.3},
	url={https://scipost.org/10.21468/SciPostPhysCodeb.4-r0.3},
}

@article{PhysRevLett.107.070601,
  title = {Time-Dependent Variational Principle for Quantum Lattices},
  author = {Haegeman, Jutho and Cirac, J. Ignacio and Osborne, Tobias J. and Pi\ifmmode \check{z}\else \v{z}\fi{}orn, Iztok and Verschelde, Henri and Verstraete, Frank},
  journal = {Phys. Rev. Lett.},
  volume = {107},
  issue = {7},
  pages = {070601},
  numpages = {5},
  year = {2011},
  month = {Aug},
  publisher = {American Physical Society},
  doi = {10.1103/PhysRevLett.107.070601},
  url = {https://link.aps.org/doi/10.1103/PhysRevLett.107.070601}
}

@book{Plonka_2001,
title={Dispersive Kinetics},
url={http://dx.doi.org/10.1007/978-94-015-9658-9},
doi={10.1007/978-94-015-9658-9},
publisher={Springer Netherlands},
author={Plonka, Andrzej},
year={2001}
}

@article{PhysRevLett.123.153001,
  title = {Purcell-Enhanced Spontaneous Emission of Molecular Vibrations},
  author = {Metzger, Bernd and Muller, Eric and Nishida, Jun and Pollard, Benjamin and Hentschel, Mario and Raschke, Markus B.},
  journal = {Phys. Rev. Lett.},
  volume = {123},
  issue = {15},
  pages = {153001},
  numpages = {6},
  year = {2019},
  month = {Oct},
  publisher = {American Physical Society},
  doi = {10.1103/PhysRevLett.123.153001},
  url = {https://link.aps.org/doi/10.1103/PhysRevLett.123.153001}
}

@book{anslyn2006modern,
  title={Modern Physical Organic Chemistry},
  author={Anslyn, E.V. and Dougherty, D.A.},
  isbn={9781891389313},
  lccn={2004049617},
  series={G - Reference,Information and Interdisciplinary Subjects Series},
  url={https://books.google.dk/books?id=gY-Sxijk_tMC},
  year={2006},
  publisher={University Science Books}
}

@article{PhysRevLett.127.250601,
  title = {Weak and Ultrastrong Coupling Limits of the Quantum Mean Force Gibbs State},
  author = {Cresser, J. D. and Anders, J.},
  journal = {Phys. Rev. Lett.},
  volume = {127},
  issue = {25},
  pages = {250601},
  numpages = {7},
  year = {2021},
  month = {Dec},
  publisher = {American Physical Society},
  doi = {10.1103/PhysRevLett.127.250601},
  url = {https://link.aps.org/doi/10.1103/PhysRevLett.127.250601}
}

@article{10.1119/10.0005944,
    author = {Nicacio, F.},
    title = {Williamson theorem in classical, quantum, and statistical physics},
    journal = {American Journal of Physics},
    volume = {89},
    number = {12},
    pages = {1139-1151},
    year = {2021},
    month = {12},
    issn = {0002-9505},
    doi = {10.1119/10.0005944},
    url = {https://doi.org/10.1119/10.0005944},
}

@article{6156597e-ed5c-3aca-b6b0-5e41642cc95f,
 ISSN = {00029327, 10806377},
 URL = {http://www.jstor.org/stable/2371633},
 author = {John Williamson},
 journal = {American Journal of Mathematics},
 number = {4},
 pages = {897--911},
 publisher = {Johns Hopkins University Press},
 title = {The Exponential Representation of Canonical Matrices},
 urldate = {2026-08-17},
 volume = {61},
 year = {1939}
}

@article{10.1063/1.462100,
    author = {Colbert, Daniel T. and Miller, William H.},
    title = {A novel discrete variable representation for quantum mechanical reactive scattering via the S‐matrix Kohn method},
    journal = {The Journal of Chemical Physics},
    volume = {96},
    number = {3},
    pages = {1982-1991},
    year = {1992},
    month = {02},
    issn = {0021-9606},
    doi = {10.1063/1.462100},
    url = {https://doi.org/10.1063/1.462100},
}

@article{10.1063/1.3097128,
    author = {Chen, Liping and Shi, Qiang},
    title = {Quantum rate dynamics for proton transfer reactions in condensed phase: The exact hierarchical equations of motion approach},
    journal = {The Journal of Chemical Physics},
    volume = {130},
    number = {13},
    pages = {134505},
    year = {2009},
    month = {04},
    issn = {0021-9606},
    doi = {10.1063/1.3097128},
    url = {https://doi.org/10.1063/1.3097128},
}

@Article{Krug2023,
author={Krug, Malte
and Stockburger, J{\"u}rgen},
title={On stability issues of the {HEOM} method},
journal={The European Physical Journal Special Topics},
year={2023},
month={Dec},
day={01},
volume={232},
number={20},
pages={3219-3226},
issn={1951-6401},
doi={10.1140/epjs/s11734-023-00972-9},
url={https://doi.org/10.1140/epjs/s11734-023-00972-9}
}

@article{PhysRevA.110.022221,
  title = {Systematic and efficient pseudomode method to simulate open quantum systems under a bosonic environment},
  author = {Zhou, L. K. and Jin, G. R. and Yang, W.},
  journal = {Phys. Rev. A},
  volume = {110},
  issue = {2},
  pages = {022221},
  numpages = {10},
  year = {2024},
  month = {Aug},
  publisher = {American Physical Society},
  doi = {10.1103/PhysRevA.110.022221},
  url = {https://link.aps.org/doi/10.1103/PhysRevA.110.022221}
}

@Article{nano11082104,
AUTHOR = {Sánchez-Barquilla, Mónica and Feist, Johannes},
TITLE = {Accurate Truncations of Chain Mapping Models for Open Quantum Systems},
JOURNAL = {Nanomaterials},
VOLUME = {11},
YEAR = {2021},
NUMBER = {8},
ARTICLE-NUMBER = {2104},
URL = {https://www.mdpi.com/2079-4991/11/8/2104},
PubMedID = {34443934},
ISSN = {2079-4991},
DOI = {10.3390/nano11082104}
}

@article{qfkp-jc7s,
  title = {Coupled {Lindblad} Pseudomode Theory for Simulating Open Quantum Systems},
  author = {Huang, Zhen and Park, Gunhee and Chan, Garnet Kin-Lic and Lin, Lin},
  journal = {Phys. Rev. Lett.},
  volume = {136},
  issue = {9},
  pages = {090403},
  numpages = {6},
  year = {2026},
  month = {Mar},
  publisher = {American Physical Society},
  doi = {10.1103/qfkp-jc7s},
  url = {https://link.aps.org/doi/10.1103/qfkp-jc7s}
}

@article{PhysRevLett.80.2661,
  title = {Density Matrix Approach to Local Hilbert Space Reduction},
  author = {Zhang, Chunli and Jeckelmann, Eric and White, Steven R.},
  journal = {Phys. Rev. Lett.},
  volume = {80},
  issue = {12},
  pages = {2661--2664},
  numpages = {0},
  year = {1998},
  month = {Mar},
  publisher = {American Physical Society},
  doi = {10.1103/PhysRevLett.80.2661},
  url = {https://link.aps.org/doi/10.1103/PhysRevLett.80.2661}
}

@article{PhysRevB.57.6376,
  title = {Density-matrix renormalization-group study of the polaron problem in the Holstein model},
  author = {Jeckelmann, Eric and White, Steven R.},
  journal = {Phys. Rev. B},
  volume = {57},
  issue = {11},
  pages = {6376--6385},
  numpages = {0},
  year = {1998},
  month = {Mar},
  publisher = {American Physical Society},
  doi = {10.1103/PhysRevB.57.6376},
  url = {https://link.aps.org/doi/10.1103/PhysRevB.57.6376}
}

@misc{rudolph2025simulatingsamplingquantumcircuits,
      title={Simulating and Sampling from Quantum Circuits with 2D Tensor Networks}, 
      author={Manuel S. Rudolph and Joseph Tindall},
      year={2025},
      archivePrefix={arXiv},
      primaryClass={quant-ph},
      url={https://arxiv.org/abs/2507.11424}, 
      note={arXiv:2507.11424}
}

@misc{githubrepo,
  author = {Vinther, Jonas},
  title = {{github.com/Vinther901/Vinther2026dynamics}},
  howpublished = {GitHub repository},
  url = {https://github.com/Vinther901/Vinther2026dynamics},
  year = {2026}
}

@ARTICLE{Keren2026-jz,
  title     = "Cavity-altered superconductivity",
  author    = "Keren, Itai and Webb, Tatiana A and Zhang, Shuai and Xu, Jikai
               and Sun, Dihao and Kim, Brian S Y and Shin, Dongbin and Zhang,
               Songtian S and Zhang, Junhe and Pereira, Giancarlo and Yao,
               Juntao and Okugawa, Takuya and Michael, Marios H and Vi{\~n}as
               Bostr{\"o}m, Emil and Edgar, James H and Wolf, Stuart and
               Julian, Matthew and Prasankumar, Rohit P and Miyagawa, Kazuya
               and Kanoda, Kazushi and Gu, Genda and Cothrine, Matthew and
               Mandrus, David and Buzzi, Michele and Cavalleri, Andrea and
               Dean, Cory R and Kennes, Dante M and Millis, Andrew J and Li,
               Qiang and Sentef, Michael A and Rubio, Angel and Pasupathy,
               Abhay N and Basov, D N",
  journal   = "Nature",
  publisher = "Springer Science and Business Media LLC",
  volume    =  650,
  number    =  8103,
  pages     = "864--868",
  month     =  feb,
  year      =  2026,
  copyright = "https://creativecommons.org/licenses/by-nc-nd/4.0"
}

@misc{xu2026vacuumdressedsuperconductivitynbnobserved,
      title={Vacuum-dressed superconductivity in {NbN} observed in a high-$Q$ terahertz cavity}, 
      author={Hongjing Xu and Andrey Baydin and Qinyan Yi and I-Te Lu and Ningxu Zhu and T. Elijah Kritzell and Jacques Doumani and Dasom Kim and Fuyang Tay and Angel Rubio and Junichiro Kono},
      year={2026},
      archivePrefix={arXiv},
      primaryClass={physics.optics},
      url={https://arxiv.org/abs/2601.08191}, 
      note={arXiv:2601.08191}
}

@ARTICLE{Shalabney2015-su,
  title     = "Coherent coupling of molecular resonators with a microcavity
               mode",
  author    = "Shalabney, A and George, J and Hutchison, J and Pupillo, G and
               Genet, C and Ebbesen, T W",
  journal   = "Nat. Commun.",
  publisher = "Springer Science and Business Media LLC",
  volume    =  6,
  number    =  1,
  pages     = "5981",
  month     =  jan,
  year      =  2015,
  copyright = "https://creativecommons.org/licenses/by/4.0"
}

@article{https://doi.org/10.1002/anie.202002527,
author = {Pang, Yantao and Thomas, Anoop and Nagarajan, Kalaivanan and Vergauwe, Robrecht M. A. and Joseph, Kripa and Patrahau, Bianca and Wang, Kuidong and Genet, Cyriaque and Ebbesen, Thomas W.},
title = {On the Role of Symmetry in Vibrational Strong Coupling: The Case of Charge-Transfer Complexation},
journal = {Angewandte Chemie International Edition},
volume = {59},
number = {26},
pages = {10436-10440},
doi = {https://doi.org/10.1002/anie.202002527},
url = {https://onlinelibrary.wiley.com/doi/abs/10.1002/anie.202002527},
year = {2020}
}

@article{10.1063/5.0077549,
    author = {Wiesehan, Garret D. and Xiong, Wei},
    title = {Negligible rate enhancement from reported cooperative vibrational strong coupling catalysis},
    journal = {The Journal of Chemical Physics},
    volume = {155},
    number = {24},
    pages = {241103},
    year = {2021},
    month = {12},
    issn = {0021-9606},
    doi = {10.1063/5.0077549},
    url = {https://doi.org/10.1063/5.0077549}
}

@article{https://doi.org/10.1002/nap2.70143,
author = {Vergauwe, Robrecht M. A. and Toppari, J. Jussi and Groenhof, Gerrit},
title = {Toward Reliable Spectroscopic Analysis of Reaction Kinetics in Polaritonic Chemistry},
journal = {Nanophotonics},
volume = {15},
number = {11},
pages = {e70143},
doi = {https://doi.org/10.1002/nap2.70143},
url = {https://onlinelibrary.wiley.com/doi/abs/10.1002/nap2.70143},
year = {2026}
}

@ARTICLE{Sun2024-tp,
  title     = "Theoretical formulation of chemical equilibrium under
               vibrational strong coupling",
  author    = "Sun, Kaihong and Ribeiro, Raphael F",
  journal   = "Nat. Commun.",
  publisher = "Springer Science and Business Media LLC",
  volume    =  15,
  number    =  1,
  pages     = "2405",
  month     =  mar,
  year      =  2024,
  copyright = "https://creativecommons.org/licenses/by/4.0"
}

@ARTICLE{Truhlar1996-rm,
  title     = "Current status of transition-state theory",
  author    = "Truhlar, Donald G and Garrett, Bruce C and Klippenstein, Stephen
               J",
  journal   = "J. Phys. Chem.",
  publisher = "American Chemical Society (ACS)",
  volume    =  100,
  number    =  31,
  pages     = "12771--12800",
  month     =  jan,
  year      =  1996
}

@ARTICLE{Hanggi1990-yf,
  title     = "Reaction-rate theory: fifty years after {Kramers}",
  author    = "H{\"a}nggi, Peter and Talkner, Peter and Borkovec, Michal",
  journal   = "Rev. Mod. Phys.",
  publisher = "American Physical Society (APS)",
  volume    =  62,
  number    =  2,
  pages     = "251--341",
  month     =  apr,
  year      =  1990,
  copyright = "http://link.aps.org/licenses/aps-default-license"
}

@ARTICLE{Lindoy2022-sf,
  title     = "Resonant cavity modification of ground-state chemical kinetics",
  author    = "Lindoy, Lachlan P and Mandal, Arkajit and Reichman, David R",
  journal   = "J. Phys. Chem. Lett.",
  publisher = "American Chemical Society (ACS)",
  volume    =  13,
  number    =  28,
  pages     = "6580--6586",
  month     =  jul,
  year      =  2022
}

@ARTICLE{Li2021-zg,
  title     = "Cavity frequency-dependent theory for vibrational polariton
               chemistry",
  author    = "Li, Xinyang and Mandal, Arkajit and Huo, Pengfei",
  journal   = "Nat. Commun.",
  publisher = "Springer Science and Business Media LLC",
  volume    =  12,
  number    =  1,
  pages     = "1315",
  month     =  feb,
  year      =  2021,
  copyright = "https://creativecommons.org/licenses/by/4.0"
}

@article{PhysRevB.104.119901,
  title = {Erratum: Universal {Lindblad} equation for open quantum systems [Phys. Rev. B 102, 115109 (2020)]},
  author = {Nathan, Frederik and Rudner, Mark S.},
  journal = {Phys. Rev. B},
  volume = {104},
  issue = {11},
  pages = {119901},
  numpages = {2},
  year = {2021},
  month = {Sep},
  publisher = {American Physical Society},
  doi = {10.1103/PhysRevB.104.119901},
  url = {https://link.aps.org/doi/10.1103/PhysRevB.104.119901}
}

@ARTICLE{Ishizaki2005-kr,
  title     = "Quantum dynamics of system strongly coupled to low-temperature
               colored noise bath: Reduced hierarchy equations approach",
  author    = "Ishizaki, Akihito and Tanimura, Yoshitaka",
  journal   = "J. Phys. Soc. Jpn.",
  publisher = "Physical Society of Japan",
  volume    =  74,
  number    =  12,
  pages     = "3131--3134",
  month     =  dec,
  year      =  2005
}

@ARTICLE{Fay2022-bc,
  title     = "A simple improved low temperature correction for the
               hierarchical equations of motion",
  author    = "Fay, Thomas P",
  journal   = "J. Chem. Phys.",
  publisher = "AIP Publishing",
  volume    =  157,
  number    =  5,
  pages     = "054108",
  month     =  aug,
  year      =  2022
}

@article{PhysRevA.86.012126,
  title = {Generalized Schrieffer-Wolff formalism for dissipative systems},
  author = {Kessler, E. M.},
  journal = {Phys. Rev. A},
  volume = {86},
  issue = {1},
  pages = {012126},
  numpages = {10},
  year = {2012},
  month = {Jul},
  publisher = {American Physical Society},
  doi = {10.1103/PhysRevA.86.012126},
  url = {https://link.aps.org/doi/10.1103/PhysRevA.86.012126}
}
\newpage

\renewcommand{\appendixname}{Supplemental Material}

\appendix

\begin{widetext}

\section*{Supplementary Information}

\startcontents[appendix]








\titlecontents{section}
  [0em]
  {\addvspace{0.9ex}\bfseries}
  {\contentslabel{1.8em}}
  {\hspace*{1.8em}}
  {\titlerule*[0.6pc]{.}\contentspage}

\titlecontents{subsection}
  [1.8em]
  {\addvspace{0.15ex}\normalfont}
  {\contentslabel{2.6em}}
  {\hspace*{2.6em}}
  {\titlerule*[0.6pc]{.}\contentspage}

\titlecontents{subsubsection}
  [3.6em]
  {\addvspace{0.1ex}\normalfont\small}
  {\contentslabel{3.4em}}
  {\hspace*{3.4em}}
  {\titlerule*[0.6pc]{.}\contentspage}

\setcounter{tocdepth}{3}
\printcontents[appendix]{}{1}{}

\setcounter{figure}{0}
\renewcommand{\thefigure}{S\arabic{figure}}
\renewcommand{\theHfigure}{supp.\arabic{figure}} 

\setcounter{section}{0}
\renewcommand{\thesection}{S\arabic{section}}
\renewcommand{\theHsection}{supp.\arabic{section}} 

\setcounter{subsection}{0}
\renewcommand{\thesubsection}{\thesection.\arabic{subsection}}
\renewcommand{\theHsubsection}{supp.\arabic{subsection}} 

\setcounter{subsubsection}{0}
\renewcommand{\thesubsubsection}{\thesubsection.\arabic{subsubsection}}
\renewcommand{\theHsubsubsection}{supp.\arabic{subsubsection}}

\setcounter{equation}{0}
\renewcommand{\theequation}{\thesection.\arabic{equation}}
\renewcommand{\theHequation}{supp.\arabic{section}.\arabic{equation}} 

\setcounter{table}{0}
\renewcommand{\thetable}{S\arabic{table}}
\renewcommand{\theHtable}{supp.\arabic{table}} 

\makeatletter
\renewcommand{\p@subsection}{}
\renewcommand{\p@subsubsection}{}
\makeatother

\section{\label{sec:GaussianEnvironments}Gaussian environments}
A generic starting point for considering a system coupled to a harmonic environment, is the mass-weighted Caldeira-Legget model of dissipation \cite{CALDEIRA1983374}: 
\begin{equation}
\label{eq:CLHamiltonian}
    \hat{H} = \hat{H}_S + \frac12\sum_j\left(\hat{p}_j^2 + \omega_j^2\left(\hat{x}_j - \frac{c_j}{\omega_j^2}\hat{S}\right)^2\right)
\end{equation}
The coupling $\frac{\omega_j^2}{2}\left(\hat{x}_j - \frac{c_j}{\omega_j^2}\hat{S}\right)^2 = \frac{\omega_j^2}{2}\hat{x}_j^2 - c_j\hat{x}_j\hat{S} + \frac{c_j^2}{2\omega_j^2}\hat{S}^2$ gives rise to the reorganization term: $\sum_j\frac{c_j^2}{2\omega_j^2}\hat{S}^2=:\lambda\hat{S}^2$. It is important, typically, to include this term as it counters unwarranted bath-induced renormalization of the system. For example, for two harmonic oscillators coupled bilinearly through their coordinates, omitting the reorganization term while holding the bare oscillator frequencies fixed causes one squared normal-mode frequency to become negative at sufficiently strong coupling, signaling an instability. Including the reorganization term keeps the quadratic Hamiltonian positive definite for all finite coupling strengths. 
For brevity, we will often absorb the reorganization term into $\hat{H}_S$.{ Throughout this Supplementary Material, $\hat{H}_S$ and $\hat{S}$ denote a generic system Hamiltonian and system coupling operator; for the model of the main text, $\hat{H}_S=\hat{H}_R$ and $\hat{S}=\hat{R}$.}

By assuming an initially factorized state and with the harmonic environment in a thermal equilibrium, one can formally write the system reduced density matrix \cite{10.1093/acprof:oso/9780199213900.001.0001}:
\begin{equation}
    \rho_S(t) = e^{-i[\hat{H}_S,\bullet] t} \mathcal{T}_S \exp\left[- \int_0^t dt_1 \int_0^{t_1} dt_2 \hat{\mathcal{K}}(t_1,t_2)\right]\rho_S(0) \label{eq:rhoS(t)}
\end{equation}
with the influence superoperator:
\begin{equation}
    \hat{\mathcal{K}}(t_1,t_2)\bullet = \left(\hat{S}(t_1)\bullet - \bullet\hat{S}(t_1)\right)\left(C(t_1-t_2)\hat{S}(t_2)\bullet-\bullet\hat{S}(t_2)C^*(t_1-t_2)\right)
\end{equation}
where $C(t_1-t_2)$ is the bath correlation function (BCF), $\hat{S}(t)=e^{i\hat{H}_St}\hat{S}e^{-i\hat{H}_St}$ the interaction picture system operator coupling to the bath and $\mathcal{T}_S$ the time-ordering operator. 
From the above, we see that the only information we need to model the environment is {exactly} the BCF. We may rewrite this in terms of the \textit{spectral density} $J(\omega)$:
\begin{equation}
\label{eq:SDDefinition}
    J(\omega) = \frac{\pi}{2}\sum_j \frac{c_j^2}{\omega_j}\delta(\omega-\omega_j)
\end{equation}
such that:
\begin{equation}
\label{eq:NoiseSpectrum}
    \int_{-\infty}^{\infty} dt C(t) e^{i\omega t} = S_\beta(\omega) = \frac{2}{1-e^{-\beta\omega}}\left(J(\omega) - J(-\omega)\right)
\end{equation}
This expression can also be found in the {literature}, written with only one $J(\omega)$ with the understanding that $J(\omega)$ is already an odd function or else you employ the odd version $J(\omega) - J(-\omega)$.

Some remarks regarding the BCF and noise spectrum:
\begin{itemize}
    \item $S_\beta(\omega)$ is real and positive and within Fermi's Golden Rule, the transition rate at frequency $\omega$ is proportional to $S_\beta(-\omega)$.
    \item $S_\beta(-\omega) = e^{-\beta\omega}S_\beta(\omega)$ known as the detailed balance condition.
    \item $C(t)$ is complex with:
    \begin{align}
        C(t) &= C^*(-t)\\
        \operatorname{Re}[C(t)] &= \int_0^\infty \frac{d\omega}{\pi} \coth(\frac{\beta\omega}{2})J(\omega)\cos(\omega t)\\
        \operatorname{Im}[C(t)] &= -\int_0^\infty \frac{d\omega}{\pi} J(\omega)\sin(\omega t)
    \end{align}
    \item The reorganization energy is given by $\sum_j\frac{c_j^2}{2\omega_j^2} = \int_0^\infty \frac{d\omega}{\pi}\frac{J(\omega)}{\omega} = -\operatorname{Im}\int_0^\infty C(t) dt $.
\end{itemize}

\section{\label{sec:GeneralSDDerivation}Basis-independent spectral density for general harmonic environments}
The above model considered a collection of uncoupled harmonic oscillators, each coupled to the system. However, it is of interest to consider more general microscopic models where the interactions can be arbitrary. This will allow us to describe the entire harmonic environment, including cavity modes, spectator modes and so on, in terms of a single spectral density (or more if more system operators are coupled to the environment). 
For now, we will absorb the reorganization term into $\hat{H}_S$ and write the system-bath Hamiltonian as:
\begin{equation}
\label{eq:GenericHarmonicBathHamiltonian}
    \hat{\mathcal{H}} = \hat{H}_S + \hat{S}\pmb{c}^T\hat{\pmb{r}} + \frac12\hat{\pmb{r}}^T H \hat{\pmb{r}}
\end{equation}
where $\hat{\pmb{r}}$ is a vector containing the $\hat{x}_j$'s and $\hat{p}_j$'s and $H$ is a positive-definite symmetric matrix that contains the parameters for the bath modes. $\pmb{c}$ is the vector of coupling constants.
We will calculate the BCF, which in this notation, is:
\begin{equation}
    C(t) = \text{Tr}[\pmb{c}^T\hat{\pmb{r}}e^{-\frac{it}{2}\hat{\pmb{r}}^T H \hat{\pmb{r}}}\pmb{c}^T\hat{\pmb{r}}e^{\frac{it}{2}\hat{\pmb{r}}^T H \hat{\pmb{r}}}\rho_B(0)]
\end{equation}
$\rho_B(0)\propto e^{-\frac\beta2\hat{\pmb{r}}^T H \hat{\pmb{r}}}$ is the initial thermal state of the bath.
The goal is to check whether the above permits the same expression for the noise spectrum in terms of a spectral density as the Caldeira-Leggett model and to find a practical expression in terms of $H$ and $\pmb{c}$.
The time-dependence can be simplified by considering:
\begin{align}
    \frac12[\pmb{c}^T\hat{\pmb{r}}, \hat{\pmb{r}}^T H \hat{\pmb{r}}] &= \frac12 c_n H_{ml}(\hat{r}_n\hat{r}_m\hat{r}_l - \hat{r}_m\hat{r}_l\hat{r}_n)\\
    &= \frac12 c_n H_{ml}([\hat{r}_n,\hat{r}_m]\hat{r}_l - \hat{r}_m[\hat{r}_l,\hat{r}_n]) \\
    &= \frac{i}{2}\pmb{c}^T\Omega H\hat{\pmb{r}} - \frac{i}{2}\hat{\pmb{r}}^T H \Omega\pmb{c} = i\pmb{c}^T\Omega H\hat{\pmb{r}}
\end{align}
where we have defined the symplectic matrix $i\Omega_{nm} = [\hat{r}_n,\hat{r}_m]$. Then we can apply the Campbell identity:
\begin{equation}
    e^{-\frac{it}{2}\hat{\pmb{r}}^T H \hat{\pmb{r}}}\pmb{c}^T\hat{\pmb{r}}e^{\frac{it}{2}\hat{\pmb{r}}^T H \hat{\pmb{r}}} = e^{\frac{it}{2}[\bullet,\hat{\pmb{r}}^TH\hat{\pmb{r}}]}\pmb{c}^T\hat{\pmb{r}} = \pmb{c}^T e^{-t\Omega H}\hat{\pmb{r}}
\end{equation}
such that:
\begin{equation}
    C(t) = \pmb{c}^T\text{Tr}[\hat{\pmb{r}}\hat{\pmb{r}}^T\rho_B(0)]e^{t H\Omega}\pmb{c}
\end{equation}
where the trace is still in the Liouville-space sense and not in the matrix sense.
Through the normal mode or Williamson decomposition \cite{6156597e-ed5c-3aca-b6b0-5e41642cc95f, 10.1119/10.0005944}, we may write $H = S^T D S$ where the diagonal matrix $D$ contains the frequencies $\{\omega_j\}$ and $S$ is a symplectic matrix ($S^T\Omega S = \Omega$). Then we can define the following Gaussian unitary:
\begin{equation}
    \hat{\mathcal{U}}\hat{\pmb{r}}\hat{\mathcal{U}}^\dagger = S\hat{\pmb{r}} \quad 
\end{equation}
such that:
\begin{equation}
    \text{Tr}[\hat{\pmb{r}}\hat{\pmb{r}}^T\rho_B(0)] = \text{Tr}[\hat{\mathcal{U}}^\dagger\hat{\pmb{r}}\hat{\pmb{r}}^T\hat{\mathcal{U}}\rho_B^D(0)] = S^{-1}\text{Tr}[\hat{\pmb{r}}\hat{\pmb{r}}^T\rho_B^D(0)] S^{-T}
\end{equation}
where $\rho_B^D(0)$ is the thermal state of the diagonal modes:
\begin{equation}
    \rho_B^D(0) \propto \prod_j e^{-\frac{\beta\omega_j}{2}(\hat{x}_j^2 + \hat{p}_j^2)} 
\end{equation}
Then, we can calculate the trace using that:
\begin{equation}\text{Tr}\left[
    \begin{pmatrix}
        \hat{x}_j \\
        \hat{p}_j
    \end{pmatrix}
    \begin{pmatrix}
        \hat{x}_j &
        \hat{p}_j
    \end{pmatrix}\rho_B^D(0)\right]
    =
    \frac12
    \begin{pmatrix}
        2n(\omega_j) + 1 & i \\
        -i & 2n(\omega_j) + 1
    \end{pmatrix}
\end{equation}
with $n(\omega_j)=1/(e^{\beta\omega_j}-1)$, such that we get:
\begin{equation}
    \text{Tr}[\hat{\pmb{r}}\hat{\pmb{r}}^T\rho_B^D(0)] = \frac12\left(i\Omega + \coth(\frac{\beta}{2}D)\right)
\end{equation}
Putting everything together, we have:
\begin{align}
    C(t) &= \frac12\pmb{c}^T S^{-1} \left(i\Omega + \coth(\frac{\beta}{2}D)\right)  S^{-T}e^{t H \Omega} \pmb{c}
\end{align}
and because of $S^{-1}\Omega S^{-T} = \Omega$ we can rewrite:
\begin{align}
    C(t) &= \frac12\pmb{c}^T\left(i\Omega + S^{-1}\coth(\frac{\beta}{2}D) S^{-T}\right) e^{t H\Omega}\pmb{c}\\
    &= \frac12\pmb{c}^T \left(i\Omega + S^{-1}\coth(\frac{\beta}{2}D) S^{-T}\right) S^Te^{t D\Omega}S^{-T}\pmb{c}\\
    &= \frac12\pmb{c}^TS^{-1}\left(i\Omega + \coth(\frac{\beta}{2}D)\right)e^{t D\Omega}S^{-T}\pmb{c}
\end{align}
Specifically for this kind of diagonal matrix $D$ from the Williamson decomposition, where the diagonal entries come in pairs, we have $\Omega D\Omega=-D$ (equivalently $[\Omega,D]=0$) such that:
\begin{equation}
    e^{tD\Omega } = \sum_{n=0}^\infty \frac{t^n}{n!} (D\Omega)^n = \cos(Dt) + \sin(Dt)\Omega
\end{equation}
Inserting this, we have:
\begin{equation}
    C(t)= \frac12\pmb{c}^TS^{-1}\left(i\Omega + \coth(\frac{\beta}{2}D)\right)(\cos(Dt) + \sin(Dt)\Omega)S^{-T}\pmb{c}
\end{equation}
We may now realize that some of these terms are zero. To see this, we consider the transpose of a scalar, which must equal itself, and utilize the commutativity of $\Omega$ and $D$ to show that it is equal to minus itself and thus equates to zero:
\begin{align}
    \pmb{c}^TS^{-1}\Omega \cos(Dt)S^{-T}\pmb{c} 
    &\overset{\text{transpose}}{=} \pmb{c}^TS^{-1}\cos(Dt)(-\Omega) S^{-T}\pmb{c} = 0\\
    \pmb{c}^TS^{-1}\coth(\frac{\beta}{2}D)\sin(Dt)\Omega S^{-T}\pmb{c} 
    &\overset{\text{transpose}}{=} \pmb{c}^TS^{-1}(-\Omega)\sin(Dt)\coth(\frac{\beta}{2}D) S^{-T}\pmb{c} = 0
\end{align}
Together with $\Omega^2=-\mathds{1}$, we therefore end up with:
\begin{equation}
    \boxed{\quad
    C(t) = \frac12\pmb{c}^TS^{-1}\left(-i\sin(Dt) + \coth(\frac{\beta}{2}D)\cos(Dt)\right)S^{-T}\pmb{c}
    \quad}
\end{equation}
We will now consider the noise spectrum, which is the Fourier transform of the bath correlation function:
\begin{equation}
    S_\beta(\omega) = \int_{-\infty}^\infty C(t) e^{i\omega t}dt
\end{equation}
By utilizing:
\begin{align}
    -i\int_{-\infty}^\infty\sin(Dt)e^{i\omega t}dt&=\pi[\delta(\omega-D) - \delta(\omega+D)]\\
    \int_{-\infty}^\infty\cos(Dt)e^{i\omega t}dt&=\pi[\delta(\omega-D) + \delta(\omega+D)]
\end{align}
where scalars such as $\omega$ are understood to be proportional to the identity matrix $\omega\mathds{1}$ and the delta function with a diagonal matrix as argument, is simply a diagonal matrix with elements $\delta(\omega-D)_{ii} = \delta(\omega-D_{ii})$, we recover:
\begin{align}
    S_\beta(\omega) &= \frac{\pi}{2}\pmb{c}^TS^{-1}\left[\delta(\omega-D)-\delta(\omega+D) + \coth(\frac{\beta}{2}D)[\delta(\omega-D) + \delta(\omega+D)]\right]S^{-T}\pmb{c}\\
    S_\beta(\omega) &= \frac{\pi}{2}\pmb{c}^TS^{-1}\left[\delta(\omega-D)-\delta(\omega+D) + \coth(\frac{\beta}{2}\omega)[\delta(\omega-D) - \delta(\omega+D)]\right]S^{-T}\pmb{c}
\end{align}
where we have used that $\coth$ is an odd function.
By defining the spectral density
\begin{align}
    J(\omega)= \frac{\pi}{2}\pmb{c}^TS^{-1}\delta(\omega-D)S^{-T}\pmb{c}
\end{align}
 we can write the noise spectrum in the familiar form:
\begin{equation}
    S_\beta(\omega) = \left(\coth\left(\frac{\beta}{2}\omega\right) + 1\right)\left(J(\omega) - J(-\omega)\right) = \frac{2}{1-e^{-\beta\omega}}\left(J(\omega) - J(-\omega)\right)
\end{equation}
Exactly like Eq.~\ref{eq:NoiseSpectrum}, except, now we have a more general presentation of the spectral density.

We will now write the spectral density in a way that does not rely on the Williamson decomposition of the matrix $H$.
Simply inserting $D=S^{-T}HS^{-1}$ does not perform the job. Instead, we will use the Sokhotski-Plemelj theorem to write the odd combination of delta functions as:
\begin{equation}
    \frac{\pi}{2}[\delta(\omega-D) - \delta(\omega+D)] =\frac{1}{2}\operatorname{Im}\left[\frac{1}{\omega-D-i0^+}-\frac{1}{\omega+D-i0^+}\right]= \operatorname{Im}\left[\frac{1}{\omega^2-D^2-\operatorname{sgn}(\omega)i0^+}\right]D
\end{equation}
We can then use that:
\begin{equation}
    -D^2 = (-D)D = \Omega D \Omega D = \Omega S^{-T}HS^{-1}\Omega S^{-T}HS^{-1} = S\Omega H\Omega HS^{-1}
\end{equation}
Thus, we can write the odd-combination of the spectral density $J^o(\omega)$ as:
\begin{equation}
    J^o(\omega):=J(\omega) -J(-\omega) 
    = \pmb{c}^TS^{-1} S\operatorname{Im}\left[\frac{1}{\omega^2 + (\Omega H)^2-\operatorname{sgn}(\omega)i0^+}\right]S^{-1} DS^{-T}\pmb{c}
\end{equation}
Lastly, $S^{-1} DS^{-T} = -S^{-1}\Omega D\Omega S^{-T} = -\Omega H \Omega$ and we arrive at:
\begin{equation}
\label{eq:GenericSD}
    \boxed{\quad
    J^o(\omega) 
    = -\pmb{c}^T\operatorname{Im}\left[\frac{1}{\omega^2 + (\Omega H)^2-\operatorname{sgn}(\omega)i0^+}\right]\Omega H \Omega\pmb{c}
    \quad
    \quad}
\end{equation}
The problem of finding the spectral density has now been turned into a matrix inversion problem. The practicality of this becomes apparent in the later examples.

We may simplify the expression further if the examples we consider are trivial in the momentum sector and only couple in the position sector. For example, in the basis $\pmb{r} = (\hat{x}_1,\hat{x}_2,\hdots,\hat{p}_1,\hat{p}_2,\hdots)^T$, we consider:
\begin{equation}
\label{eq:SimpleInMomentumSector}
    \Omega = \begin{pmatrix}
        0 & \mathds{1} \\
        -\mathds{1} & 0
    \end{pmatrix}
    \quad
    H = \begin{pmatrix}
        X& 0\\
        0 & \mathds{1}
    \end{pmatrix}
    \quad \pmb{c} = \begin{pmatrix}
        \pmb{c}_x\\
        0
    \end{pmatrix}
\end{equation}
Then $(\Omega H)^2 = \begin{pmatrix}
    -X & 0\\
    0 & -X
\end{pmatrix}$ and $\Omega H \Omega = -\begin{pmatrix}
        \mathds{1} & 0\\
        0 & X
    \end{pmatrix}$ which results in:
\begin{equation}
\label{eq:SDX}
    \boxed{\quad
    J^o(\omega) = \pmb{c}_x^T\operatorname{Im}\left[\frac{1}{\omega^2 - X - \operatorname{sgn}(\omega)i0^+}\right]\pmb{c}_x
    \quad}
\end{equation}

\subsection{Spectral density of interacting lossy harmonic modes}
Consider $N$ bosons that may couple directly to the system, as described by a vector $\pmb{c}$, and interact among themselves through coupling constants contained in a matrix $\chi$. $\chi$ should be positive-definite. In addition, each boson may be coupled to its own structured bath. Such a scenario is visualized in Fig.~\ref{fig:ModelDichotomy_SI}.
\begin{figure}[h]
    \centering
    \includegraphics[width=0.7\linewidth]{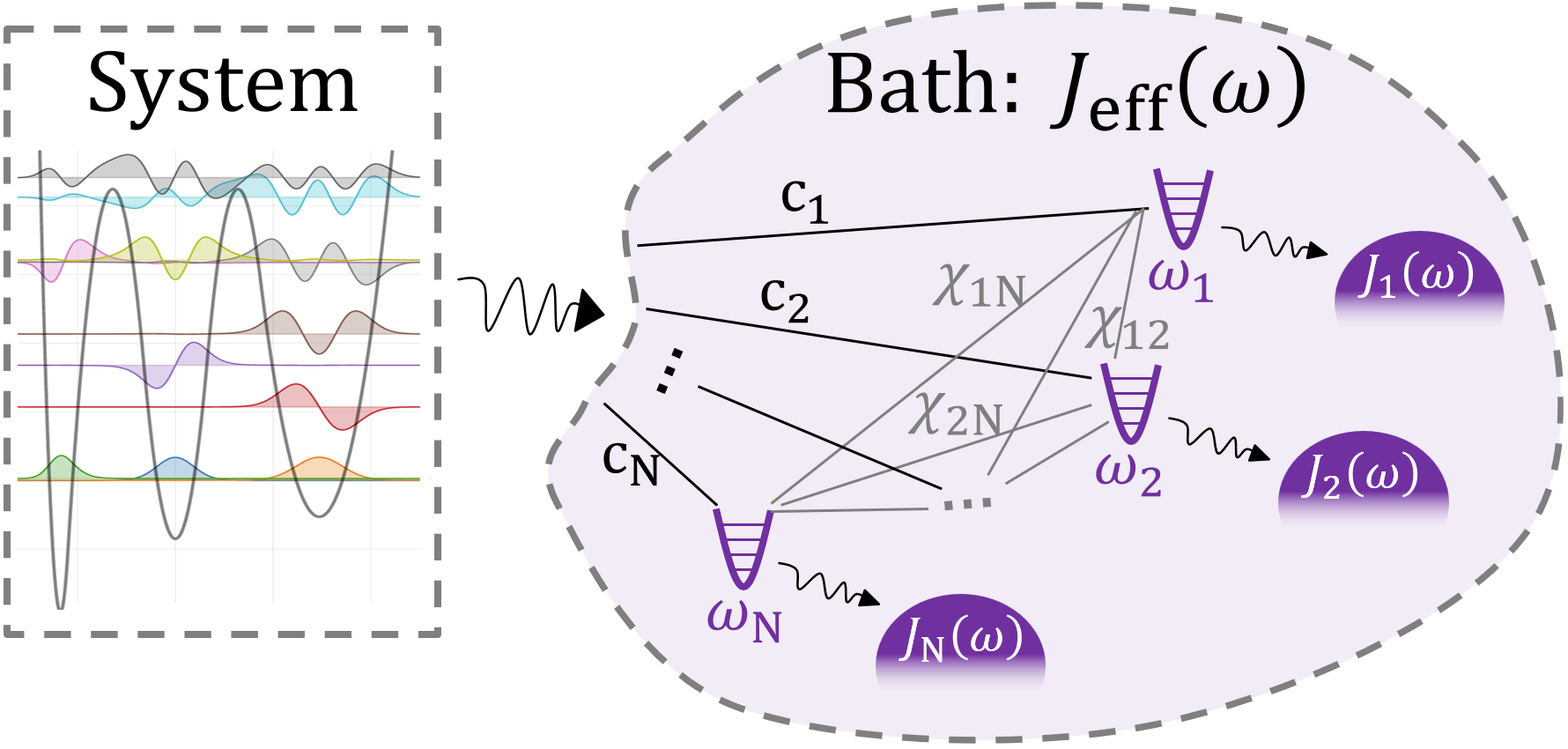}
    \caption{
    Schematic of the efficient modeling of an interacting harmonic environment in terms of an effective spectral density for the full environment (Eq.~\ref{eq:SDInteractingLossyHE}). The example used for 'the system' is the triple well PES from Sec.~\ref{subsec:competingpathways}.}
    \label{fig:ModelDichotomy_SI}
\end{figure}
The resulting Hamiltonian is:
\begin{equation}
\label{eq:FullHamiltonian}
    H = H_R+\lambda\hat{R}^2 + \hat{R}\sum_n^N c_n\hat{x}_n + \sum_n^N\left[\frac{\hat{p}_n^2}{2} +\omega_n^2\frac{\hat{x}_n^2}{2}\right] + \sum_{n\neq m}^N\frac{\chi_{nm}}{2}\hat{x}_n\hat{x}_m + \sum_n^N H_\text{CL}^{(n)}
\end{equation}
$H_R$ is the system Hamiltonian and $\hat{R}$ is the system operator that couples to the harmonic environment. The mass-weighted coordinates of the harmonic environment $\hat{x}_n$, are mutually interacting and they each experience their own {harmonic} environment through the inclusion of the $H_\text{CL}^{(n)}${:}
\begin{equation}
\label{eq:AdditionalCL}
    H_\text{CL}^{(n)} = \sum_l\frac{\hat{p}_{n,l}^2}{2}+\frac{\omega_{n,l}^2}{2}\left(\hat{q}_{n,l}-\frac{c_{n,l}}{\omega_{n,l}^2}\hat{x}_n\right)^2
\end{equation}
which gives rise to the spectral densities $J_n(\omega) = \frac{\pi}{2}\sum_l \frac{c_{n,l}^2}{\omega_{n,l}}\delta(\omega-\omega_{n,l})$ that characterize the local harmonic environment for the $n$'th oscillator $\hat{x}_n$.

This Hamiltonian (Eq.~\ref{eq:FullHamiltonian}), with formally infinitely many degrees of freedom, is of the form Eq.~\ref{eq:GenericHarmonicBathHamiltonian}. In the appropriate basis, one can then write the coupling matrix of the full harmonic environment, $X$ (Eq.~\ref{eq:SimpleInMomentumSector}) in the following block form:

    
    
    

    
\begin{equation}
X = 
    \begin{pmatrix}
        \chi & C^T\\
        C & W
    \end{pmatrix}
\end{equation}
where $W$ is diagonal and contains all the squared frequencies of the $J_i(\omega)$ dofs, $C$ is a matrix with $N$ columns that couples each boson $i$ to its respective bath $J_i(\omega)$ and $\chi$ is the general symmetric $N\times N$ matrix that couples the lossy bosons. The system only couples to the $\chi$ part, such that Eq.~\ref{eq:SDX} becomes:
\begin{equation}
    \boxed{\quad
    \resizebox{.9\hsize}{!}{$
    J_\text{eff}(\omega)=\operatorname{Im}\begin{pmatrix}
        c_1\\
        c_2\\
        \vdots\\
        c_N
    \end{pmatrix}^T
    \begin{pmatrix}
        \omega^2 - \omega_1^2-iJ_1^o(\omega)-\tilde{\mathcal{R}}_1(\omega) & -\chi_{12} & \hdots & -\chi_{1N}\\
        -\chi_{21} & \omega^2 - \omega_2^2-iJ_2^o(\omega)-\tilde{\mathcal{R}}_2(\omega) &\hdots &-\chi_{2N} \\
        \vdots & \vdots&\ddots&\vdots\\
        -\chi_{N1} & -\chi_{N2}& \hdots& \omega^2 - \omega_N^2-iJ_N^o(\omega)-\tilde{\mathcal{R}}_N(\omega)
    \end{pmatrix}^{-1}
    \begin{pmatrix}
        c_1\\
        c_2\\
        \vdots\\
        c_N
    \end{pmatrix}
    $}
    \quad}
    \label{eq:SDInteractingLossyHE}
\end{equation}
on account of the inversion of a block matrix $\begin{pmatrix}
    {A} & {B}\\
    {C} & {D}
\end{pmatrix}^{-1} = \begin{pmatrix}
    ({A - BD^{-1}C})^{-1} & ...\\
    ... & ...
\end{pmatrix}$. The Schur complement produced entries along the diagonal, which we rewrote by again applying the Sokhotski-Plemelj theorem:
\begin{equation}
    \sum_jc_{n,j}^2\frac{1}{\omega^2-\omega_{n,j}^2-\operatorname{sgn}(\omega)i0^+} = \sum_jc_{n,j}^2\left[\operatorname{sgn}(\omega)i\pi\delta(\omega^2-\omega_{n,j}^2)+\mathcal{P}\left(\frac{1}{\omega^2-\omega_{n,j}^2}\right)\right] = iJ_n^o(\omega)+\mathcal{R}_n(\omega)
\end{equation}
We applied the identity: $\operatorname{sgn}(\omega)\delta(\omega^2 - \nu^2) = \frac{1}{2\nu}\left[\delta(\omega-\nu)-\delta(\omega+\nu)\right]$ (for $\nu >0$) and defined the principal value integral:
\begin{align}
    \mathcal{R}_n(\omega) &= \frac{2}{\pi}\mathcal{P}\int_0^\infty \frac{\omega'J_n(\omega')}{\omega^2-\omega'^2}d\omega'\\
    &=\mathcal{P}\int_{-\infty}^\infty\frac{d\omega'}{\pi}\frac{\omega'J_n^o(\omega')}{\omega^2-\omega'^2}\quad
    \\
    &=\mathcal{P}\int_{-\infty}^\infty\frac{d\omega'}{2\pi}J_n^o(\omega')\left(\frac{1}{\omega-\omega'} - \frac{1}{\omega+\omega'}\right)\\
    &=\mathcal{P}\int_{-\infty}^\infty\frac{d\omega'}{\pi}\frac{J_n^o(\omega')}{\omega-\omega'}
\end{align}
Lastly, we reintroduced the neglected reorganization energies by the following consideration:
\begin{equation}
    \hat{H} = \dots + \frac{\hat{p}_n^2}{2}+\frac{\omega_n^2}{2}\hat{x}_n^2+\hat{x}_n^2\overbrace{\sum_j \frac{c_{n,j}^2}{2\omega_{n,j}^2}}^{\int_0^\infty \frac{d\omega}{\pi}\frac{J_n(\omega)}{\omega}}+\dots
    \quad\Rightarrow\quad\omega_n^2\rightarrow\omega_n^2+2\int_0^\infty \frac{d\omega}{\pi}\frac{J_n(\omega)}{\omega}
\end{equation}
which means that we may incorporate them into $\mathcal{R}_n(\omega)$ in the following way:
\begin{equation}
    \tilde{\mathcal{R}}_n(\omega) = \mathcal{R}_n(\omega)+2\int_0^\infty \frac{d\omega'}{\pi}\frac{J_n(\omega')}{\omega'}=\omega\mathcal{P}\int_{-\infty}^\infty\frac{d\omega'}{\pi}\frac{J_n^o(\omega')}{(\omega-\omega')\omega'}
\end{equation}
It will be a helpful reminder, that if the pole structure of the bath spectral density is simple e.g. if $J_n^o(\omega)=\frac{f(\omega)}{g(\omega)}$ where $g(\omega)$ accounts for the simple poles, then:
\begin{equation}
    \tilde{\mathcal{R}}(\omega) = -2\omega\operatorname{Im}\left(\sum_{\omega_p}\frac{f(\omega_p)}{(\omega-\omega_p)\omega_pg'(\omega_p)}\right)
\end{equation}
where $\omega_p$ are the zeros of $g(\omega)$ in the upper-half of the complex plane. 
For the Drude-Lorentz spectral density
\begin{equation}
    J^{(DL)}(\omega)=\frac{2\lambda\gamma\omega}{\omega^2+\gamma^2}
    \quad\quad
    \tilde{\mathcal{R}}^{(DL)}(\omega)=\frac{2\lambda\omega^2}{\omega^2+\gamma^2}
\end{equation}
For the following section, and in the main text in general, we use $J(\omega)$ and its odd-version $J^o(\omega)=J(\omega)-J(-\omega)$ interchangeably.
\subsection{Spectral density of interacting cavity and spectator mode, averaged over molecule orientation}
\label{sec:AngleAveragedSpectralDensity}
We consider the spectral density in Eq.~\ref{eq:spectraldensity} while ignoring the phonon bath:
\begin{equation}
    J(\omega) = \operatorname{Im}\frac{\Gamma_c(\omega)c_Q^2+\eta_R^2\Gamma_Q(\omega) 2\omega_c^3\eta_c^2+4\eta_Q\eta_R(\omega^2-i\alpha\omega)c_Q\omega_c\eta_c^2+4\eta_Q^2\eta_R^2(\omega^2-i\alpha\omega)\omega_c^2\eta_c^4}{\Gamma_c(\omega)\Gamma_Q(\omega) - 2(\omega^2-i\alpha\omega)\omega_c\eta_c^2\eta_Q^2}
\end{equation}
where $\Gamma_c(\omega) = \omega^2-\omega_c^2-i\alpha\omega$ and $\Gamma_Q(\omega)=\omega^2-\omega_Q^2-\left(i+\frac{\omega}{\gamma_Q}\right)J^{(DL)}_Q(\omega)$.

We will now introduce $\eta_R=\tilde{\eta}_R\cos(\theta-\phi)$ and $\eta_Q=\tilde{\eta}_Q\cos(\phi)$ where $\theta$ is the angle between the {reaction coordinate} and spectator mode{ dipole moments} and $\phi$ is the angle between the cavity polarization and the spectator mode dipole moment.
We are interested in averaging over $\phi$
\begin{equation}
    \langle J(\omega)\rangle_\phi= \frac{1}{2\pi}\int_0^{2\pi}J(\omega)d\phi 
\end{equation}
We apply $\cos(\theta-\phi) = \cos(\theta)\cos(\phi)+\sin(\theta)\sin(\phi)$ and utilize that the average over odd functions equals zero which together with polynomial division gives:
\begin{align}
\label{eq:AngleAveraging}
    \left\langle\frac{1}{a-b\cos^2(\phi)}\right\rangle_\phi &=\frac{1}{a\sqrt{1-b/a}}\\
    \left\langle\frac{\eta_R^2}{a-b\cos^2(\phi)}\right\rangle_\phi&=\tilde{\eta}_R^2\left(\frac{\sin^2(\theta)}{a\sqrt{1-b/a}} + \frac{\cos(2\theta)}{b}\left(\frac{1}{\sqrt{1-b/a}}-1\right)\right)\\
    \left\langle\frac{\eta_R\eta_Q}{a-b\cos^2(\phi)}\right\rangle_\phi&=\tilde{\eta}_R\tilde{\eta}_Q\frac{\cos(\theta)}{b}\left(\frac{1}{\sqrt{1-b/a}}-1\right)\\
    \left\langle\frac{\eta^2_R\eta^2_Q}{a-b\cos^2(\phi)}\right\rangle_\phi&=\tilde{\eta}^2_R\tilde{\eta}^2_Q\left(\frac{\sin^2(\theta)}{b}\left(\frac{1}{\sqrt{1-b/a}}-1\right) + \cos(2\theta)\left(\frac{a}{b^2}\left(\frac{1}{\sqrt{1-b/a}}-1\right)-\frac{1}{2b}\right)\right)
\end{align}
with $a = \Gamma_c(\omega) \Gamma_Q(\omega)$ and $b=2\tilde{\eta}_Q^2(\omega^2-i\alpha\omega)\omega_c\eta_c^2$ and a branch cut along the negative real axis.

\section{Reaction coordinate potentials and vibrational state details}
\label{sec:RxNPES}
We consider a model system which consists of a reaction coordinate on a one dimensional potential energy surface.
\begin{equation}
    H_R = \frac{\hat{P}^2}{2M} + V(\hat{R})
\end{equation}
\subsection{Double well}
One way to parameterize the symmetric double well potential is as a quartic: $V_\text{sym}(\hat{R}) = E_b\left(1 - \left(\frac{\hat{R}}{R_0}\right)^2\right)^2$
However, we are interested in a parameterization where we have control of the product-, reactant- and barrier frequencies independently. To this end, we consider a piecewise function of quadratic polynomials.
\begin{equation}
    V(R) =
    \begin{cases}
        \frac{1}{2}M\omega^2_l(R+R_0^l)^2 -E_l& R<R_l \\
        E_b - \frac{1}{2}M\omega^2_0 R^2 & R_l \leq R \leq R_r \\
        \frac{1}{2}M\omega^2_r (R+R_0^r)^2 -E_r& R_r < R
    \end{cases}
\end{equation}
where we only have to specify the frequencies and offsets, with the location of the parabolas and the regions given by:
\begin{align}
    R_0^l &= \sqrt{\frac{2(\omega^2_0+\omega^2_l)(E_b + E_l)}{M\omega^2_0 \omega^2_l}}\\
    R_l &= -\frac{\omega^2_l}{\omega^2_0 + \omega^2_l}R_0^l\\
    R_0^r &= -\sqrt{\frac{2(\omega^2_0+\omega^2_r)(E_b + E_r)}{M\omega^2_0 \omega^2_r}}\\
    R_r &= -\frac{\omega^2_r}{\omega^2_0 + \omega^2_r}R_0^r
\end{align}
The definition of $\omega_l$ and $\omega_r$ {differs} from that used in the main text, where they referred to the energy spacing between states in the left and right well, respectively. Similarly, $\omega_0$ has also been used to describe the transition frequency of the symmetric double well or the spacing between the states in the middle well of the triple well PES described below. The meaning should be clear from context.


\subsubsection{Derivation}
We require any contiguous pair of parabolas to touch at a single point:
\begin{align}
    &\quad a(R+b)^2-c = C-A(R+B)^2\\
    \Rightarrow &\quad 0= (a+A)R^2 + 2(ab+AB)R+ab^2+AB^2-c-C
\end{align}
Where a,b,c and A,B,C denote the parameters of the well and barrier, respectively. We require a unique solution for $R$, which is the case when the discriminant is zero (for real parameters):
\begin{equation}
    R = -\frac{ab+AB}{a+A} \Leftrightarrow (ab+AB)^2=(a+A)(ab^2 + AB^2 -c -C)
\end{equation}
Furthermore, we would like to solve for either b or B since the other parameters are fixed. To this end, the rightmost identity can be rewritten as:
\begin{equation}
    (b-B)^2 = \frac{(a+A)(c+C)}{aA}
\end{equation}
Associating a,b,c and A,B,C with the parameters of the potential results in the expressions above.

\subsection{Triple well}
In order to study competing pathways, we consider a potential energy surface with three wells. We can utilize the same parameterization as above, which results in:
\begin{equation}
\label{eq:TripleWellPotential}
V(R) =
\begin{cases}
\frac{1}{2}M\omega^2_{ll}(R+R_0^{ll})^2-E_{ll} & -\infty < R \leq R_{ll} \\
E_l - \frac{1}{2}M\omega^2_l(R+R_0^l)^2 & R_{ll} < R \leq R_{l} \\
\frac{1}{2}M\omega^2_0R^2-E_0 & R_l < R \leq R_r \\
E_r-\frac{1}{2}M\omega^2_r(R+R_0^r)^2 & R_r < R \leq R_{rr} \\
\frac{1}{2}M\omega^2_{rr}(R+R_0^{rr})^2 - E_{rr} & R_{rr} < R < \infty
\end{cases}
\end{equation}
with
\begin{align}
    R_0^l &= \sqrt{\frac{2(\omega^2_0+\omega^2_l)(E_0+E_l)}{M\omega^2_0\omega^2_l}}\\
    R_l &=-\frac{\omega^2_l}{\omega^2_0+\omega^2_l}R_0^l\\
    R_0^{ll} &= R_0^l+\sqrt{\frac{2(\omega^2_{ll}+\omega^2_l)(E_{ll}+E_l)}{M\omega^2_{ll}\omega^2_l}} \\
    R_{ll} &= -\frac{\omega^2_lR_0^l + \omega^2_{ll}R_0^{ll}}{\omega^2_l + \omega^2_{ll}}\\
    R_0^r&=-\sqrt{\frac{2(\omega^2_0+\omega^2_r)(E_0+E_r)}{M\omega^2_0\omega^2_r}}\\
    R_r &= -\frac{\omega^2_r}{\omega^2_0+\omega^2_r}R_0^r\\
    R_0^{rr}&=R_0^r-\sqrt{\frac{2(\omega^2_{rr}+\omega^2_r)(E_{rr}+E_r)}{M\omega^2_{rr}\omega^2_r}}\\
    R_{rr} &= -\frac{\omega^2_rR_0^r + \omega^2_{rr}R_0^{rr}}{\omega^2_r + \omega^2_{rr}}
\end{align}

The ground state of the different wells can then be made approximately degenerate, by specifying the offset energy as $E=\frac{\omega_\text{well}}{2}$. This was done for {the }triple well potential in {Fig.}~\hyperref[fig:competing_pathways]{\ref*{fig:competing_pathways}a}, with some additional fine-tuning to get the excited-state degeneracy seen there.
Alternatively, which was done for the double wells in {Fig.}~\ref{fig:asymmetric_rates}, one can instead specify $E_{l,r}=0$ and add a linear term $\frac{1}{2}\frac{\omega_l - \omega_r}{R_0^l-R_0^r}\hat{R}$ to the potential. Probably, these approaches are equivalent even though we do not attempt to show that here.

\subsection{Localized states}
\label{sec:StateLocalization}
In order to analyze the dynamics in terms of intra- and inter-well dynamics, we consider localized states which are not eigenstates of the reaction coordinate Hamiltonian $H_R=\sum_n E_n|n\rangle\langle n|$. We obtain these localized states by diagonalizing the reaction coordinate operator in the sub-space of the relevant delocalized energy-states, e.g.:
\begin{equation}
    \begin{pmatrix}
        \langle n_1|\hat{R}| n_1\rangle & \langle n_1|\hat{R}| n_2\rangle & \langle n_1|\hat{R}| n_3\rangle\\
        \langle n_2|\hat{R}| n_1\rangle & \langle n_2|\hat{R}| n_2\rangle & \langle n_2|\hat{R}| n_3\rangle\\
        \langle n_3|\hat{R}| n_1\rangle & \langle n_3|\hat{R}| n_2\rangle & \langle n_3|\hat{R}| n_3\rangle
    \end{pmatrix}
    =
    O
    \overbrace{D}^\text{diag}
    O^T
\end{equation}
The localized states $|l_{1,2,3}\rangle$ are then given by:
\begin{equation}
    |l_i\rangle = \sum_{j=1,2,3}O_{ji} |n_j\rangle
\end{equation}
In {Fig.}~\ref{fig:EstablishMethod}, the two ground states and the two excited states were hybridized to produce the localized states shown.
In {Fig.}~\ref{fig:asymmetric_rates}, even though the eigenstates were already approximately localized due to the broken symmetry of the potentials, we still hybridized the ground states and also the two states closest to $\text{Frequency}=1200\,\text{cm}^{-1}$ to obtain the localized states shown.
In {Fig.}~\hyperref[fig:competing_pathways]{\ref*{fig:competing_pathways}a} the three excited states around $1570\,\text{cm}^{-1}$ were localized with this process.

\subsection{Numerics}
For the simulations, the reaction coordinate was expressed in the eigen basis of the system Hamiltonian (without the reorganization term).
The system Hamiltonian was numerically diagonalized after expressing it in the Discrete Variable Representation (DVR) based on the sinc-function \cite{10.1063/1.462100}. Sinc-DVR resembles that of discretizing the position basis at $N$ points spaced by $\Delta R$, where operators that are a function of $\hat{R}$ remain diagonal in this representation. However, instead of a finite-difference representation of the momentum operator, in sinc-DVR, the momentum operator is instead represented as:
\begin{align}
    (\hat{P})_{nm} &=
    \begin{cases}
        0 & n=m\\
        \frac{-i(-1)^{n-m}}{\Delta R(n-m)} & n \neq m
    \end{cases}\\
    (\hat{P}^2)_{nm} &=
    \begin{cases}
        \frac{\pi^2}{3\Delta R^2} & n=m\\
        \frac{2(-1)^{n-m}}{\Delta R^2(n-m)^2} & n \neq m
    \end{cases}
\end{align}
$N$ and $\Delta R$ was chosen such that convergence was observed in the eigenvalue of the relevant states. 



\section{\label{sec:HEOMDetails}Details on hierarchical equations of motion}

We refer to our choice of method for simulating the dynamics as HEOM \cite{10.1063/5.0011599}, but instead of viewing the equations of motion as a hierarchy of coupled differential equations, we treat it as the system coupled to a collection of bosons. See e.g. {Refs.}~\citenum{PhysRevResearch.2.043058, PhysRevB.110.195148, PhysRevLett.129.230601, xu2023universalframeworkquantumdissipationminimally, w3nw-hbjc}.
For example, the master equation \cite{xu2023universalframeworkquantumdissipationminimally, w3nw-hbjc, PhysRevB.110.195148}:
\begin{align}
\label{eq:QuasiLindblad}
    \frac{d\varrho}{dt} = \mathcal{L}\varrho = -i\Bigg[\hat{H}_S &+ \sum_{k}\operatorname{Im}\left(z_k\right)\hat{a}_k^\dagger\hat{a}_{k} + \sum_{k}\operatorname{Re}\left(\sqrt{d_k}\right)\hat{S}\left(\hat{a}_k+\hat{a}_k^\dagger\right),\varrho\Bigg] \nonumber \\
    &+ 2\sum_{k}\operatorname{Re}\left(z_k\right)\left(\hat{a}_{k}\varrho\hat{a}_k^\dagger - \frac12 \left\{\hat{a}_k^\dagger\hat{a}_{k},\varrho\right\}\right) \\
    &- 2\sum_{k}\operatorname{Im}\left(\sqrt{d_k}\right)\left(\hat{a}_k\varrho\hat{S} - \frac12 \left\{\hat{S}\hat{a}_k,\varrho\right\}\right) \nonumber \\
    &- 2\sum_{k}\operatorname{Im}\left(\sqrt{d_k}\right)\left(\hat{S}\varrho\hat{a}_k^\dagger - \frac12 \left\{\hat{a}_k^\dagger\hat{S},\varrho\right\}\right) \nonumber 
\end{align}
constitutes an open quantum system for the system $S$ with a harmonic environment of $K$ bosons with BCF:
\begin{equation}
    C(t) = \sum_k^K d_ke^{-z_kt}
\end{equation}
provided that the initial state factorizes as $\varrho(t=0) = \rho_S(0)\otimes^K |0\rangle\langle 0|$.
The noise spectrum is consequently:
\begin{equation}
\label{eq:EffectiveNoiseSpectrum}
    S_\beta(\omega) = \sum_k^K 2\frac{\operatorname{Re}(d_kz_k^*) - \operatorname{Im}(d_k)\omega}{\operatorname{Re}(z_k)^2 + (\omega - \operatorname{Im}(z_k))^2}
\end{equation}
In order to better understand the generator in Eq.~\ref{eq:QuasiLindblad}, consider $\operatorname{Im}(\sqrt{d_k})=0$. In such cases, the structure is equivalent to a Lindbladian (a generator for a completely positive, trace preserving dynamical semigroup). This would constitute a physical setup, where the system is coupled to bosons with Markovian loss and the noise spectrum is decomposed as a sum of Lorentzians. However, in general, this will not be an efficient decomposition and $K$ would need to be too large to be practical. Allowing $\operatorname{Im}(\sqrt{d_k}) \neq 0$ alleviates this problem, since the terms in Eq.~\ref{eq:EffectiveNoiseSpectrum} are now able to acquire negative values and may cancel {each other} in a manner that assists the decomposition \cite{PhysRevLett.129.230601}. Allowing this comes at the cost of the two terms in Eq.~\ref{eq:QuasiLindblad}, which breaks the complete positivity of the generator. This means that, even though Eq.~\ref{eq:QuasiLindblad} is trace-preserving ($\text{Tr}[\dot{\varrho}]=0$) the full density matrix $\varrho(t)$ may acquire negative eigenvalues and no longer corresponds to a physical setup. However, crucially, the reduced density matrix $\rho_S(t) = \text{Tr}_K\left[\varrho(t)\right]$ is still a physical density matrix\cite{xu2023universalframeworkquantumdissipationminimally, w3nw-hbjc, PhysRevB.110.195148}{.}

Ref.~\citenum{Krug2023} pointed to a potential problem. Namely, when cutoffs of the auxiliary modes are inevitably introduced for numerical implementation, the evolution can diverge in a way that is seemingly not just remedied by increasing the cutoff. However, Ref.~\citenum{PhysRevB.110.195148} notes that the presence of the divergence depends on the choice of 'gauge' when decomposing $C(t-t')$. This gauge-freedom is made manifest in the master equation by applying a similarity transformation using $\kappa_k^{\hat{a}_k^\dagger\hat{a}_k}$, which does not change the initial and final conditions for the evolution, but simply transforms $\hat{a}_k\rightarrow\hat{a}_k/\kappa_k$ and $\hat{a}_k^\dagger\rightarrow\kappa_k\hat{a}^\dagger_k$. The $\kappa_k$'s can then be chosen such that norm of the CP-breaking term is minimized (but non-zero), which, in that case, remedied the divergence. In the case of a single system operator interacting with the environment, i.e. a scalar valued BCF, the optimal gauge is known and Eq.~\ref{eq:QuasiLindblad} is written in this gauge.
The ubiquitous and successful application of these CP-breaking formulations suggests that these concerns are not dire and also, in our experience, Eq.~\ref{eq:QuasiLindblad} has yet to exhibit divergent behavior.


Nonetheless, a lack {of} understanding how and when such divergent behavior arises and because it may be related to the CP-breaking of these types of methods, could motivate the search for Markovian embeddings that {preserve} complete-positivity. There are already numerous formulations that achieve this (see {Refs.}~\citenum{10.1063/1.3490188, 10.1063/1.4866769, nano11082104, PhysRevA.110.022221, qfkp-jc7s} to name a few).
However, an overview of the drawbacks for each and every method, such as computational complexity and robustness, has yet to be developed. Thus, the tried and tested HEOM formulations remain our choice of method.



Another useful version of HEOM is obtained from Eq.~\ref{eq:QuasiLindblad} by applying a similarity transformation under $e^{\hat{a}_k\bullet\hat{a}_k^\dagger}$, which has the following effects \cite{xu2023universalframeworkquantumdissipationminimally, w3nw-hbjc}: 
\begin{align}
    \rho_S(t) = \text{Tr}_K\left[\varrho(t)\right] &\rightarrow\langle \pmb{0}|\varrho(t)|\pmb{0}\rangle\\
    \hat{a}_k^\dagger\bullet & \rightarrow \hat{a}_k^\dagger\bullet+ \bullet\hat{a}_k^\dagger \\
    \bullet\hat{a}_k & \rightarrow  \bullet\hat{a}_k +\hat{a}_k\bullet 
\end{align}
with $\hat{a}_k\bullet$ and $\bullet\hat{a}_k^\dagger$ unchanged.
Now, the reduced density matrix for the system can be obtained by simply projecting the auxiliary modes onto the vacuum \cite{PhysRevLett.129.230601, xu2023universalframeworkquantumdissipationminimally, w3nw-hbjc}. This is helpful for both the hierarchical approaches, where you can now read-off the density matrix from the lowest tier and also tensor networks, where it allows you to place left and right dofs of the auxiliary modes independently of each other, since the trace could otherwise introduce loops that in the worst case could be difficult to contract.
The price for this benefit, is that the master equation:
\begin{equation}
\label{eq:FPHEOM_SM}
    \dot{\varrho}(t) =  
    -i\left[\hat{H}_S,\varrho(t)\right] + \sum_k^K\Bigg\{
    -z_k\hat{a}_k^\dagger\hat{a}_k\varrho(t) -i\sqrt{d_k}\left(\hat{a}_k + \hat{a}_k^\dagger\right)\hat{S}\varrho(t) -i\sqrt{d_k^*}\hat{S}\varrho(t) \hat{a}^\dagger_k + H.C.\Bigg\}
    %
\end{equation}
{is} no longer trace {preserving}:
\begin{equation}
    \text{Tr}\left[\dot{\varrho}(t)\right] =\sum_k -2\operatorname{Re}(z_k)\text{Tr}\left[\hat{a}_k^\dagger\hat{a}_k\varrho(t)\right] -i\text{Tr}\left[\left(\sqrt{d_k}\hat{a}_k^\dagger -\sqrt{d_k^*}\hat{a}_k\right)\hat{S}\varrho(t)\right]
\end{equation}
For weak to moderate coupling, this does not pose a problem and on the contrary, we experienced that simulating the dynamics with Eq.~\ref{eq:FPHEOM_SM} required fewer computational resources than Eq.~\ref{eq:QuasiLindblad} (e.g. due to a lower bond-dimension). However, we did not carry out a precise comparison and it remains anecdotal information.
Nonetheless, for strong coupling scenarios the trace can grow uncontrollably and one should resort to the trace preserving formulation, Eq.~\ref{eq:QuasiLindblad}.

\subsection{Decomposing the bath correlation function}
\label{sec:FittingBCF}
We will now describe how to obtain the parameters $d_k$ and $z_k$. 
There exists multiple methods to fit a given bath correlation function by a sum of complex exponentials. We rely on the AAA-algorithm \cite{doi:10.1137/23M1570508, Driscoll2025-sm} to efficiently acquire the decomposition in Eq.~\ref{eq:BCFSumOfExponentials}.
In essence, $S_\beta(\omega)$ is decomposed as a sum of simple poles $p_n$ with residues $r_n$ within a window of frequencies $\omega\in [{\omega_l},{\omega_h}]$, such that:
\begin{equation}
    d_k = -i\omega_-r_k
    \quad\quad
    z_k = i(\omega_-p_k +\omega_+)
\end{equation}
where $\omega_\pm = ({\omega_h} \pm {\omega_l})/2$ and $k$ labels the poles in the lower half of the complex plane. The decomposition can usually be carried out to numerical accuracy with $K$ on the order of $10-100$, depending on the intricacies of the noise spectrum.

The reorganization energy in terms of the effective spectral density can be written as:
\begin{align}
    \lambda 
    &=\frac{1}{\pi}\int_0^\infty d\omega \frac{J(\omega)}{\omega} \\
    &=\int_0^\infty \frac{d\omega}{2\pi}(1 - e^{-\beta\omega})\frac{S_\beta(\omega)}{\omega} \\
    &=\mathcal{P}\int_{-\infty}^\infty \frac{d\omega}{2\pi}\frac{S_\beta(\omega)}{\omega} \\
    &= \mathcal{P}\int_{-\infty}^{\infty}\frac{d\omega}{2\pi}\int_{-\infty}^\infty dt C(t)\frac{e^{i\omega t}}{\omega} \\
    &=\frac{i}{2}\int_{-\infty}^\infty dt C(t)\text{sgn}(t) \\
    &= -\int_0^\infty dt\operatorname{Im}\left[C(t)\right]
\end{align}
by using that $S_\beta(\omega) = \frac{2}{1 - e^{-\beta\omega}}J(\omega) = e^{\beta\omega}S_\beta(-\omega)=\int_{-\infty}^\infty dt C(t) e^{i\omega t}$ and $C(t) = C^*(-t)$.
Then the reorganization energy for our emulating bath is easily obtained as Eq.~\ref{eq:HEOMReorganizationEnergy}.

\subsection{\label{sec:EfficientMarkovCorrection}Effective Markovian correction}
As detailed in Sec.~\ref{sec:HEOMDetails}, the noise spectrum is decomposed into contributions from each auxiliary mode. We have found that some of these auxiliary modes constitute a Markovian bath. Thus, instead of treating these degrees of freedom explicitly, we can incorporate them as a jump operator on the system degree of freedom. This can be made precise using the Universal Lindblad Equation (ULE) by Nathan and Rudner~\cite{PhysRevB.102.115109, PhysRevB.104.119901}. Their rigorous analysis proves that a Gaussian environment constitutes a Markovian bath once the characteristic correlation time of the bath $\tau$ {remains} small compared to the timescale of bath-induced evolution $\Gamma^{-1}$. Once $\Gamma\tau\ll1$ the Markovian dynamics are accurately described by introducing the jump operator $\hat{L} = \sum_{n,m}\sqrt{S_\beta(E_m - E_n)}R_{nm}|n\rangle\langle m|$ for the system. They\cite{PhysRevB.104.119901} also provide an expression for the Hermitian Lamb shift, which for Markovian noise {spectra} (e.g. $S_\beta(\omega)\sim\text{constant}$) {cancels} the reorganization term: 
\begin{equation}
    H_\text{Lamb Shift} = -\sum_{n,m,l} \mathcal{P}\int_{-\infty}^\infty \frac{d\omega}{2\pi}\frac{\sqrt{S_\beta(\omega+E_n-E_l)S_\beta(\omega+E_m-E_l)}}{\omega}R_{nl}R_{lm}|n\rangle\langle m|\approx-\int_{-\infty}^\infty\frac{d\omega}{2\pi}\frac{S_\beta(\omega)}{\omega} \hat{R}^2 = -\lambda \hat{R}^2
\end{equation}
Thus, for the auxiliary modes that are deemed Markovian, we simply remove their contribution to the reorganization energy.

In their exposition, they have scaled the system operator to have unit spectral norm, such that the complete error bound is $||\hat{R}||^2\Gamma\tau$, which becomes trivial for the unbounded operator $\hat{R}$. However, the system Hilbert space is anyways truncated when carrying out the dynamical simulation, so applying their results to the system after it has been projected to a finite-dimensional Hilbert space $\hat{R}\rightarrow \hat{P}\hat{R}\hat{P}$ ensures that the error induced by compressing the Markovian auxiliary modes into the system jump operator $\hat{L}$ is bounded by $||\hat{P}\hat{R}\hat{P}||^2\Gamma\tau$, when comparing to the uncompressed dynamical simulation.
Assuming the dynamical simulation is converged with regards to the projection error, the error induced by the Markovian compression should not scale with increasing the projected space, as the formal error bound $||\hat{P}\hat{R}\hat{P}||^2\Gamma\tau$ otherwise suggests. Instead, in the following discussion, we treat $||\hat{P}\hat{R}\hat{P}||$ as the characteristic length scale of the reaction coordinate under the dynamics, which is $\mathcal{O}(1)\times\text{a.u.}$.

It should be noted that $E_n$ and $|n\rangle$ refer to the energies and eigenstates of the system dressed by the residual environment, which may be difficult to compute exactly. 
However, as discussed by Nathan and Rudner, the jump operator can be evaluated through an interaction-picture expansion. Writing $H=H_0+V$, we treat the evolution generated by the tractable reference Hamiltonian $H_0$ exactly and incorporate the residual interaction $V$ perturbatively. In our case, $H_0=H_R+\lambda\hat{R}^2+H_B$ contains the system Hamiltonian, the reorganization term, and the free Hamiltonian of the residual bath, while $V$ contains the coupling to that bath. A low-order truncation is justified when the interaction-picture perturbation $e^{iH_0t}Ve^{-iH_0t}$ changes the coupling operator ($\hat{R}$ in our case) only weakly over the timescale of the Markovian bath.
In the numerical examples considered, we find that the zeroth-order construction of $\hat{L}$, which retains the evolution under $H_0$ exactly, yields quantitatively accurate results in the regime of validity of the ULE, $\Gamma\tau\ll1$. Practically, this means that we may approximate the $E_n$ and $|n\rangle$ that are used to compute $\hat{L}$ by $H_R + \lambda \hat{R}^2=\sum_n E_n|n\rangle\langle n|$.
This additional approximation is motivated by the perturbative argument above and supported by our numerical results, but is not covered by the original ULE error bound. Although $\Gamma\tau\ll1$ alone does not guarantee the accuracy {of }our zeroth-order truncation, the overall method remains controllable by the inclusion of higher orders.

While Nathan and Rudner~\cite{PhysRevB.102.115109} control the Markovian approximation in terms of $\Gamma\tau$, they also define $\Gamma_0\tau_0=4\int_0^\infty t |C(t)|dt$ 
which is more practical to work with than their expression for $\Gamma\tau$. However, as they explain, $\Gamma_0\tau_0\leq\Gamma\tau$ 
and while $2\Gamma_0\tau_0$ can be used to bound the correction to the Bloch-Redfield equation \cite{agerskov2026markovianquantummasterequations} they are only able to bound the correction to their master equation using $\Gamma\tau$ \cite{PhysRevB.102.115109}.
Interestingly, for a single exponential with a real prefactor $C(t)=|d|e^{-zt}$, we can compute $\Gamma\tau=\frac{2}{\pi}8\frac{|d|}{\operatorname{Re}(z)^2}$ (using integrals of the zeroth modified Bessel function of the second kind) which compared to $2\Gamma_0\tau_0=8\frac{|d|}{\operatorname{Re}(z)^2}$ is smaller by a factor $\frac{2}{\pi}$. 
Thus, as a practical guiding principle, we may assess the Markovian-ness of the auxiliary modes in terms of:
\begin{equation}
    2\Gamma_0\tau_0 = 8\int_0^\infty t |C(t)|dt = 8\int_0^\infty t \left|\sum_kd_ke^{-z_kt}\right|dt 
    \leq8\sum_k\int_0^\infty t \left|d_ke^{-z_kt}\right|dt = 8\sum_k\frac{|d_k|}{\operatorname{Re}(z_k)^2}
\end{equation}
and use this to supply candidates for the Markovian compression.
However, for the approximation to be controlled, the final delegation of auxiliary modes to the Markovian compression should be informed by the exact error bound by numerically evaluating $\Gamma\tau$, as the two error bounds need not be comparable in general.

Lastly, we note that if the noise spectrum of the proposed Markovian bath $S_M(\omega)=\sum_{k\in K_\text{M}} 2\operatorname{Re}\left(\frac{d_k}{z_k-i\omega}\right)$ does not remain nonnegative, $S_M(\omega) \ngeq0$, it is in conflict with the assumptions of Nathan and Rudner~\cite{PhysRevB.102.115109} and their results do not immediately apply. Compression methods based on adiabatic elimination~\cite{Ishizaki2005-kr, Fay2022-bc, PhysRevA.86.012126} generally do not rely on such a semi-positivity requirement and possibly the results of Nathan and Rudner, could also be extended to such cases.
However, developing such an extension and additionally deriving a bound for the error of approximating the residual-bath-dressed system, in order to elevate our compression method from controllable to generally rigorously bounded, remains outside the scope of this work.


We complement the discussion above, by exemplifying the validity of our approach, when simulating the symmetric double well interacting with the phonon bath, for different values of $2\Gamma_0\tau_0$. The resulting reaction rates are shown in Fig.~\ref{fig:MarkovianCorrectionExample}. 
The phonon bath has been fitted to high-accuracy (tolerance of $\sim10^{-14}$) and the number of auxiliary modes is $|K_\text{fit}|=19$. However, many of these modes basically constitute a Markovian environment, so choosing $K_\text{M}$ of these such that $2\Gamma_0\tau_0\approx0.01$, reduces the number of auxiliary modes that need to be simulated explicitly to $|{K}|=|K_\text{fit}\setminus K_\text{M}|=2$. That is, the evolution induced by $|K_\text{M}|=17$ of these auxiliary modes can be captured by the system jump operator $\hat{L}$, as seen by the resulting reaction rate $k(t)=\frac{\dot{P}_r(t)}{1-2P_{{r}}(t)}$ (Eq.~\ref{eq:ForwardReactionRate}). The resulting evolution is nearly indistinguishable from the case where all the auxiliary modes have been simulated explicitly, $2\Gamma_0\tau_0=0$ and $|{K}|=19$.
However, increasing $|K_\text{M}|=18$ and thus $2\Gamma_0\tau_0\approx0.06$, begins to introduce a deviation on the calculated reaction rate $k$. Based on this, we have chosen to truncate $K_\text{fit}$ to ${K}$ by choosing $K_\text{M}$ such that $\Gamma\tau$ is at most $0.01$ (in practice often smaller) for all our simulations. This allows us to substantially reduce the number of explicit auxiliary modes while staying true to the dynamics.

The results for this example are summarized in Table~\ref{tab:MarkovianCompressionSummary}{.}
\begin{table}[h]
    \centering
  \begin{tabular}{c|c||c|c||c}
          Number of explicit modes, $|K|$ \quad& \quad$|K_\text{M}|$\quad\quad&\quad\quad${4\sum_k^M}\frac{|d_k|}{\operatorname{Re}(z_k)^2} $\quad\quad& \quad\quad\quad\quad$\Gamma\tau$\quad\quad\quad\quad & Absolute relative deviation in $k$\\\hline\hline
          19&0&0&0&$0\%$\\ \hline
  2&17& ${0.00504}$&$0.00598$ & $0.102\%$\\\hline
  1&18& ${0.03045}$&$0.03643$ & $1.215\%$\\\hline
  0&19& ${7.3125}$&$5.50957$ & $25.8932\%$\\\hline
  \end{tabular}
    \caption{
    Comparison of different levels of compression for the dynamics of the symmetric double well interacting with the phonon bath. The heuristic for "Markovianity" ${4\sum_k^M}\frac{|d_k|}{\operatorname{Re}(z_k)^2}$ is compared to the proper error bound $\Gamma\tau$ and the deviation in the reaction rate $|k-k_{max}|/k_{max}$ when compared to the fully explicit simulation $k_{max}$ is presented.
    }
    \label{tab:MarkovianCompressionSummary}
\end{table}
\begin{figure}[h]
    \centering
    \includegraphics[width=0.6\linewidth]{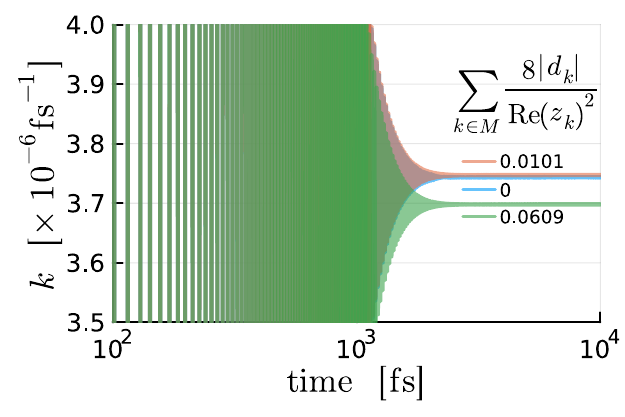}
    \caption{
    Comparison of $k(t)=\frac{\dot{P}_r(t)}{1-2P_r(t)}$ (Eq.~\ref{eq:ForwardReactionRate}) for different levels of Markovian compression of the phonon bath. See Table~\ref{tab:MarkovianCompressionSummary} for an overview of the results. The observed oscillations correspond to the inherent time scale of the phonon bath $\gamma_R^{-1}\approx \beta\approx25~\text{fs}$ and originates from the quenched dynamics since the system is initialized in a product state with the phonon bath.
    The simulations were carried out using tensor networks, and for this environment, the number of physical indices needed to represent the environment could be reduced from $2\times 19=38$ to $2\times2=4$ without a noticeable effect on the observed reaction rate. 
    This should be compared to reducing the number of indices by performing a less accurate decomposition of the bath correlation function in terms of only two exponentials, which would result in a larger deviation in the observed rate.}
    \label{fig:MarkovianCorrectionExample}
\end{figure}

\section{\label{sec:ReactionRateAnalysis}Reaction rate analysis}
We define the reaction rate constant by following e.g. {Refs.}~\citenum{10.1063/1.3097128, 10.1063/1.445581, 10.1063/1.2772265}
For the double well potential, we define the left well as the reactant region with partial concentration $P_r$ and the right well as the product region with partial concentration $P_p = 1-P_r$\\
$P_r$ is defined as an expectation value of the projection onto the left well:
\begin{equation}
    P_r(t) = \text{Tr}\left[\hat{P}_r\rho(t)\right]
    \quad\text{with}\quad \hat{P}_r = \int_{-\infty}^0 dR |R\rangle\langle R|
\end{equation}
Past some transient behavior, we can assume the expectation values follow the rate equation:
\begin{equation}
\label{eq:DoubleWellRateEquation}
    \dot{P}_r(t) = -kP_r(t) + k_b P_p(t)
\end{equation}
where $k$ is the forward reaction rate and $k_b$ is the backward reaction rate, which determine the equilibrium distribution $\frac{k_b}{k} = \frac{P_r(t\rightarrow\infty)}{P_p(t\rightarrow\infty)}$. Using this, one can isolate $k$ in Eq.~\ref{eq:DoubleWellRateEquation}:
\begin{equation}
\label{eq:ForwardReactionRate}
    k = \frac{\dot{P}_r(t)}{-P_r(t) + \frac{P_r^{eq}}{P_p^{eq}}(1-P_r(t))}
\end{equation}
The time derivative of $P_r(t)$ can be calculated either by finite difference:
\begin{equation}
    \label{eq:dotP_FD}
    \dot{P}_r(t) \approx \frac{P_r(t+\Delta t) - P_r(t)}{\Delta t}
\end{equation}
or by using the equations of motion, $\dot{\rho}=\mathcal{L}[\rho]$:
\begin{equation}
    \label{eq:dotP_EoM}
    \dot{P}_r(t) = \text{Tr}\left[\hat{P}_r\dot{\rho}(t)\right] = \text{Tr}\left[\hat{P}_r\mathcal{L}[\rho(t)]\right] = \text{Tr}\left[\mathcal{L}^\dagger[\hat{P}_r]\rho(t)\right]
\end{equation}
Where $\mathcal{L}^\dagger$ is the adjoint of the super-operator $\mathcal{L}$.

Furthermore, for symmetric potential energy surfaces, it is straightforward to show that the equilibrium distribution is uniform $P_r^{eq} = P_p^{eq} = \frac12$. Consider e.g. the joint system-bath parity operation
$\hat{S} : \begin{pmatrix}
    R\\
    B
\end{pmatrix}\rightarrow \begin{pmatrix}
    -R \\
    -B
\end{pmatrix}$, which leaves $H = H_R + \hat{R}\hat{B}+H_B$ invariant. Then:
\begin{equation}
    \text{Tr}\left[e^{-\beta\hat{H}}\hat{P}_r\right] = \text{Tr}\left[\hat{S}e^{-\beta\hat{H}}\hat{S}\hat{P}_r\right] = \text{Tr}\left[e^{-\beta\hat{H}}\hat{S}\hat{P}_r\hat{S}\right] = \text{Tr}\left[e^{-\beta\hat{H}}\left(\mathds{1} - \hat{P}_r\right)\right]
\end{equation}
However, for general potential energy surfaces, it is helpful to consider alternatives to Eq.~\ref{eq:ForwardReactionRate} for retrieving the forward reaction rate.
One approach is to fit the solution of the rate equation to the expectation value of the simulation:
\begin{equation}
\label{eq:PopulationClosedForm}
    \quad P_r(t) = \frac{k_b}{k+k_b} + \left(P_r(t_0) - \frac{k_b}{k+k_b}\right)e^{-(k+k_b)(t-t_0)} 
\end{equation}
where $t_0$ is typically chosen to be at a later time than the initialization of the simulation ($t=0$) past the transients.
Another approach is to consider a linear fit of Eq.~\ref{eq:DoubleWellRateEquation} in terms of $k$ and $k_b$ e.g. through the pseudo-inverse of the time-series data:
\begin{align}
    \begin{pmatrix}
        \dot{P}_r(t_0)\\
        \dot{P}_r(t_1)\\
        \dot{P}_r(t_2)\\
        \vdots
    \end{pmatrix}
    &=
    \begin{pmatrix}
        -P_r(t_0) & 1-P_r(t_0)\\
        -P_r(t_1) & 1-P_r(t_1)\\
        -P_r(t_2) & 1-P_r(t_2)\\
        \vdots & \vdots
    \end{pmatrix}
    \begin{pmatrix}
        k\\
        k_b
    \end{pmatrix}\\
    \Rightarrow 
    \begin{pmatrix}
        k\\
        k_b
    \end{pmatrix} &= \begin{pmatrix}
        -P_r(t_0) & 1-P_r(t_0)\\
        -P_r(t_1) & 1-P_r(t_1)\\
        -P_r(t_2) & 1-P_r(t_2)\\
        \vdots & \vdots
    \end{pmatrix}^{+}\begin{pmatrix}
        \dot{P}_r(t_0)\\
        \dot{P}_r(t_1)\\
        \dot{P}_r(t_2)\\
        \vdots
    \end{pmatrix}
\end{align}
However, the transients in $P_r(t)$ can become long lived in $\dot{P}_r(t)$ due to the time-derivative increasing the amplitude of the fluctuations. Therefore, it can be advantageous to "average out" the transients by considering the integral of Eq.~\ref{eq:DoubleWellRateEquation} before performing the linear fit of $k$ and $k_b$.
\begin{equation}
\label{eq:ReactionRateFormula}
    \begin{pmatrix}
        k\\
        k_b
    \end{pmatrix} = \begin{pmatrix}
        -\int_{t_0}^{t_1} P_r(t)dt & \int_{t_0}^{t_1} (1-P_r(t))dt\\
        -\int_{t_0}^{t_2} P_r(t)dt & \int_{t_0}^{t_2} (1-P_r(t))dt\\
        -\int_{t_0}^{t_3} P_r(t)dt & \int_{t_0}^{t_3} (1-P_r(t))dt\\
        \vdots & \vdots
    \end{pmatrix}^{+}\begin{pmatrix}
        P_r(t_1)-P_r(t_0)\\
        P_r(t_2)-P_r(t_0)\\
        P_r(t_3)-P_r(t_0)\\
        \vdots
    \end{pmatrix}
\end{equation}
where the integrals can be performed as a Riemann sum. By simulating for long enough times and calculating the reaction rates for different $t_0$, we make sure that $k$ has converged upon a stationary value. By comparing the different approaches, we found Eq.~\ref{eq:ReactionRateFormula} to be the most robust with fastest convergence, in the general (asymmetric potential) setting.

\subsection{Triple well reaction rates}
In this setting, we consider the left and right {wells as} distinct product regions with partial concentrations $P_l$ and $P_r$, respectively. The middle well is the reactant region denoted by $P_0$. Similar to the above, these are defined as expectation values of (notice the sign-convention of Eq.~\ref{eq:TripleWellPotential}, e.g. $-R_0^l<0<-R_0^r$):
\begin{align}
    \hat{P}_l &= \int_{-\infty}^{-R_0^l}dx|x\rangle\langle x|\\
    \hat{P}_0 &= \int_{-R_0^l}^{-R_0^r}dx|x\rangle\langle x|\\
    \hat{P}_r &= \int_{-R_0^r}^{\infty}dx|x\rangle\langle x|
\end{align}
for which the rate equation becomes:
\begin{equation}
\label{eq:ThreeWellRateEquation}
\begin{pmatrix}
    \dot{P}_l(t)\\
    \dot{P}_0(t)\\
    \dot{P}_r(t)
\end{pmatrix}
=
\begin{pmatrix}
    P_0(t) & 0 & -P_l(t) & 0\\
    -P_0(t) & -P_0(t) & P_l(t) & P_r(t)\\
    0 & P_0(t) & 0 & -P_r(t)
\end{pmatrix}
\begin{pmatrix}
    k_{0\rightarrow l}\\
    k_{0\rightarrow r}\\
    k_{l\rightarrow 0}\\
    k_{r\rightarrow 0}
\end{pmatrix} 
\end{equation}
Similar to Eq.~\ref{eq:ReactionRateFormula}, the reaction rates are calculated by forming a matrix from the discrete time-series and solving the linear problem.
\section{Probability currents: transition rate between states}
\label{sec:RateBetweenStates}

We are interested in how the rate of change in the population of the different system states depend on the environment. Using Eq.~\ref{eq:FPHEOM}, we can consider the time derivative of $\hat{P}_n=|n\rangle\langle n|$, in terms of the reduced system density matrix $\rho_S(t) = \langle \pmb{0}|\rho(t)|\pmb{0}\rangle$. The instantaneous rate of change of $\rho_S(t)$ is (in our case $\hat{H}_S=\hat{H}_R$ and $\hat{S}=\hat{R}$):
\begin{align}
    \dot{\rho}_S(t)&=-i\left[\hat{H}_S + \sum_k\lambda_k \hat{S}^2,\rho_S(t)\right]-i\sum_k\left[\hat{S},\sqrt{d_k}\rho_k(t) + \sqrt{d^*_k}\rho_k^\dagger(t)\right]+\sum_i L_i\rho_S(t)L_i^\dagger - \frac12\left\{L_i^\dagger L_i,\rho_S(t)\right\}
\end{align}
with $\lambda_k=-\operatorname{Im}\left(\frac{d_k}{z_k}\right)$ and $\rho_k(t) =\langle\pmb{0}|\hat{a}_k\varrho(t)|\pmb{0}\rangle$ \footnote{if the trace-preserving version of the master equation (Eq.~\ref{eq:QuasiLindblad}) was instead used, then $\rho_k(t) = \text{Tr}_B[\hat{a}_k\varrho(t)]$} and where we accommodate multiple markovian correctional terms $L_i$, which will be useful when we later subdivide the noise spectrum into distinct components.
The Heisenberg equation of motion for $\hat{P}_n$ is then:
\begin{align}
\dot{P_n}(t)
     &= i\text{Tr}\left[
     [\hat{H}_S,\hat{P}_n]\rho_S(t)
     \right]
     + i\sum_k \left(\lambda_k \text{Tr}\left[
     [\hat{S}^2,\hat{P}_n]\rho_S(t)
     \right]
     + \text{Tr}\left[
     [\hat{S},\hat{P}_n]\left(\sqrt{d_k}\rho_k(t) + \sqrt{d^*_k}\rho_k^\dagger(t)\right)
     \right]\right)\nonumber\\
     &+ \sum_i \left(\text{Tr}\left[L_i^\dagger\hat{P}_nL_i\rho_S(t)\right]
     -\frac12\text{Tr}\left[\{L_i^\dagger L_i,\hat{P}_n\}\rho_S(t)\right]\right)
     \label{eq:PnHeisenbergEoM}
\end{align}
Utilizing References \citenum{PhysRevE.93.012128, schumacher2016probabilitycurrentthermodynamicsopen} one can imbue the interpretation of a probability current flowing from the state $n$ to $m$:
\begin{equation}
    \dot{P_n}(t) = -\sum_m j_{n\rightarrow m}(t)
\end{equation}
We will expand on this interpretation by decomposing the probability current further. Specifically, we will assign the different terms in Eq.~\ref{eq:PnHeisenbergEoM} to different contributions $j^{(i)}_{n\rightarrow m}(t)$ of the probability current $j_{n\rightarrow m}(t) = \sum_i j^{(i)}_{n\rightarrow m}(t)$.
To do this, we consider a grouping of the transitions in terms of those induced by the system Hamiltonian and those induced by coupling to the harmonic environment. Furthermore, the harmonic environment is sub-divided into contributions from the distinct spectral densities:
\begin{equation}
    J(\omega) = \sum_i J^{(i)}(\omega)
    \quad\quad\Rightarrow\quad\quad
    C(t) = \sum_i\sum_k d_k^{(i)}e^{-z_k^{(i)}t}
\end{equation}
where the $d_k^{(i)}$ and $z_k^{(i)}$ parameters are obtained by decomposing the $i$'th noise spectrum independently from the others.
In our case of Eq.~\ref{eq:SimpleSD}, we have two distinct spectral densities such that we define three different contributions to the probability currents (Eq.~\ref{eq:probabilitycurrents}).
We denote the probability current due to $H_S$ as \textit{Tunneling}, the contributions from the Drude-Lorentz spectral density of the phonons as \textit{Solvent} and the contributions associated with the spectral density of the lossy cavity is simply referred to as \textit{Cavity}.
The probability currents are then defined as follows:
\begin{align}
    \partial_t \langle n|\rho_S(t)|n \rangle &= -\sum_m j_{n\rightarrow m}(t)\\
    j_{n\rightarrow m}(t) &= j_{n\rightarrow m}^\text{Tunneling}(t) + j_{n\rightarrow m}^\text{Solvent}(t) + j_{n\rightarrow m}^\text{Cavity}(t)\\
    j_{n\rightarrow m}^\text{Tunneling}(t)&= 2\operatorname{Im}\left((H_S)_{mn}(\rho_S(t))_{nm}\right)\\
    j_{n\rightarrow m}^\text{Solvent}(t) &= \sum_{k\in\text{solvent}}2\operatorname{Im}\left(S_{mn}\left[\sqrt{d_k}(\rho_k(t))_{nm}+\sqrt{d_k^*}(\rho_k(t))^*_{mn}\right]\right)+\lambda_k2\operatorname{Im}\left((S^2)_{mn}(\rho_S(t))_{nm}\right)\\
    &\quad\quad+\frac12\left( \left(L_s\rho_S(t)\right)_{mn} \left(L_s^\dagger\right)_{nm} + \left(L_s\right)_{mn} \left(\rho_S(t)L_s^\dagger\right)_{nm}\right)\\
    &\quad\quad-\frac12\left( \left(\rho_S(t)L_s^\dagger\right)_{mn} \left(L_s\right)_{nm} + \left(L_s^\dagger\right)_{mn} \left(L_s\rho_S(t)\right)_{nm}\right)\\
    j_{n\rightarrow m}^\text{Cavity}(t) &= \sum_{k\in\text{cavity}}2\operatorname{Im}\left(S_{mn}\left[\sqrt{d_k}(\rho_k(t))_{nm}+\sqrt{d_k^*}(\rho_k(t))^*_{mn}\right]\right)+\lambda_k2\operatorname{Im}\left((S^2)_{mn}(\rho_S(t))_{nm}\right)\\
    &\quad\quad+\frac12\left( \left(L_c\rho_S(t)\right)_{mn} \left(L_c^\dagger\right)_{nm} + \left(L_c\right)_{mn} \left(\rho_S(t)L_c^\dagger\right)_{nm}\right)\\
    &\quad\quad-\frac12\left( \left(\rho_S(t)L_c^\dagger\right)_{mn} \left(L_c\right)_{nm} + \left(L_c^\dagger\right)_{mn} \left(L_c\rho_S(t)\right)_{nm}\right)
\end{align}
where $L_s$ and $L_c$ refer to the jump-operator for the Markovian correction (Sec.~\ref{sec:EfficientMarkovCorrection}) of the solvent degrees of freedom and lossy cavity degrees of freedom, respectively.

In the main text, when we refer to the value of a probability current, it is the value of $j_{n\rightarrow m}(t)$ at $t=6\,\text{ps}$. At this time, the transients have died out and $j_{n\rightarrow m}(t)$ has become largely time-independent. However, $j_{n\rightarrow m}(t)$ still depends on the populations and coherences of $\rho(t)$, so it slowly decays to zero until the equilibrium distribution is reached.

\subsection{Relating probability currents to reaction rates}
\label{sec:ProbCurrentsAsReactionRate}
The probability currents are inherently a basis-dependent quantity. However, if we choose a basis for which the projection operators onto the distinct wells are approximately diagonal, such as the localized basis used in this work, we may approximate observables such as the reactant population as:
\begin{equation}
    P_{r}(t)=\text{Tr}\left(\hat{P}_r\rho(t)\right) \approx \sum_n^\text{reactant}\rho_{nn}(t)
\end{equation}
where the sum is only over states in the reactant well. For now, let us consider the double well PES. Then, with Eq.~\ref{eq:probcurrents}, we can relate the change in the reactant population to the probability currents:
\begin{equation}
    -\dot{P}_r(t)\approx\sum_n^\text{reactant}\sum_mj_{n\rightarrow m}(t)=\sum_n ^\text{reactant} \sum_m^\text{product} j_{n\rightarrow m}(t)
\end{equation}
where the last equality follows from the fact that $j_{n\rightarrow m}$ is anti-symmetric, such that the intra-reactant contributions cancel.
Then, if we are in a regime where the linear rate equations (Eq.~\ref{eq:DoubleWellRateEquation} in this example) describe the dynamics, we can relate the total probability current that goes across the dividing surface of the reactant and product population to the linear kinetics across that surface, leading to the natural expression:
\begin{equation}
    \sum_n ^\text{reactant} \sum_m^\text{product} j_{n\rightarrow m}(t) \approx kP_r(t)-k_bP_p(t)
\end{equation}
These arguments can be extended to more elaborate reaction networks, such that the above expression also applies to e.g. the left-product$\leftrightarrow$reactant and right-product$\leftrightarrow$reactant dividing surfaces of the triple-well PES.
Thus, at early times where $P_r(t)\approx1$ and $P_p(t)\approx0$, the sum of probability currents going from states in the reactant well to states in the product well is approximately equal to the forward reaction rate:
\begin{equation}
    \sum_n ^\text{reactant} \sum_m^\text{product} j_{n\rightarrow m}(t) \approx k
\end{equation}


\subsection{FGR as perturbative limit of probability currents}
\label{sec:FGR}
Here, we will gain some intuition about the probability currents introduced in Sec.~\ref{sec:RateBetweenStates} by considering the Born-Markov-Secular approximation and relating the resulting expression to the well known Fermi's Golden Rule.
We will begin by considering a generic system-bath setup:
\begin{equation}
    H = H_S + \hat{S}\hat{B} + H_B
\end{equation}
Since we consider the perturbative limit we have rightfully neglected the reorganization term in $H_S$. The resulting bath contribution to the probability current will be:
\begin{equation}
    j_{n\rightarrow m}^{B}(t) = 2\operatorname{Im}\left(S_{mn}\langle n|\text{Tr}_B(\hat{B}\rho(t))|m\rangle\right)
\end{equation}
With $\rho(t)$ the full system plus bath density matrix. Thus, we seek to approximate $\text{Tr}_B(\hat{B}\rho(t))$. In the interaction picture $\rho_I(t) = e^{i(H_S+H_B)t}\rho(t)e^{-i(H_S+H_B)t}$ we may write:
\begin{equation}
    \rho_I(t) = \rho_I(t_0) - i\int_{t_0}^tdt'\left[\hat{S}(t')\hat{B}(t'),\rho_I(t')\right]
\end{equation}
Then, under the Born-Markov approximation $\rho_I(t') \approx\rho_I^S(t)\otimes\rho_B$ with $\rho_B\propto e^{-\beta H_B}$:
\begin{equation}
\label{eq:MarkovianIntegral}
    \text{Tr}_B(\hat{B}\rho(t))\approx
    -i\int_{t_0}^te^{-iH_S(t-t')}\hat{S}e^{iH_S(t-t')}\rho_S(t)C(t-t')dt' + H.C.
\end{equation}
where we've used $\text{Tr}_B\left(\hat{B}(t)\rho_B\right) = 0$ and $C(t-t')=\text{Tr}_B\left(\hat{B}(t)\hat{B}(t')\rho_B\right)$.
We can then write the probability current as:
\begin{align}
    j_{n\rightarrow m}^{B}(t) &\approx 2\operatorname{Re}\left[-\int_0^{t-t_0} C_S^{nm}(t,\tau)C(\tau)d\tau + \int_0^{t-t_0} \bigg(C_S^{mn}(t,\tau)C(\tau)\bigg)^*d\tau\right]\\
    C_S^{nm}(t,\tau)&=\langle m|\hat{S} |n\rangle\langle n|e^{-iH_S\tau}\hat{S}e^{iH_S \tau}\rho_S(t)|m\rangle
\end{align}
When $|n\rangle$ and $|m\rangle$ are eigenstates of $H_S$, we get:
\begin{equation}
\label{eq:FGR_MatElement}
    \langle n|e^{-iH_S(t-t')}\hat{S}e^{iH_S(t-t')}\rho_S(t)|m\rangle
    =\sum_le^{-i(E_n - E_l)(t-t')}S_{nl}\rho^S_{lm}(t)
\end{equation}
and we can invoke the secular approximation to argue that only the diagonal part of the density matrix contributes to the integral, such that $\sum_{l}S_{nl}\rho^S_{lm}(t)\approx S_{nm}\rho^S_{mm}(t)$. If we furthermore introduce the noise spectrum $C(\tau)=\int_{-\infty}^\infty\frac{d\omega}{2\pi}e^{-i\omega\tau}S_\beta(\omega)$, we recover:
\begin{align}
    j_{n\rightarrow m}^{B}(t) &=\int_{-\infty}^\infty S_\beta(\omega)\left[\mathcal{A}^{mn}(\omega,t) - \mathcal{A}^{nm}(\omega,t)\right]d\omega\\
    \mathcal{A}^{nm}(\omega,t)&=\frac{1}{\pi}\operatorname{Re}\left[\int_0^{t-t_0}C_S^{nm}(t,\tau)e^{-i\omega \tau}d\tau\right]\\
    &=|\langle m|\hat{S}|n\rangle|^2\rho_{mm}^S(t)\times \underbrace{\frac{1}{\pi}\operatorname{Re}\left[\int_0^{t-t_0}e^{-i(\overbrace{E_n-E_m}^{\omega_{nm}})\tau}e^{-i\omega \tau}d\tau\right]}_{\int_{-(t-t_0)}^{t-t_0}\frac{d\tau}{2\pi}e^{-i(\omega_{nm}+\omega)\tau}} \label{eq:SystemResponseFGR}
\end{align}
such that in the long time limit $t\gg t_0$:
\begin{equation}
\label{eq:FGR_ProbCurrent}
    j_{n\rightarrow m}^{B} \approx |S_{nm}|^2S_\beta(\omega_{nm})\rho^S_{nn} - |S_{nm}|^2S_\beta(-\omega_{nm})\rho^S_{mm}
\end{equation}
That is, the probability current from $n$ to $m$ becomes the population in $n$ times the FGR rate from $n$ to $m$ minus the population in $m$ times the FGR rate from $m$ to $n$ (the two rates are related by $S_\beta(-\omega)=e^{-\beta\omega}S_\beta(\omega)$, see Sec.~\ref{sec:GaussianEnvironments}). Thus, when the probability current is subject to the same approximations as FGR, the resulting expression is consistent with FGR.

\subsubsection{Broadening of transition frequencies due to phonon bath}
\label{subsec:Broadening}
The derivation above (and FGR in general) relied on $|n\rangle$ and $|m\rangle$ being eigenstates of $H_S$. However, in the present case, we are interested in intra-well transitions between localized states, which are not eigenstates of $H_S$. The reason that FGR can still be applied as is, is because a secondary bath, in this case the solvent, renders these localized states metastable by suppressing the tunneling between the constituent eigenstates (this can be seen in Fig.~\ref{fig:TripleWellEquilibria} where the bath-dressed states look like the localized states in Fig.~\hyperref[fig:competing_pathways]{\ref*{fig:competing_pathways}a}, even though no such post-processing localization took place when visualizing the states in Fig.~\ref{fig:TripleWellEquilibria}).
Furthermore, the secondary bath broadens the transition frequencies of the system by inducing a finite lifetime of constituent states.

To incorporate these ideas into the result above, we may view $H_S$ as an enlarged system that not only contains the system degrees of freedom, but now also incorporates e.g. the phonon degrees of freedom.
In order to recover a succinct expression for $C_S^{nm}(t,\tau)$ for this enlarged system, we make the following considerations. The Markov approximation simply says that, when evaluating the integral~\ref{eq:MarkovianIntegral}, we may consider the dynamics of $\rho_S(t-\tau)\approx e^{iH_S\tau}\rho_S(t)e^{-iH_S\tau}$ to be closed under $H_S$ (i.e. we disregard the perturbing bath when calculating the induced transition rates). In the current case where $H_S$ contains the phonon bath, we will not only assume that the dynamics are closed under $H_S$ but that it will evolve towards the thermal equilibrium distribution of the system plus phonon bath $\rho^{eq}_S$. Such that in the long time limit and within this approximation $C^{nm}_S(\tau) \approx \text{Tr}\left[e^{iH_S\tau}|m\rangle\langle m|\hat{S}|n\rangle\langle n|e^{-iH_S\tau}\hat{S}\rho^{eq}_S\right]$ which resembles the dipole-dipole correlation function $C_S(\tau)$,
except it is restricted to the pathway between $n$ and $m$:
$C_S(\tau) = \text{Tr}\left[e^{iH_S \tau}\hat{S} e^{-iH_S\tau}\hat{S}\rho_S^{eq}\right]=\sum_{nm}C_S^{nm}(\tau)$.
Thus, the simplest way to incorporate dissipation, is to broaden the delta function in Eq.~\ref{eq:SystemResponseFGR} by including decay:
\begin{equation}
    \mathcal{A}^{nm}(\omega,t) = |\langle m|S|n\rangle|^2\rho_{mm}^S(t)\underbrace{\operatorname{Re}\int_0^\infty\frac{d\tau}{\pi}e^{-i(\omega_{nm}+\omega)\tau-{\Gamma}\tau}}_{\frac{1}{\pi}\frac{\Gamma}{(\omega_{nm}+\omega)^2+\Gamma^2}}
\end{equation}
where the inverse lifetime ${\Gamma}$ and the possibly renormalized transition frequency $\omega_{nm}$, can be read-off from the part of $C_S(\tau)$ that is due to the $n\rightarrow m$ transition.
In our example, Fig.~\ref{fig:FanoRates}, we did not attempt to find ${\Gamma}$ and $\omega_{nm}$ a priori, but simply found ${\Gamma}=30\,\text{cm}^{-1}$ and $|\omega_{nm}|=1205\,\text{cm}^{-1}$ to best explain the results across all $\theta$ and $\omega_c$. The value for ${\Gamma}$ {aligns} well with what has earlier been found for this system \cite{Ying2024}. However, the bare transition frequency is $\approx1195\,\text{cm}^{-1}$ indicating that the phonon bath not only broadens the transition but also shifts it slightly. 

\section{Distribution of reaction rates and comparison to isotropic averaging}
\label{sec:DistributionOfRates}
For the symmetric double well PES, we can write the difference of the population in the two wells (Eq.~\ref{eq:PopulationClosedForm}) as $\Delta P(t)=\Delta P_0 e^{-2kt}$. Based on the agreement seen in Fig.~\ref{fig:FanoRates}, we can use the FGR rate in Eq.~\ref{eq:FGR} to study what the reaction rate profile would look like for a given molecule orientation. 
We can then imagine an ensemble of molecules with random orientations and since the orientation affects the reaction, this yields a distribution of reaction rates \cite{Plonka_2001}.
Thus, instead of the average reaction rate $\langle k(\phi)\rangle_\phi$, which was studied in Fig.~\ref{fig:FanoRates}, the primary experimentally relevant entity is arguably:
\begin{equation}
    \Delta P(t) = \langle \Delta P(t,\phi)\rangle_\phi = \Delta P_0 \int p(\Omega)e^{-2k(\Omega)t}d\mu
\end{equation}
where we have changed notation to accommodate different orientational models, such that $p(\Omega)$ is the distribution of the molecule orientation relative to the cavity polarization in terms of the solid angle $\Omega$ with measure $d\mu$.
In the 2-dimensional coplanar model (Sec.~\ref{subsec:Environment-driven effects}), $\eta_Q\propto \cos(\phi)$ and $\eta_R\propto\cos(\theta-\phi)=\cos(\theta)\cos(\phi)+\sin(\theta)\sin(\phi)$ with $\int p(\Omega)d\mu=\frac{1}{2\pi}\int_0^{2\pi}d\phi$. 

For an isotropic distribution in three dimensional space $\eta_Q\propto\cos(\phi)$ and $\eta_R\propto\cos(\theta)\cos(\phi)+\sin(\theta)\sin(\phi)\cos(\psi)$ with $ \int p(\Omega)d\mu = \frac{1}{4\pi}\int_0^\pi \sin(\phi)d\phi\int_0^{2\pi}d\psi$.

Since the reaction rate {depends} on the orientation, the distribution of reaction rates $p(k)$ can be defined by a change of variables:
\begin{equation}
    \Delta P(t) = \Delta P_0 \int_{k_\text{min}}^{k_\text{max}} p(k)e^{-2kt}dk
\end{equation}
As an example, consider the harmonic environment without the spectator mode (Eq.~\ref{eq:SimpleSD}) and an isotropically distributed orientation. Using Eq.~\ref{eq:FGR}, as was done in \cite{ying_resonance_2023}, we schematically recover: $k(\Omega)=k_0 +\cos^2(\phi)k_1$, where $k_0$ is the reaction rate outside the cavity and $k_1$ is the maximal contribution from the cavity. Using a change-of-variables, one recovers the following distribution of reaction rates: $p(k)=\frac{1}{2\sqrt{k_1(k-k_0)}}$ which naturally reproduces the average reaction rate: $\int_{k_0}^{k_0+k_1}kp(k)dk = k_0 + \frac{1}{3}k_1 = \langle k(\Omega)\rangle_\Omega$.

In this kind of simplified scenario, where we consider an ensemble of molecules whose reaction rate is described by the distribution $p(k)$, we could imagine an observer fitting the time-series data $\Delta P(t)$ with a single exponential. 
Using an unweighted time-domain least-squares fit with the ansatz $\Delta P(t)=e^{-2k_\text{fit}t}$, the observer would find a rate $k_\text{fit}$ that lies between the result obtained for a short time window $\langle k\rangle$ and the result obtained from considering the full time window $\langle k^{-1}\rangle^{-1}$ \cite{Plonka_2001}, i.e. $\langle k^{-1}\rangle^{-1}\leq k_\text{fit}\leq \langle k\rangle$.

The effect of a distribution of rates can perhaps be seen in the experimental data presented in e.g. Fig. 2E of Ref.~\citenum{doi:10.1126/science.aau7742}. While the error bars make it difficult to make a precise comparison, the mean values suggest that there is an initial fast decay and a subsequent slower decay. A fit would then yield a value that lies in between these two rates.

Now, for the example in Sec.~\ref{subsec:Environment-driven effects}, we would like to understand what the distribution of rates looks like for both the 2- and 3-dimensional orientational distributions. 
Figure~\ref{fig:DistributionOfRates} {shows} these distributions, based on Eq.~\ref{eq:FGR}, alongside the resulting fast rate $\langle k\rangle$ and the slow rate $\langle k^{-1}\rangle ^{-1}$. We see that these two rates are comparable, so, in this example, $\langle k\rangle$ is representative of what one could expect to obtain from a single-exponential fit to the decay of reactant from an ensemble of molecules with random orientations.
Furthermore, the angular distributions lead to qualitatively similar results, so even though $p(\phi)=\frac{1}{2\pi}$ was used in Sec.~\ref{subsec:Environment-driven effects}, the discussion there also applies to the scenario where the molecule orientations are distributed isotropically in 3-dimensions.

\begin{figure}[h]
    \centering
    \includegraphics[width=1.0\linewidth]{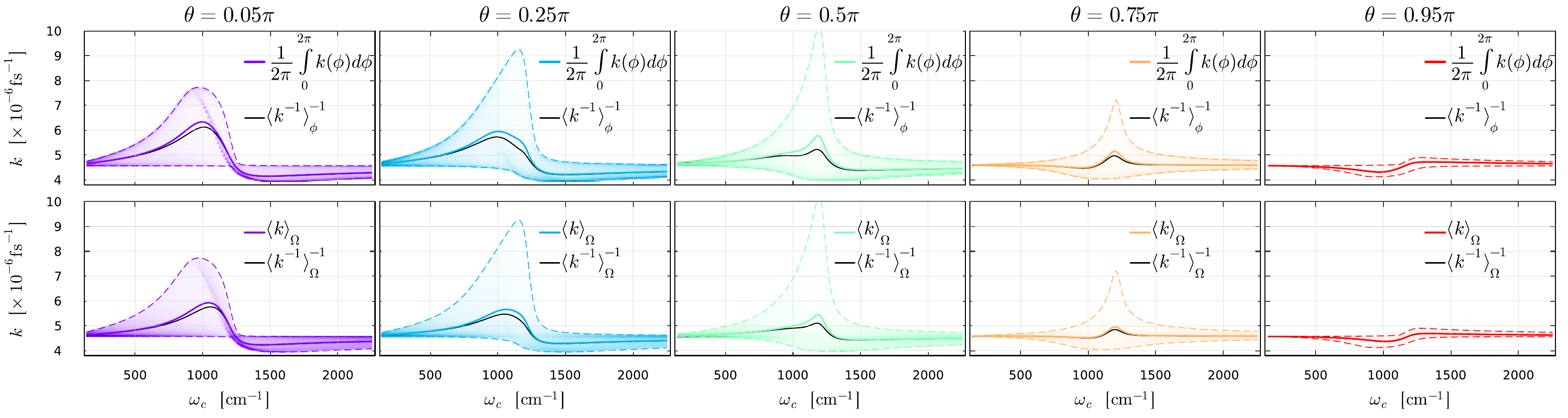}
    \caption{Distribution of reaction rates $p(k)$ for an ensemble of molecules with random orientations, visualized by shaded rate profiles. The rate is calculated from Eq.~\ref{eq:FGR} and the distribution of the molecule orientations is either uniform in the plane containing the cavity polarization (top panel) or isotropic in 3-dimensions (lower panel). The dashed lines represent the minimum and maximum rates and the full colored line is the mean $\langle k\rangle$ while the black line is the harmonic-mean rate $\langle k^{-1}\rangle ^{-1}=\left(\int_0^\infty p(k)k^{-1}dk\right)^{-1}$. Fitting a single exponential with fixed initial amplitude and zero offset to the decay curve of such an ensemble, would yield a reaction rate in between these two limits that depend on the time window considered.}
    \label{fig:DistributionOfRates}
\end{figure}

\section{Equilibrium state of the reaction coordinate}
\label{sec:MeanForceGibbsState}
For an N-level system coupled to a dissipative environment, such as the one considered in this work and under fairly general conditions \cite{10.1116/5.0073853}, it has been proven that the long-time asymptotics of system observables $\mathcal{O}_S$ is given by:
\begin{equation}
    \lim_{t\rightarrow \infty} \text{Tr}\left[e^{-i(H_S+H_{SB}+H_B)t}\rho_S(0)\otimes \frac{e^{-\beta H_B}}{\text{Tr}e^{-\beta H_B}}e^{i(H_S+H_{SB}+H_B)t}\mathcal{O}_S\right] = \text{Tr}_S\left[\tau_\text{MF}\mathcal{O}_S\right]
\end{equation}
where $\tau_\text{MF}$ is the \textit{mean force Gibbs state}:
\begin{equation}
    \tau_\text{MF} = \text{Tr}_B\left[\frac{e^{-\beta(H_S+H_{SB}+H_B)}}{\text{Tr}\left[e^{-\beta(H_S+H_{SB}+H_B)}\right]}\right]
\end{equation}
While proving the same for infinite-dimensional systems with discrete spectra remains an open problem, we have utilized a finite dimensional representation of our system in the dynamical simulation and may apply the above result to describe our observations. 
The mean force Gibbs state has been calculated perturbatively in the bath coupling \cite{PhysRevLett.127.250601}:
\begin{equation}
\label{eq:PerturbativeMFG}
    \tau_\text{MF}=\tau + \tau_\text{MF}^{(2)} + \hdots
\end{equation}
where $\tau\propto e^{-\beta H_S}$ is the system Gibbs state and $\tau_\text{MF}^{(2)}$ is the leading correction. Whether this perturbative expansion applies is governed by:
\begin{equation}
    \epsilon := \sqrt{\left|\beta\sum_{nm}|S_{nm}|^2\frac{e^{-\beta E_n}}{\sum_le^{-\beta E_l}}D_\beta(E_n-E_m)\right|}\ll 1
\end{equation}
where $\hat{S}$ is the system operator that couples to the bath and the function $D_\beta(\nu)$, which is also used in the expression for $\tau_\text{MF}^{(2)}$, is:
\begin{equation}
    D_\beta(\nu)=\mathcal{P}\int_0^\infty \frac{d\omega}{\pi} J(\omega)\left(\frac{\nu\coth(\beta\omega/2)+\omega}{\omega^2-\nu^2}-\frac{1}{\omega}\right)=\frac{\nu}{2}\mathcal{P}\int_{-\infty}^\infty \frac{d\omega}{\pi} \frac{S_\beta(\omega)}{\omega(\omega-\nu)}
\end{equation}
which for the decomposed noise spectra (Eq.~\ref{eq:BCFSumOfExponentials}) reduces to:
\begin{equation}
    D_\beta(\nu)=\nu\sum_k\operatorname{Im}\left[\frac{d_k}{z_k(\nu+iz_k)}\right]
\end{equation}
For the systems and noise spectra considered in this study $\epsilon \sim 0.12$. 
In case it is of interest to apply this perturbative methodology for stronger coupling, the method known as \textit{reaction coordinate mapping} (see Ref.~\citenum{Nazir2018}) could prove useful to reduce the perturbative error. The basic idea is to include strongly coupled degrees of freedom in $H_S$ and perturbatively expand in the residual couplings. 

By calculating $\tau_\text{MF}$ to second order using the results of Ref.~\citenum{PhysRevLett.127.250601}, we evaluate the population inside each well $P^{eq}_{\text{left}/\text{middle}/\text{right}}$ which, for the double well PES, we can compare to the reaction rates through $\frac{k_\rightarrow}{k_\leftarrow}=\frac{P_\text{right}^{eq}}{P_\text{left}^{eq}}$. Furthermore, we can visualize $\tau_\text{MF}$ in terms of its eigendecomposition $\tau_\text{MF}=UDU^\dagger$ where $U$ contains the equilibrium eigenstates of the mean-force Hamiltonian and $\omega_{nm} =-\frac{1}{\beta}\left(\ln D_n - \ln D_m\right)$ defines the mean-force energy gaps.

\subsection{Mean force Gibbs state for the triple well, Fig.~\ref{fig:competing_pathways}}
\label{sec:MFGTripleWell}
In {Fig.}~\ref{fig:TripleWellEquilibria}, the effective system (described above) is visualized for the triple well PES coupled to the phonon bath. Compared to Fig.~\hyperref[fig:competing_pathways]{\ref*{fig:competing_pathways}a}, we see that all the excitations have been somewhat lowered in energy. However, the left well has been favored due to a more significant lowering of the energy of its excited state. More importantly, this is also reflected in the energy of the left {well's} ground state which has been decreased, while the energy of the right well ground state has been increased, slightly.

These observations are consistent with the interpretation that 
the environment observes the position ($\hat{S}=\hat{R}$) of the reaction coordinate, which localizes the wavefunction. The reason this favors the left well is because it has lower potential energy. Colloquially, the kinetic energy is anyway increased by interacting with the environment, so it might as well save some potential energy by residing more in the left well.

We quantify this effect in the second part of {Fig.}~\ref{fig:TripleWellEquilibria}, where we also include the effect of the cavity on the equilibrium population of each well. We see a minor correction from the phonon bath to the prediction given by the Gibbs state $\tau$ (dotted lines). However, the presence of the cavity has nearly no effect on the equilibrium distribution in the parameter regime considered. Especially, we do not see any resonant behavior. The cavity seems to only enhance the effect described above, where the localization of the reaction coordinate favors the deeper left well.
Exploratory calculations with a non-linear dipole of the form $\hat{R}\rightarrow2\hat{R}e^{-|\hat{R}|/2}$ showed the same qualitative trend.

\begin{figure}[h]
    \centering
    \includegraphics[width=0.4\linewidth]{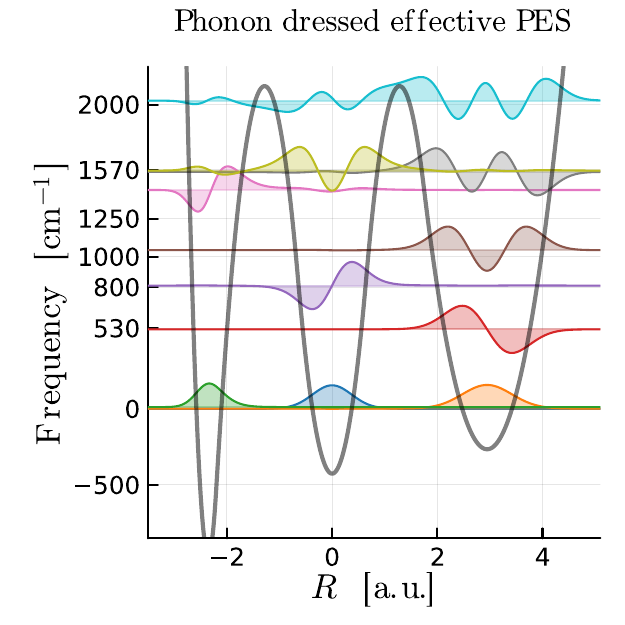}
    \hspace{1cm}
    \includegraphics[width=0.45\linewidth]{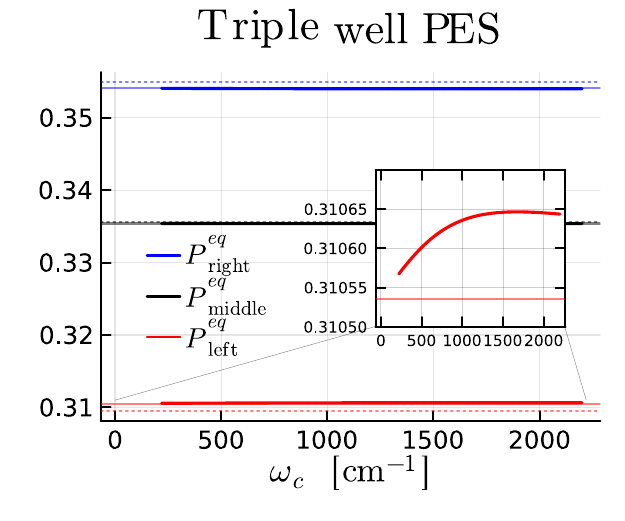}
    \caption{\textit{(left)} Visualization of the phonon-dressed vibrational states of the reaction coordinate according to the diagonalization of $\tau_\text{MF}$ {described} in Sec.~\ref{sec:MeanForceGibbsState}. \textit{(right)} Comparison of the population in the left (red), middle (black) and right (blue) wells at equilibrium. The dotted lines are the predictions from the Gibbs state $\tau\propto e^{-\beta H_R}$. The solid lines include the correction from the structured environment via Eq.~\ref{eq:PerturbativeMFG}, for which the horizontal line is for outside the cavity while the thicker line includes the cavity contribution.}
    \label{fig:TripleWellEquilibria}
\end{figure}
\begin{figure}[h]
    \centering
    \includegraphics[width=0.9\linewidth]{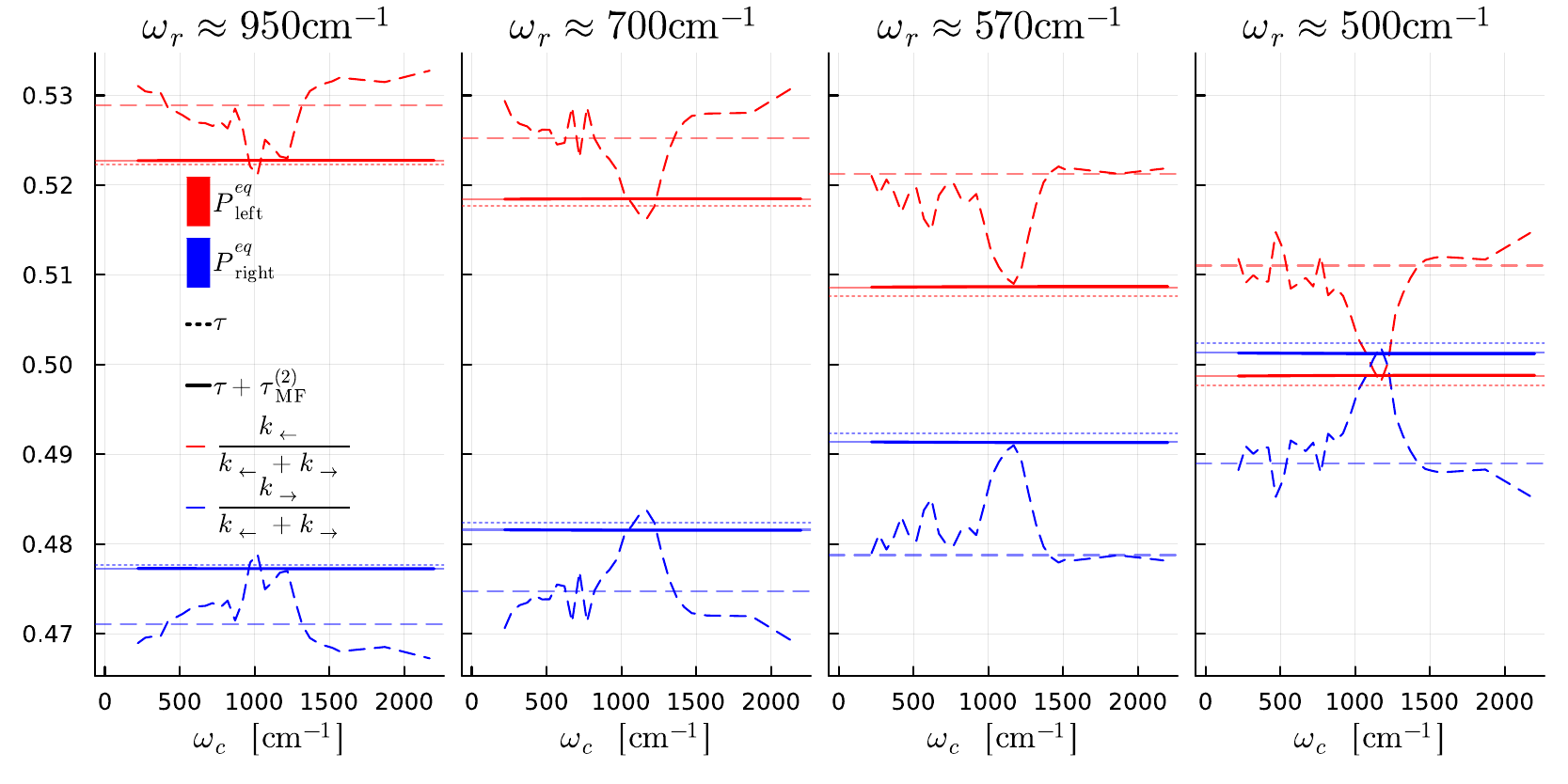}
    \caption{Comparison of the population in the left (red) and right (blue) wells at equilibrium for the double wells in Fig.~\ref{fig:asymmetric_rates}. The dotted lines are the predictions from the Gibbs state $\tau\propto e^{-\beta H_R}$. The solid lines include the correction from the structured environment via Eq.~\ref{eq:PerturbativeMFG}, for which the horizontal line is for outside the cavity while the thicker line includes the cavity contribution to the noise spectrum.
    The dashed lines represent the empirical prediction from the normalized reaction rates.}
    \label{fig:DoubleWellEquilibria}
\end{figure}

\subsection{Mean force Gibbs state for the double wells, Fig.~\ref{fig:asymmetric_rates}}
For the double well systems, we can compare the predictions of the mean force Gibbs state to those obtained from the dynamical simulation $P^{eq}_{\text{left}/\text{right}}=\frac{k_{\leftarrow/\rightarrow}}{k_\leftarrow + k_\rightarrow}$. This is plotted in {Fig.}~\ref{fig:DoubleWellEquilibria} alongside the distribution determined by the Gibbs state $\tau$ and its leading bath-induced correction $\tau_\text{MFG}^{(2)}$. The $\omega_r$ refers to the frequency of the right well and the ordering of the subfigures from left to right is exactly the same as in Fig.~\ref{fig:asymmetric_rates}.

Again, we see the same behavior that the narrow well is favored because it has lower potential energy once the wavefunction is localized. 

It is difficult to establish quantitative agreement to the numerical predictions via the reaction rates, because the numerical noise is magnified from the mutual division of $k_\leftarrow$ and $k_\rightarrow$. These numerical predictions display an $\omega_c$ dependence, which could be an artifact from this kind of short-time extrapolation to long-time asymptotics. For a proper comparison, one could consider obtaining the steady state directly from the dynamical equation, Eq.~\ref{eq:FPHEOM}.

\section{Kinetic and thermodynamic control of the triple well system}
\label{sec:KineticAndThermodynamicControl}
The triple well PES gives rise to reversible parallel reactions, which can be described by the rate equation:
\begin{equation}
\label{eq:ReversibleParallelRateEquation}
\begin{pmatrix}
    \dot{P}_l\\
    \dot{P}_0\\
    \dot{P}_r
\end{pmatrix}
=
\begin{pmatrix}
    -k_{l0}&k_{0l}& 0\\
    k_{l0} & -(k_{0l} + k_{0r}) & k_{r0}\\
    0 & k_{0r} & -k_{r0}
\end{pmatrix}
\begin{pmatrix}
    P_l\\
    P_0\\
    P_r
\end{pmatrix}
\end{equation}
with $P_0(t=0)=1$ and $k_{ij}:=k_{i\rightarrow j}$, i.e. this is the same rate equation as Eq.~\ref{eq:ThreeWellRateEquation}.
In order to preserve probability, one of the eigenvalues of the matrix {is} zero and the associated eigenvector is the steady state distribution:
\begin{equation}
    \lim_{t\rightarrow \infty} \begin{pmatrix}
        P_l(t)\\
        P_0(t)\\
        P_r(t)
    \end{pmatrix}
    \propto
    \begin{pmatrix}
        k_{0l}k_{r0}\\
        k_{l0}k_{r0}\\
        k_{l0}k_{0r}
    \end{pmatrix}
\end{equation}
Here, we see why we should not expect a modification of the steady state distribution from simply manipulating the left or right reaction rates, within the linear rate equation model. Because, if either $k_{0l}$ or $k_{0r}$ is scaled by a factor $\alpha$, the reverse {rates} $k_{l0}$, $k_{r0}$ {are} also scaled by $\alpha$ since they are linked by detailed balance. Then, every well distribution acquires the same scaling, which disappears when accounting for normalization.
A simple way to affect the equilibrium distribution, within this model, would be to change the temperature which affects the detailed balance condition and thus the equilibrium populations. The regime in which the product distribution is determined by these equilibrium populations, rather than by the rates of formation, is referred to as thermodynamic control \cite{anslyn2006modern}. 
Note that the above argument is somewhat circular, in that the equilibrium distributions determine the detailed balance condition, so if this is held-fixed, of course the equilibrium distributions will also remain fixed. Instead, Sec.~\ref{sec:MeanForceGibbsState} describes the principled way to study cavity-modified thermodynamics of the reaction coordinate, that does not rely on linear rate equations and simplified arguments.

Nonetheless, the timescales of how the distribution reaches the above steady state {are} determined by the remaining two negative eigenvalues $-\gamma_\pm$ of the rate matrix:
\begin{align}
\label{eq:ReversibleParallelReactionTimescales}
    \gamma_\pm &= \frac{K_l+K_r}{2} \pm \frac{\Delta}{2}\\
    \text{with}\quad K_l &= k_{0l} + k_{l0} \text{,} \quad K_r = k_{0r} + k_{r0}\\
    \text{and}\quad\Delta &=\sqrt{(K_l+K_r)^2 - 4(K_lK_r - k_{0l}k_{0r})}
\end{align}
In the irreversible limit $k_{l0}\approx0\approx k_{r0}$ the two decay rates become a slow $\gamma_-\approx 0$ and a fast mode $\gamma_+\approx k_{0l} + k_{0r}$. During the time $\mathcal{O}(\gamma_+^{-1})$ the middle well has depleted into both the left and the right well. The population in either well is determined by the branching ratio $P_l(t\sim \gamma_+^{-1}) \approx \frac{k_{0l}}{k_{0l}+k_{0r}}$, $P_r(t\sim \gamma_+^{-1}) \approx \frac{k_{0r}}{k_{0l}+k_{0r}}$ which constitutes the kinetic yield. The kinetic yield is straightforwardly affected by changes in the reaction rates. The regime in which the product distribution is governed by relative reaction rates rather than equilibrium populations, is generally referred to as kinetic control \cite{anslyn2006modern}.
Nonetheless, as long as the reaction is not completely irreversible ($k_{l0}\neq0\neq k_{r0}$), the distribution will eventually reach the steady state. Accounting for the non-zero backward rates, the kinetic yield becomes $P_l(t\sim \gamma_+^{-1}) \approx\frac{k_{0l}(\gamma_+-k_{r0})}{\gamma_+(\gamma_+-\gamma_-)}$, $P_r(t\sim \gamma_+^{-1}) \approx\frac{k_{0r}(\gamma_+-k_{l0})}{\gamma_+(\gamma_+-\gamma_-)}$.
\begin{figure}[h]
    \centering
    \includegraphics[width=1.0\linewidth]{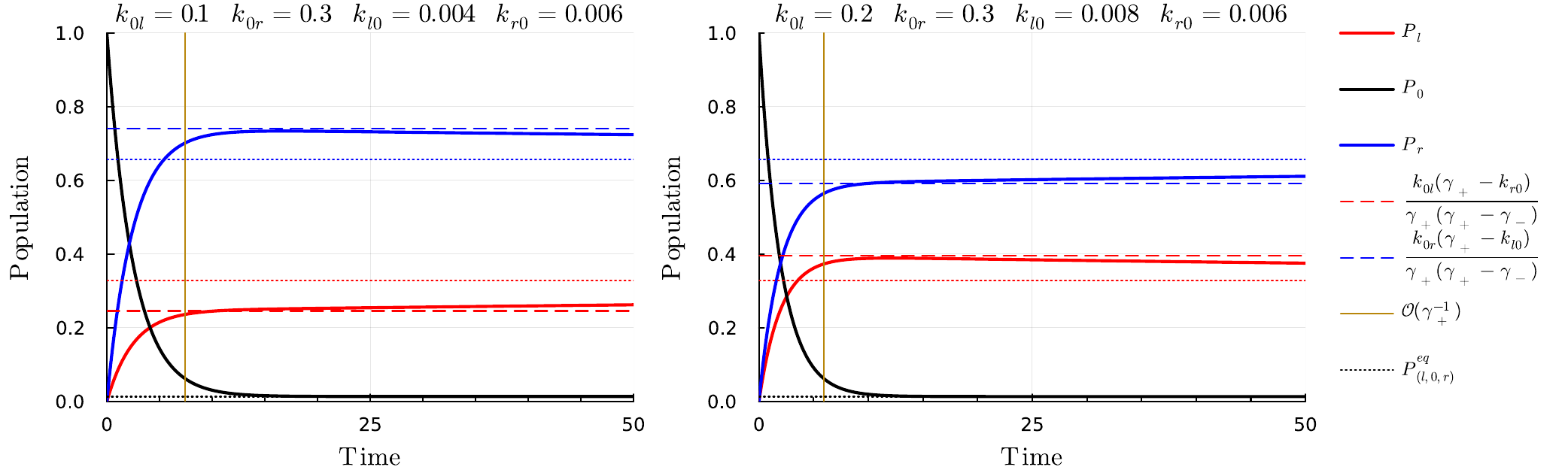}
    \caption{Toy example of the population dynamics for the triple well rate equation $P_l \overset{k_{0l}}{\underset{k_{l0}}{\leftrightharpoons}} P_0\overset{k_{0r}}{\underset{k_{r0}}{\rightleftharpoons}}P_r$ with dimensionless units.
    The dotted lines are the equilibrium distributions while the dashed lines are the kinetic yield.}
    \label{fig:ReversibleParallelReactionExamples}
\end{figure}
Figure~\ref{fig:ReversibleParallelReactionExamples} shows two examples of the dynamics of the population for the reversible parallel rate equation, Eq.~\ref{eq:ReversibleParallelRateEquation}. The horizontal lines represent the equilibrium distribution (dotted) and the kinetic yield (dashed).
 The kinetic yield best {describes} the distribution of the products in the intermediate time window  after the fast transient and before appreciable reverse redistribution $\mathcal{O}(\gamma_+^{-1})\lesssim t \lesssim\mathcal{O}(\gamma_-^{-1})$, with $\gamma_\pm$ defined in Eq.~\ref{eq:ReversibleParallelReactionTimescales}. 
 The kinetic yield is affected by the doubling of the rates between the left well and the reactant well whereas the equilibrium distribution remains unaffected. In both examples, the populations slowly evolve towards the equilibrium distribution.

\section{The phonon bath and Kramers turnover}
\label{sec:KramersTurnover}
The phonon spectral density is the same for all simulations. It is of the Drude Lorentz form with 
\begin{equation}
    J_R^{(DL)}(\omega) = \frac{2\lambda_R\gamma_R\omega}{\gamma_R^2+\omega^2}
    \quad\quad
    \lambda_R=\frac{\eta_\nu M \gamma_R}{2}\times1000\,\text{cm}^{-1} \approx \eta_\nu\times 0.00381157\,\text{a.u.}
    \quad\quad
    \gamma_R=200\,\text{cm}^{-1}
\end{equation}
The reaction rate for the symmetric double well coupled to the phonon bath{was} calculated for different {dimensionless }coupling strengths $\eta_\nu$, as seen in Fig.~\ref{fig:KramersTurnover}.
\begin{figure}[h]
    \centering
    \includegraphics[width=0.5\linewidth]{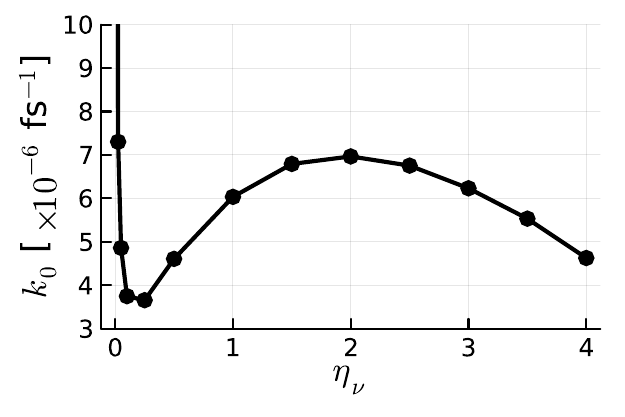}
    \caption{
    Reaction rates for the symmetric double well coupled to the phonon bath for different coupling strengths $\eta_\nu$. For all other simulations $\eta_\nu=0.1$ was used (third visible dot from the left).    
    }
    \label{fig:KramersTurnover}
\end{figure}
In the limit of vanishing $\eta_\nu$ the state of the system would oscillate between the wells at a rate $\sim600\times10^{-6}\,\text{fs}^{-1}$. By increasing $0<\eta_\nu<0.25$, the coupling to the bath greatly suppresses the tunneling. In this regime, the reaction is limited by the sparse energy transactions with the environment, so by increasing further $0.25<\eta_\nu<2$, the bath helps drive the transition until the crossover $\eta_\nu\approx2$. From this point, the reaction rate decreases due to the increased localization from strongly coupling to the environment.


\subsection{Tensor network simulations in the strong coupling regime}
In obtaining the above results, we looked into different tensor network techniques to alleviate the numerical hurdles that arise in the strong coupling regime $3\lesssim \eta_\nu$. The primary obstacles are that the cutoff of the auxiliary modes need to be increased together with the bond-dimension of the tensor network ansatz. 

In an attempt to compress the local Hilbert space dimensions, we applied the Optimized Boson Basis or Local Basis Optimization approach \cite{PhysRevLett.80.2661}.
In short, the three legged tensors of the MPS, which all have high local dimension in this scenario, are decomposed such that the physical index is separated from the two virtual bond indices. This introduces an intermediate index that connects the two tensors, and the goal is to substantially compress this new bond, such that the computational cost of the MPS representation is reduced.
\begin{center}
\includegraphics[width=0.3\linewidth, trim=1cm 3cm 1cm 0cm, clip]{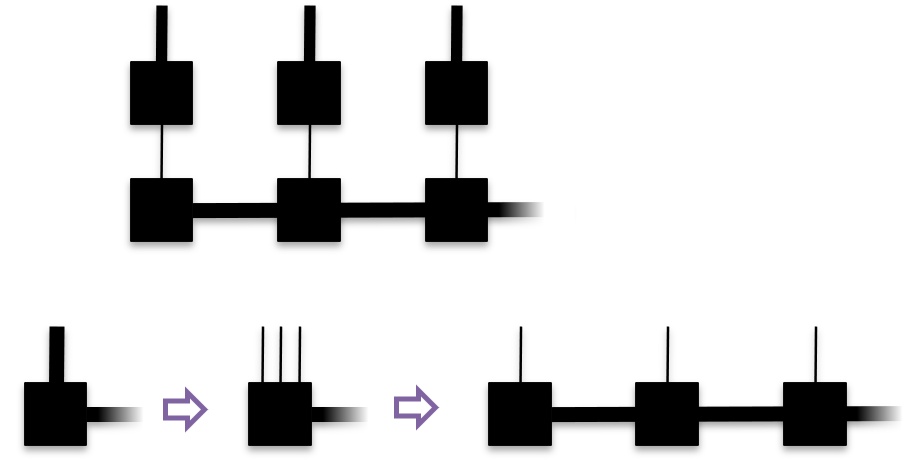}
\end{center}
However, our experience is that such a compression is not achievable for this scenario, and that the bond that connects the two new tensors had to be too large to render the approach useful.

Another approach we tested is the pseudosite method \cite{PhysRevB.57.6376}. The bosonic occupation number is represented in binary form, such that a bosonic degree of freedom with a local cutoff of e.g. $64=2^6$ is mapped exactly onto six two-dimensional pseudosites. The joint state of these six pseudosites therefore encodes the Fock state of the original bosonic mode. Bosonic operators can be represented in terms of operators acting on the pseudosites, allowing the original bosonic site to be treated within a standard tensor-network framework. By distributing the original large local Hilbert space over several two-dimensional sites, standard tensor-network compression can be applied to the resulting pseudosite representation, potentially reducing the computational cost.
\begin{center}
\includegraphics[width=0.3\linewidth, trim=0cm 0cm 0cm 5cm, clip]{figures/SI/StrongCouplingTNapproaches.png}
\end{center}
Doing this for each bosonic degree of freedom, we again experienced that this did not reduce the computational cost because the resulting tensor network could not compress enough while still keeping the necessary accuracy.

In the end, we simply simulated the strong-coupling scenarios using 1-site TDVP with GPU acceleration \cite{rudolph2025simulatingsamplingquantumcircuits}.
Methods such as the reaction coordinate mapping (see Ref.~\citenum{Nazir2018}) might also prove {useful} for simulating the dynamics of such scenarios.


\section{Additional asymmetric double well probability currents}
Figure~\ref{fig:all_asymmetric_currents} provides a complete overview of the probability currents for the systems shown in Fig.~\ref{fig:asymmetric_rates}.
\begin{figure}[h]
    \centering
    \includegraphics[width=0.495\linewidth, trim={0.3cm 0cm 0.8cm 0cm}, clip]{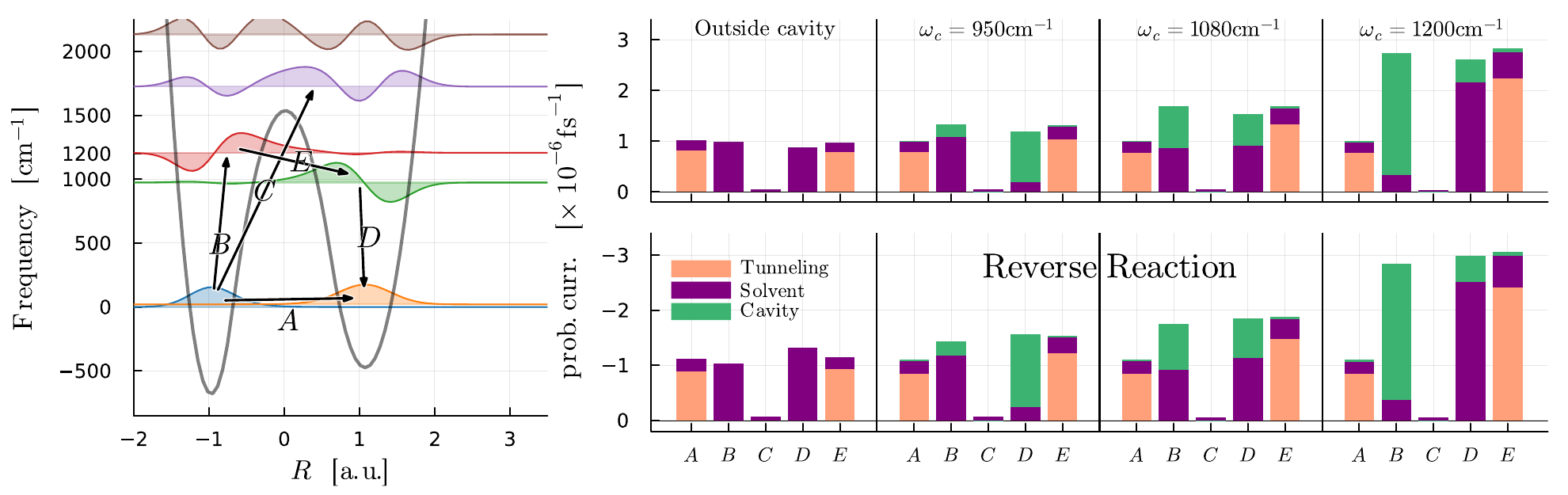}
    \hfill
    \includegraphics[width=0.495\linewidth, trim={0.3cm 0cm 0.8cm 0cm}, clip]{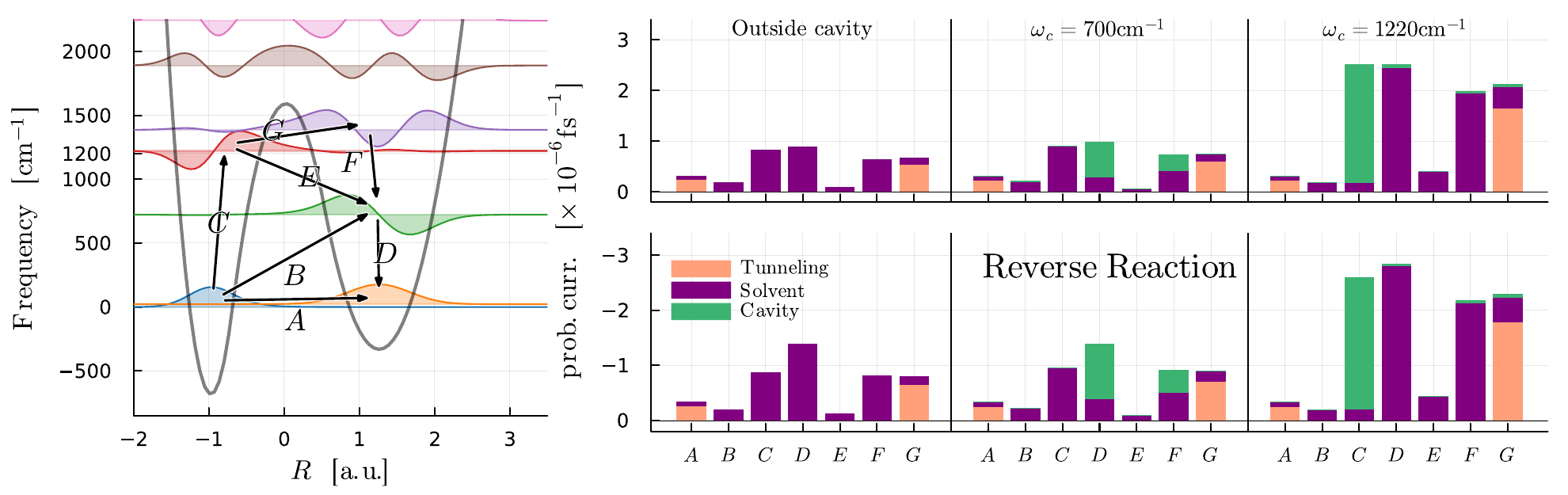}
    \vfill
    \includegraphics[width=0.495\linewidth, trim={0.3cm 0cm 0.8cm 0cm}, clip]{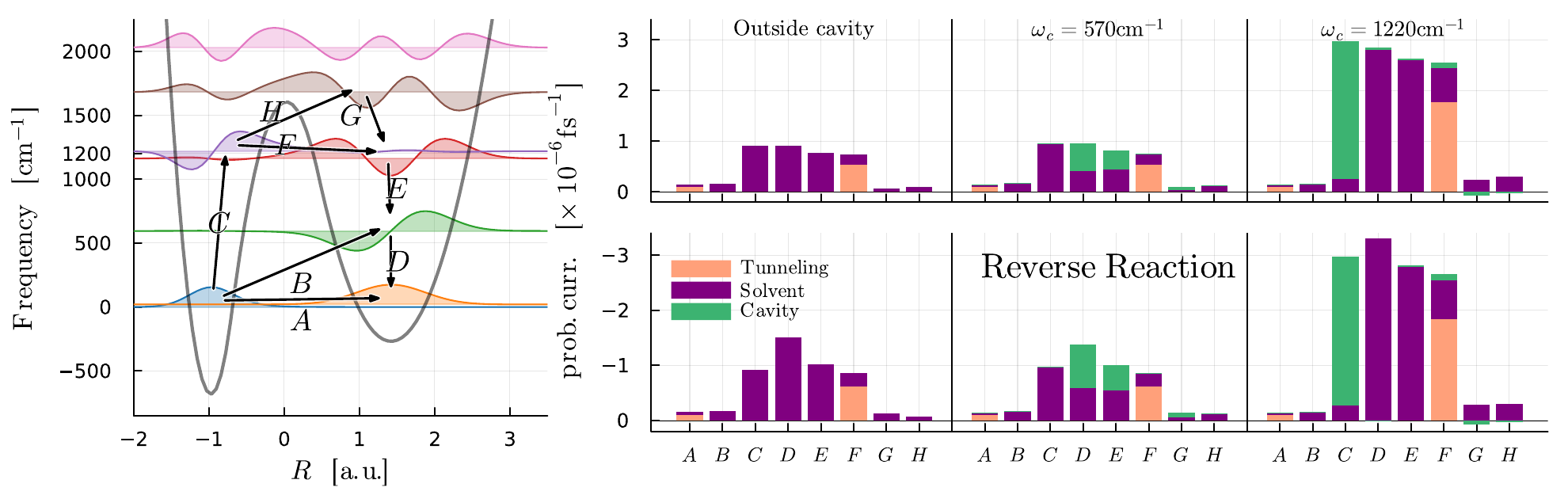}
    \hfill
    \includegraphics[width=0.495\linewidth, trim={0.3cm 0cm 0.8cm 0cm}, clip]{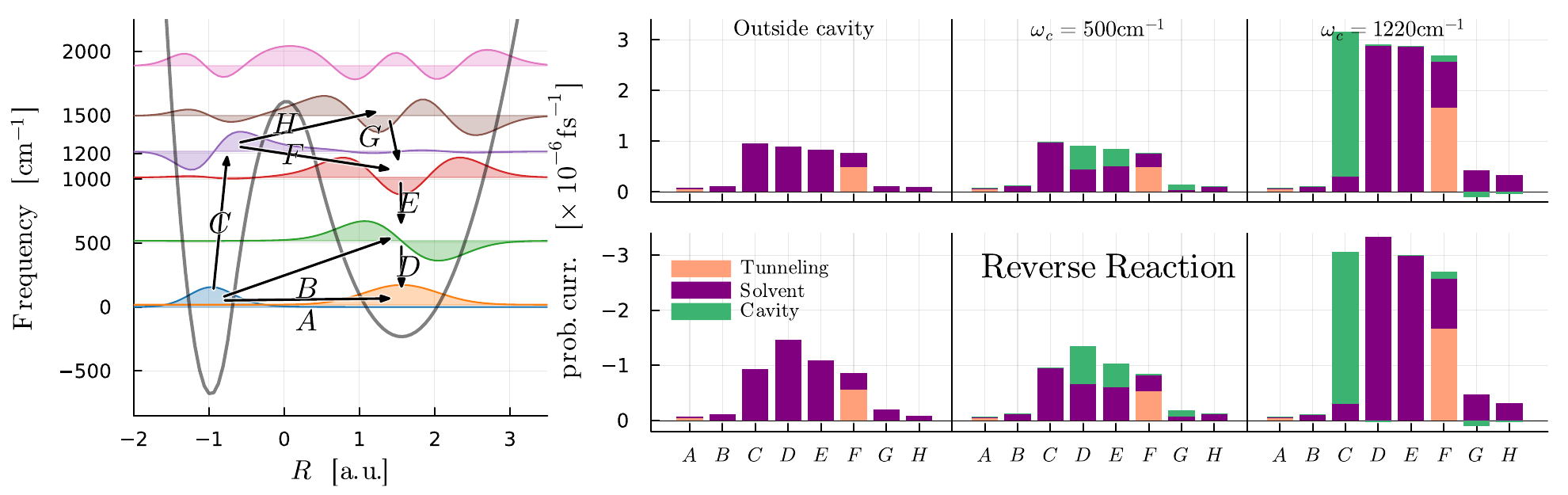}
    \caption{Probability currents for the asymmetric double wells. A negative value for the probability current means that the current goes in the opposite direction of the arrow. All currents with magnitude larger than $0.05\times10^{-6}\,\text{fs}^{-1}$ are shown.}
    \label{fig:all_asymmetric_currents}
\end{figure}






\section{Overview of parameter values}
\begin{table}[h!]
    \centering
    \begin{tabular}{lc}
        Description & Value range\\
        \hline
        \hline
        \multicolumn{2}{c}{\textbf{Simulation parameters: }} \\
        \hline
        \hline
        System Hilbert space dimension & 12 to 30 \\
        Fock space cutoff &  4 to 75 \\
        Maximum value of $\Gamma\tau$ (for the Markovian compression, Sec.~\ref{sec:EfficientMarkovCorrection}) & 0.01 with typical values of $\mathcal{O}(0.001)$ \\
        bond dimension & 4 to 600 \\
        Tolerance of the Krylov update for the local propagation in TDVP & 1e-8\\        
        Time step & 0.25fs\\
    \end{tabular}
    \caption{{Additional} simulation parameters. Convergence has been carefully checked for these. The extra large Fock space cutoff and bond dimension was mainly used for the strong coupling scenarios in Sec.~\ref{sec:KramersTurnover}.
    Additional information: The probability currents quoted throughout the figures are $j_{n\rightarrow m}(t)$ evaluated at $t=6\,\text{ps}$. The initial state of the reaction coordinate is the localized ground state in the reactant well and the total simulation time was $10\,\text{ps}$.
    }
\end{table}

\begin{table}[h!]
    \centering
    \begin{tabular}{llcc}
        \hline
        Symbol & Description & Value & Unit \\
        \hline
        \multicolumn{4}{c}{\textbf{Reaction Coordinate PES:}} \\
        \hline
        \hline
        \multicolumn{4}{c}{\textbf{Symmetric Double Well}} \\
        \hline
        $M$             & Effective mass                 & 1836& a.u. \\
        $E_b$           & Barrier height                 & 2120& cm$^{-1}$ \\
        $\omega_0$           & Barrier frequency                 & 866.020& cm$^{-1}$ \\
        $\omega_l=\omega_r$           & Reactant and product well frequency & 1414.177& cm$^{-1}$ \\
        \hline
        \multicolumn{4}{c}{\textbf{Asymmetric Double Well}} \\
        \hline
        $E_b$           & Barrier height                 & 2120& cm$^{-1}$ \\
        $\omega_0$           & Barrier frequency                 & 866.020& cm$^{-1}$ \\
        $\omega_l$           & Reactant well frequency & 1414.177& cm$^{-1}$ \\
        $\omega_r$           & Product well frequency & \{999.914, 707.046, 577.301, 500.077\}& cm$^{-1}$ \\
        \hline
        \multicolumn{4}{c}{\textbf{Triple Well}} \\
        \hline
        $E_{ll}$           & Left well offset                 & 960.660& cm$^{-1}$ \\
        $E_l$           & Left barrier height                 & 2120& cm$^{-1}$ \\
        $E_0$           & Middle well offset                 & 426.963& cm$^{-1}$ \\
        $E_r$           & Right barrier height                 & 2120& cm$^{-1}$ \\
        $E_{rr}$           & Right well offset                 & 266.695& cm$^{-1}$ \\
        $\omega_{ll}$           & Left well frequency                 & 2121.320 & cm$^{-1}$ \\
        $\omega_l$           & Left barrier frequency                 & 866.02 & cm$^{-1}$ \\
        $\omega_0$           & Middle well frequency                 & 853.927& cm$^{-1}$ \\
        $\omega_r$           & Right barrier frequency                 & 866.02 & cm$^{-1}$ \\
        $\omega_{rr}$           & Right well frequency                 & 533.389 & cm$^{-1}$ \\
        \hline
        \hline
        \multicolumn{4}{c}{\textbf{Harmonic Environment: }} \\
        \hline
        \hline
        $T$             & Temperature                    & 300& K \\
        $\eta_c$        & Light-matter interaction strength & 0.1& --- \\
        $\tilde{\eta}_R$& Reaction coordinate-cavity coupling& 1.0&---\\
        $\tilde{\eta}_Q$& Spectator mode{-}cavity coupling& 0.1&$\text{a.u.}$\\
        $c_Q$& Reaction coordinate-spectator mode coupling& $2.2326\times10^{-5}$ & a.u.\\
        $\omega_Q$& Spectator mode frequency& 1170&$\text{cm}^{-1}$\\
        \hline
        \multicolumn{4}{c}{\textbf{Cavity loss parameters}} \\
        \hline
        $\tau_c$        & Cavity lifetime                & 100& fs \\
        $\alpha$& Cavity Markovian parameter& $(1-e^{-\beta\omega_c})/\tau_c$& $\text{fs}^{-1}$\\
        \hline
        \multicolumn{4}{c}{\textbf{Reaction coordinate Drude Lorentz bath}} \\
        \hline
        $\gamma_R$    & Characteristic frequency & 200& cm$^{-1}$ \\
        $\lambda_R$& Reorganization energy & 83.654&$\text{cm}^{-1}$\\
        \hline
        \multicolumn{4}{c}{\textbf{Spectator mode Drude Lorentz bath}} \\
        \hline
        $\lambda_Q$& Reorganization energy& 0.147&$\text{cm}^{-1}$\\
        $\gamma_Q$& Characteristic frequency & 6000&$\text{cm}^{-1}$\\
        \hline
        \multicolumn{4}{c}{\textbf{Bath decomposition hyper-parameters}} \\
        \hline
        $[{\omega_l}, {\omega_h}]$& Frequency window & [-0.99,1]&a.u.\\
        Tolerance & Criterion for convergence & below $10^{-10}$&---\\
        Lawson iterations & post-processing step & 20 &---\\ 
        \hline
        \multicolumn{4}{c}{\textbf{Parameters used for calculating eq.~\ref{eq:FGR}}} \\
        \hline
        $k_0$& Phonon-only reaction rate & 3.79442657 & $10^{-6}\text{fs}^{-1}$\\ 
        $|\langle e|\hat{R}|g\rangle|^2$& Transition matrix element & 0.215 & $\text{a.u.}$\\
        Prefactor $\frac{1}{2}\rightarrow0.38$& \textit{Post hoc} rescaling coefficient & 0.38 & -\\
        $\omega_0$& Transition frequency in $\mathcal{A}(\omega)$ & 1205 & $\text{cm}^{-1}$\\
        $\Gamma$& Broadening in $\mathcal{A}(\omega)$ & 30 & $\text{cm}^{-1}$\\
    \end{tabular}
    \caption{Overview of simulation parameters.
    Note that $\omega_l$, $\omega_0$ and $\omega_r$ in the main text {refer} to the energy spacing between states of the associated well. Here, they refer to the curvature of the parabolas that define the potential energy surface in Sec.~\ref{sec:RxNPES}.
    }
\end{table}

\stopcontents[appendix]

\clearpage
\end{widetext}

\end{document}